\documentclass[12pt,dvipsnames]{article}

\usepackage[
  top=0.9in,
  bottom=0.9in,
  left=0.9in,
  right=0.85in
]{geometry}
 
\usepackage{lipsum}
\usepackage{amsmath, amssymb, amsthm, mathtools}
\usepackage{physics}
\usepackage{graphicx}
\usepackage{romannum}
\usepackage{url}
\usepackage{slashed}
\usepackage{bbm}
\usepackage[numbers]{natbib}
\usepackage{tensor}
\usepackage{mathrsfs}
\usepackage{braket}
\usepackage{authblk}

\usepackage{array}
\usepackage[colorlinks=true,allcolors=blue]{hyperref}
\setcitestyle{square}

\usepackage{orcidlink}

\makeatletter
\patchcmd{\@citex}{;}{,}{}{}
\makeatother

\usepackage{booktabs}
\usepackage{tikz}

\usetikzlibrary{arrows.meta}

\usepackage{eufrak}
\usepackage{accents}

\usepackage{comment}

\usepackage[stretch=10,shrink=10,step=1]{microtype}

\usepackage{tocloft}

\author[1]{P Samuel Wesley\orcidlink{0009-0005-4708-1463}}
\author[2]{Tejinder P. Singh\orcidlink{0000-0002-1862-1505}}
\author[3]{Jos\'e M. Isidro\orcidlink{0000-0002-0720-9945}}
\affil[1]{{\it Department of Theoretical Physics, University of Madras-Alumnus, Chennai 600025, India}; samwesprem7@gmail.com}
\affil[2]{{\it Tata Institute of Fundamental Research, Mumbai 400005, India}; tpsingh@tifr.res.in}
\affil[3]{{\it Instituto Universitario de Matem\'atica Pura y Aplicada, Universidad Polit\'ecnica de 
Valencia, Valencia 46022, Spain};
joissan@mat.upv.es}

\title{Gravity and electroweak sector from symmetry breaking of an $SO(3,3)$ BF theory}

\allowdisplaybreaks 

\usepackage{lmodern}
\usepackage{microtype}

\newcommand{\Autad}{\mathrm{Aut}_{\mathrm{adm}}}
\newcommand{\Derad}{\mathrm{Der}_{\mathrm{adm}}}
\newcommand{\Diff}{\mathrm{Diff}}

\newcommand{\Lie}{\mathcal{L}}

\newtheorem*{principle}{Principle}
\newtheorem*{conjecture}{Conjecture}

\begin{document}
\maketitle
\pagenumbering{arabic}

\begin{abstract} 
\noindent An $SO(3,3)$ BF-type gauge theory is formulated on a six-dimensional spacetime of split signature $(3,3)$, interpreted as the pre-electroweak-symmetry-breaking phase. A MacDowell--Mansouri-type symmetry breaking to $SU(2)\times SU(2)$ is implemented, and the corresponding stabilizer and coset structures are computed. The curvature decomposes into chiral sectors, and effective tetrads are introduced using components of the higher-dimensional connection. The resulting left and right sectors are formulated as constrained BF/Plebanski-like theories with appropriate simplicity and reality conditions. The six-dimensional theory yields two overlapping four-dimensional Lorentzian sectors of opposite signature, related via gluing constraints across their intersection. 
In the first sector, the selfdual two-forms ($\Sigma^{(+)}$) satisfy simplicity constraints that select the non-degenerate branch and reproduce Einstein gravity.
Subsequently, the $SU(2)_R\times U(1)_{Y{\rm dem}}\to U(1)_{\rm dem}$ breaking pattern is outlined which admits an ultra-soft regime consistent with current phenomenological bounds under sufficiently suppressed couplings. 
In the second sector, the antiself dual two-forms  ($\Sigma^{(-)}$) satisfy analogous simplicity constraints, realizing weak gauge dynamics as gravity on the opposite-signature sector. Subsequently, the $SU(2)_L\otimes U(1)_Y$ electroweak symmetry is realized within the Yang--Mills branch of the BF theory which incorporates the standard Higgs mechanism $SU(2)_L\otimes U(1)_Y \to U(1)_{\mathrm{EM}}$, recovering the conventional electroweak $W^\pm$, $Z$, and photon spectrum. 
The findings indicate that gravitation and the weak interaction may result as distinct low-energy phases of a common $SO(3,3)$ gauge theory. The construction provides a constrained setting wherein general relativity appears as an emergent classical branch of an underlying renormalizable gauge theory. From a phenomenological perspective, the long-range $U(1)_{\rm dem}$ and ultra-soft right-handed vectors place the model under strong equivalence-principle and fifth-force constraints, offering concrete avenues for experimental scrutiny.
The proposed framework addresses the asymmetries present in earlier gravi–weak models by treating both $SU(2)$ chiral sectors symmetrically at the six-dimensional level and assigning a geometric interpretation to weak isospin while remaining compatible with known low-energy physics.
\end{abstract}

\tableofcontents
\section{Introduction}
\subsection*{Background and Motivation}
\noindent Gravitation and the gauge interactions of the Standard Model are formulated within strikingly different theoretical frameworks, and understanding their possible common origin remains a persistent open problem in fundamental physics. General relativity (GR) provides an exceptionally successful classical description of gravitation as spacetime geometry; however, its naive quantization yields a perturbatively non-renormalizable theory, indicating that Einstein gravity should be regarded as an effective field theory valid below the Planck scale. By contrast, the Standard Model is a renormalizable quantum gauge theory based on internal symmetries, with the electroweak sector exhibiting spontaneous symmetry breaking via the Higgs mechanism. The conceptual divide between these frameworks is further accentuated by the hierarchy between the electroweak scale ($\sim10^{2}\; \mathrm{GeV}$) and the Planck scale ($\sim10^{19}\; \mathrm{GeV}$), as well as by the absence of experimentally accessible probes of quantum gravitational phenomena. These considerations motivate intermediate approaches: rather than pursuing a complete unification of all interactions at once, one may seek partial unifications that elucidate structural connections between gravity and gauge theories, clarify the origins of chirality and symmetry breaking, and provide a more constrained framework for quantization. Such approaches aim to identify common geometric or algebraic principles that may underlie both gravitation and selected sectors of the Standard Model.

A particularly suggestive avenue is the striking similarities between gravitation and weak interactions, which has prompted several researchers to explore if these two interactions may result from a unified underlying framework. First, the weak interaction is intrinsically parity-violating: only left-handed fermions transform nontrivially under $SU(2)_L$. Because parity is a symmetry of spacetime itself, it is conceptually striking that an internal Yang--Mills gauge symmetry depends on parity, unless  it is, in some sense, also a space-time symmetry. Second, like gravity, the weak force is universal in the sense that it acts on all quark and lepton families (albeit chirally), in contrast with QCD (which acts solely on quarks) and electromagnetism (which acts only on electrically charged particles). A third suggestive similarity is that both gravity and the Fermi four-point interaction are characterized by dimensionful coupling constants. This is unlike the coupling constants for the unbroken symmetries of strong force and electromagnetism, namely $SU(3)_{\rm color}$ and $U(1)_{\rm em}$, which are dimensionless.  The gravitational constant $G_N= \hbar c / m_{Pl}^2$ has dimensions of squared length (in natural units $\hbar=c=1)$, whereas the Fermi constant $G_F$  of Fermi's four-point weak interaction theory is given by $G_F/(\hbar c)^3 = \sqrt2  g^2/8m_W^2 c^2$. Here $m_W$ is the mass of the $W$ boson, while $g$ is the weak isospin, and hence dimensionless, implying that $G_F$ has dimensions of squared length, just like $G_N$. In natural units, $G_F/G_N\sim m_P^2/m_W^2$, suggesting that a cosmological scaling down (in an inflationary epoch) of the Higgs mass, from its original Planck scale value, could be responsible for the mass hierarchy---we have provided evidence for this idea in a prior work \cite{Singh:2023fpb}. Furthermore, the fact that both $G_F$ and $G_N$ are dimensionful makes the Fermi theory as well as GR perturbatively
non-renormalisable. This parallel reinforces the perspective that both GR and the weak interaction are low-energy effective field theories, emerging from a more complete high-energy description. 

In contrast with these similarities, the stark differences between gravity and the weak interaction are significant. First, the weak force is short range ($\sim 10^{-17}$ cm) and is mediated by massive spin-1 gauge bosons, whereas gravity is long range and is mediated by a massless spin-2 field. Second, there is no experimental evidence yet that gravitation is a parity violating interaction. Third, GR is formulated as the (pseudo-) Riemannian geometry of spacetime, whereas the electroweak interaction is a Yang-Mills gauge theory on an internal bundle over the underlying 4D spacetime.

These contrasts do not preclude a unified description at a deeper level. One promising perspective is to regard gravity itself as a broken phase of an enlarged gauge theory: one can package coframe and spin connection into a single connection and recover Einstein dynamics after a symmetry reduction. In such a setting, different components of the enlarged connection can naturally exhibit different infrared behavior: one sector yields long-range gravitational dynamics, whereas another survives as a Yang--Mills field whose short range is governed by a mass gap generated in the broken phase. Therefore, the conventional graviweak problem is reformulated as a symmetry-breaking question: find an enlarged gauge structure whose broken phase reproduces GR in one sector and weak isospin in another.

The present work pursues a similar strategy and aims to demonstrate that the weak interaction and gravitation described by general relativity can be unified into one BF theory on a six dimensional spacetime with signature (3,3). This space-time describes the evolving universe prior to the electroweak symmetry breaking.

\subsection*{Literature Survey}

 Attempts to relate gravitation and weak interactions predate the modern gauge-theoretic graviweak program and resulted in several conceptually distinct lines of inquiry. One of the earliest attempts was by Jordan \cite{Jordangwa, Jordangwb}, who also explored constructions based on a 6D spacetime with (3,3) signature. Subsequently, closely related ideas appear in Kaluza--Klein-type approaches, wherein one requires two additional dimensions \cite{Kerner} beyond 4D. However, these additional dimensions typically have been assumed to be space-like, because the internal degrees of freedom must encode the appropriate group manifold. From this perspective, the dimensionality of the internal space is not arbitrary: for the $SU(2)$ group, the simplest homogeneous space is the 2-sphere $S^2\simeq CP^1$, with $dim = 2$. For $SU(3)$, one encounters higher-dimensional complex projective spaces such as $CP^2$, with $dim = 4$. In general, the $SU(n)$ acts on the non-zero vectors of $C^n$ up to the phase (or the center of $SU(n)$) then we get $[C^n-\{0\}]/U(1)$
which is the complex projective space $CP^{n-1}$.
A different, though related, motivation was articulated by Onofrio \cite{Oa, Ob}, who proposed the possibility of weak interaction as a short distance manifestation of gravity. Furthermore, Chisholm and Farwell \cite{chisholm1987electroweak, chisholm1989unified} proposed a spin gauge theory based on the real Clifford algebra $Cl(2,6)$ wherein Lorentz and electroweak symmetries are unified within an eight-dimensional framework. In their construction, the vierbein emerges dynamically from the algebraic structure, and fermions obey a generalized Dirac equation coupled to both gravitational and electroweak gauge fields.

The modern graviweak literature originates from the observation that chiral couplings of fermions to gravity naturally single out one $SU(2)$ subalgebra of the complexified Lorentz algebra. 
In this spirit, Nesti and Percacci proposed a ``bottom-up'' construction in which the weak $SU(2)$
is related to (or embedded into) the anti-self-dual part of the complexified Lorentz algebra, and the
combined symmetry can be organised into a gravi--weak group such as $SO(4,\mathbb{C})$ (in conjunction with
extensions that accommodate realistic fermion content and larger unification groups)
\cite{nesti2008gravi,nesti2009standard}.

\noindent {\it Woit's Euclidean twistor unification.}
A conceptually distinct, but ``gravi--weak-like'', proposal has been advanced by Woit
in the context of Euclidean twistor geometry \cite{Woit:2021EuclideanTwistorUnification}.
The starting point is four-dimensional \emph{Euclidean} signature spacetime, with local
rotation group $Spin(4)\simeq SU(2)_L\times SU(2)_R$, treated as fundamental.
In this framework, gauging the $SU(2)_R$ factor yields a chiral spin-connection formulation
of gravity, while gauging the $SU(2)_L$ factor yields the $SU(2)$ gauge fields of the weak
interactions. Reconstructing a Lorentz-signature theory by analytic continuation requires
introducing an additional degree of freedom that selects an ``imaginary time'' direction;
this field can be identified with the Higgs field and simultaneously provides a mechanism
to treat one $SU(2)$ factor as spacetime symmetry and the other as an internal symmetry.

Woit further argues that these structures are naturally encoded in Euclidean twistor geometry:
after conformal compactification $\mathbb R^4\to S^4\simeq \mathbb H\mathbb P^1$, the relevant arena
is the projective twistor space $PT=\mathbb{C}P^3$, a $\mathbb{C}P^1$ bundle over $S^4$.
A point in the $\mathbb{C}P^1$ fibre corresponds to a choice of complex structure on the tangent
space, and the Higgs field selecting the imaginary-time direction is naturally taken as a field
living on $PT$. A notable feature of this construction is that $\mathbb{C}P^3$ comes equipped with
internal $U(1)\times SU(3)$ symmetry at each point, providing the remaining Standard Model internal
symmetries in a geometric way; correspondingly, the relevant gauge bundles and connections are
taken to live on twistor space rather than directly on spacetime.

Although technically and conceptually different from BF/Plebanski-type unification mechanisms, this Euclidean-twistor perspective shares with gravi--weak programmes the central theme that chirality and the special factorisation properties of four-dimensional geometry can allow one $SU(2)$ to be interpreted as gravitational/spacetime structure and the other as the weak internal gauge symmetry. By contrast, in the present work, we begin from a six-dimensional split-signature BF theory and obtain two Lorentzian spacetime sectors via symmetry breaking; nevertheless, Woit's proposal highlights an alternative route in which the Higgs degree of freedom is tied directly to the choice of time direction and to the Euclidean--Lorentzian transition.

A complementary perspective is to treat gravity itself as a gauge theory in a Higgs/broken phase,
and to seek unified actions whose symmetry breaking simultaneously generates an effective soldering form (vierbein/tetrad) and Yang--Mills sectors.

A broad class of such models is naturally expressed in BF \cite{celada2016bf} or Plebanski formulations \cite{plebanski1977separation}.
Smolin showed that a simple extension of the Plebanski action to a larger gauge group can yield a unified
framework for gravity and Yang--Mills fields after symmetry breaking \cite{Smolin2009PlebanskiYM}.
Lisi, Smolin and Speziale further developed an action principle in which gravity, gauge fields and Higgs
degrees of freedom result from a single enlarged connection with an appropriate breaking pattern
\cite{LisiSmolinSpeziale2010Unification}.
Related gravi--weak constructions incorporating the weak $SU(2)$ fields directly to a chiral part of the
spacetime connection were explored in \cite{AlexanderMarcianoSmolin2014Chirality},
and Spin$(3,3)$/Spin$(4,4)$-based gravi--weak models within extended Plebanski frameworks were analyzed in
\cite{DasLaperashviliTureanu2013Graviweak}. More recent developments in graviweak unification include Konitopoulos et al. \cite{konitopoulos2024unification}, Alexander et al. \cite{alexander2025spontaneous, alexander2025gravigut}, and Kurihara \cite{kurihara2025yang}.
For a wider synthesis of four-dimensional (non-Kaluza--Klein) unification attempts, refer to the topical review by Krasnov and Percacci \cite{KrasnovPercacci2018Review}.

Despite the conceptual appeal of existing gravi-weak models, some limitations persist. Many constructions remain strictly four-dimensional and must treat the two chiral $SU(2)$ sectors asymmetrically, often assigning geometric significance to one while suppressing or constraining the other in an ad hoc manner. In higher-dimensional approaches, additional dimensions are typically assumed to be purely spacelike, and the geometric interpretation of weak interactions is not always placed on the same footing as gravity. Moreover, the mechanisms by which distinct spacetime sectors, if present, might interact or remain compatible are often left implicit.

\subsection*{Contributions of the study}
Motivated by the gaps identified above, the present work develops a gravi-weak unification framework that advances beyond existing approaches in several key respects. To the best of our knowledge, this is the first construction wherein gravitation and the weak interaction acquire a genuine geometric interpretation as external symmetries, formulated as explicitly resulting from a single BF-type gauge theory on a six-dimensional spacetime of split signature $(3,3)$.
The primary contributions and objectives of this study can be summarized as follows:

\begin{enumerate}
    \item To address the asymmetric treatment of chiral $SU(2)$ sectors in four-dimensional models, we formulate a unified BF/MacDowell--Mansouri framework with gauge group $SO(3,3)$ on a split-signature $D=6$ spacetime. This framework provides a symmetric starting point for both gravitational and the weak interaction degrees of freedom.
    \item Building on chiral formulations of gravity, we implement an explicit symmetry breaking $SO(3,3)\to SU(2)\times SU(2)$ and compute the corresponding stabilizer/coset structures  required for MacDowell--Mansouri-type curvature decompositions.
    \item Advancing beyond prior four-dimensional frameworks, we derive a two-sector framework from a $D=6$ spacetime, which results in two overlapping $D=4$ spacetime sectors of opposite Lorentz signatures. One sector reproduces the familiar gravitational spacetime described by self-dual variables (Region $\mathcal{I}$) and the second sector supports an $SU(2)_{L}$ weak sector described by anti-self-dual variables (Region $\mathcal{II}$).
    \item Unlike earlier approaches in which weak interactions remain purely internal, we articulate a gravity-like geometric interpretation for the weak interaction: the anti-self-dual $SU(2)$ connection on the opposite-signature spacetime sector is identified with weak isospin, placing gravity and the weak interaction on parallel geometric footing.
    \item To relate the two spacetime sectors without collapsing them into a single metric geometry, we introduce interface or gluing constraints that consistently relate the two sectors along their overlap. 
\end{enumerate}
These contributions collectively establish a coherent framework in which the weak interaction and gravitation emerge from a common six-dimensional gauge structure, directly addressing the conceptual and technical gaps highlighted in the literature. The realization of these results is developed explicitly in the subsequent sections.

Technically, we start from an $SO(3,3)$ connection and implement a MacDowell--Mansouri-type symmetry
breaking to two $SU(2)$ sectors, corresponding to ``right''
and ``left'' components.
We compute the corresponding curvature decomposition, insert the non-vanishing stabilizer into the
MacDowell--Mansouri action, and expand the action to isolate the nonvanishing terms.
We then focus on the left and right sector actions and introduce effective tetrads so that each sector
can be rewritten in a Plebanski-like $GL(2,\mathbb{C})$ (BF-type) form where the simplicity/metricity
constraints can be expressed compactly (e.g.\ as matrix constraints).
This sets the stage for a clean constraint analysis, including additional compatibility constraints
associated with the overlap/interface region.

Our present study of gravi-weak unification forms part of our overall $E_8 \times \omega E_8$ unification program \cite{Kaushik, Singh:2024ven, Singh:2025xxv}, wherein each $E_8$ branches into $SU(3)\times E_6$. The two $E_6$ describe the matter and field content of the standard model and of general relativity, whereas the two extra $SU(3)$s relate to the 6D spacetime and to the internal symmetry space \cite{Singh2026Preprint}. 

\paragraph{Organization.}
\hyperref[sec:6dbfframework]{Section~\ref*{sec:6dbfframework}} describes the gravi-weak framework as an $SO(3,3)$ gauge theory on a six-dimensional manifold of split signature $(3,3)$, formulated using a BF-type action, and explains how symmetry breaking and chiral decomposition results in two emergent four-dimensional sectors that reproduce gravity and the weak interaction.
\hyperref[sec:6dmmso33]{Section~\ref*{sec:6dmmso33}} presents the implementation of the MacDowell--Mansouri construction within the $SO(3,3)$ framework and demonstrates how symmetry breaking can be implemented geometrically via algebraic projection rather than a scalar Higgs potential, thereby avoiding the boundedness issues of renormalisable $D=6$ scalar theories. A representative toy model breaking $SO(3,3)\to SO(3,1)$ is constructed explicitly to exhibit the connection decomposition, curvature splitting, and the emergence of Einstein–Hilbert–type dynamics from a projected curvature-squared action.
\hyperref[sec:twocopiescurvatureexpansion]{Section~\ref*{sec:twocopiescurvatureexpansion}} presents the explicit algebraic and geometric decomposition underlying the symmetry breaking $SO(3,3) \rightarrow SU(2)_{R}\times SU(2)_{L}$, including the stabilizer–coset structure, connection splitting, curvature expansion, and transformation properties of the coset fields. Subsequently, the MacDowell–Mansouri mechanism is implemented at the level of the action: non-vanishing terms in the expanded action are consolidated and the left/right sectors are isolated.
In \hyperref[sec:firstandsecondspacetimesectors]{Section~\ref*{sec:firstandsecondspacetimesectors}}, we construct the two emergent four-dimensional sectors in the post-symmetry breaking phase, formulating Region $\mathcal{I}$ as a self-dual Plebanski gravity theory and Region $\mathcal{II}$ as an anti-self-dual sector identified with weak isospin, each defined using BF-type actions supplemented by simplicity and reality constraints. Subsequently, we analyze the mixed-signature 2D interface and derive its presymplectic structure, corner constraint algebra, and first-class gluing conditions mediated by edge modes that consistently match the chiral data across the two spacetimes.
\hyperref[sec:couplngconstantactionandu1]{Section~\ref*{sec:couplngconstantactionandu1}} presents the complete 4D effective actions by relating the bare parameters of the theory to the physical constants, expresses the obtained actions in terms of the physical constants $G_{N}$, $\Lambda_{\mathcal I}$, $\Lambda_{\mathcal{II}}$, $g_{w}$, and $G_{F}$, and provides the extension of the $SU(2)$ framework to the $SU(2)\times U(1)$ gauge structure. In
\hyperref[sec:discussionsection]{Section~\ref*{sec:discussionsection}}, we clarify the geometric–gauge equivalence of the weak interaction, distinguish the various symmetry-breaking mechanisms (MacDowell–Mansouri, Lorentz–Higgs, and electroweak Higgs), and analyze the role of a possible $SU(2)_R \times U(1)_{Y_{\rm dem}} \to U(1)_{\rm dem}$ breaking in the ultra-soft gravidem regime. We then discuss a speculative conformal gluing relation between the two spacetime sectors and its cosmological implications, compare our framework with prior gravi–weak proposals, and conclude with a summary of experimental and observational constraints. Finally,
\hyperref[sec:outlook-conclusions]{Section~\ref*{sec:outlook-conclusions}} provides the conclusions and outlooks by summarizing the $SO(3,3)$ gravi--weak construction and outlines scope for future research.\\

\noindent A recurring challenge in presenting a new theoretical construction is that it must be evaluated against well-established and highly successful standard frameworks, within which most practitioners are deeply fluent. As a result, aspects of the present formulation may initially appear structurally different from conventional formulations. To facilitate accessibility and provide additional conceptual and technical context, we have included extended motivations and supplementary discussions in the Appendices. These are intended to clarify the structural choices underlying the construction and to support the reader in navigating its relation to more standard approaches.
\hyperref[sec:motivationforsdasd]{Appendix~\ref*{sec:motivationforsdasd}} provides the motivation for using the selfdual and antiself dual formalisms. \hyperref[sec:motivationforsu2gravity]{Appendix~\ref*{sec:motivationforsu2gravity}} discusses the motivation for employing $SU(2)$ for describing gravitational theories. 
\hyperref[sec:significanceof6Dspacetime]{Appendix~\ref*{sec:significanceof6Dspacetime}} comprehensively explains the significance of the (3,3) signature.
\hyperref[sec:6dsymmetricphase]{Appendix~\ref*{sec:6dsymmetricphase}} reviews the Clifford Algebra and discusses chirality in SO(3,3). \hyperref[sec:algebraautomorphisms]{Appendix~\ref*{sec:algebraautomorphisms}} introduces fundamental automorphism invariance and its relation to emergent diffeomorphism and gauge symmetries.

\pagebreak
\section{Theory in D=6}\label{sec:6dbfframework}

The central premise of the proposed gravi--weak theory is that gravitation and the weak interaction originate from a single gauge theory with structure group $SO(3,3)$ defined on a six-dimensional manifold $\mathcal{M}_6$ of split signature $(3,3)$. The familiar four-dimensional spacetime description results only after a symmetry-breaking mechanism selects two complementary chiral sectors and dynamically generates effective metrics.\\

At the kinematical level, the gauge algebra $\mathfrak{so}(3,3)$ admits a decomposition,
$\mathfrak{so}(3,3) \longrightarrow \mathfrak{su}(2)_L \oplus \mathfrak{su}(2)_R \oplus \mathfrak{p}$,
where the stabilizer $\mathfrak{su}(2)_L \oplus \mathfrak{su}(2)_R$ plays the role of a chiral gauge symmetry, and the nine-dimensional coset $\mathfrak{p}$ transforms as a bifundamental representation under $SU(2)_L \times SU(2)_R$. In the broken phase, this coset sector encodes generalized tetrad variables that couple two emergent four-dimensional geometries. Therefore, rather than introducing independent gravitational and weak sectors by hand, both result from different chiral components of a single six-dimensional gauge structure.\\

A key conceptual point is that the $D=6$ theory is \textit{\textbf{pregeometric}} (more details are provided in \hyperref[sec:algebraautomorphisms]{Appendix~\ref*{sec:algebraautomorphisms}}). Prior to symmetry breaking, there is no dynamical spacetime metric, no distinguished notion of time versus space, and no a priori decomposition into four-dimensional submanifolds. The fundamental variables are gauge-theoretic connections, higher-degree forms, and algebra-valued constraints. Geometry emerges only after imposing simplicity-type constraints that reduce the \textit{\textbf{topological BF-type theory}} (\hyperref[sec:6dtopologicalbfaction]{Section~\ref*{sec:6dtopologicalbfaction}}) to a sector in which effective vierbeins and induced metrics appear. In this sense, spacetime is not assumed but dynamically generated.\\

Following symmetry breaking $SO(3,3) \longrightarrow SU(2)_R \times SU(2)_L$, the two chiral sectors yield two overlapping four-dimensional spacetimes with opposite Lorentz signatures. One sector, described by self-dual $SU(2)_R$ Plebanski two-forms $\Sigma^{(+)i}$ satisfying simplicity constraints, reproduces standard Einstein gravity on a $(3,1)$-signature spacetime (Region $\mathcal{I}$), and is modified by a $U(1)_{DEM}$ force, which we call darkelectro-grav,  at large distances. The complementary sector, described by anti-self-dual $SU(2)_L$ two-forms $\Sigma^{(-)i}$, yields a formally identical geometric structure on a $(1,3)$-signature spacetime (Region $\mathcal{II}$). In this construction, the weak interaction is interpreted geometrically as gravity in the opposite-signature sector. The two emergent spacetime sectors intersect along a two-dimensional $(1,1)$ interface as $[ - - (- +) + + ]$, across which gluing constraints are imposed via first-class boundary (edge-mode) fields, ensuring consistency of the unified description.\\

From the four-dimensional perspective of the gravitational sector, the additional timelike directions associated with the second spacetime appear as internal gauge directions. This reproduces the conventional $SU(2)_L$ gauge structure of the weak interaction, while simultaneously embedding it into a higher-dimensional geometric framework. Therefore, gravitation and the weak interaction are placed on identical structural footing: each can be described either as Riemannian geometry of an emergent spacetime or as a chiral component of an $SU(2)$ gauge symmetry inherited from $SO(3,3)$. Gravitation is the right-handed counterpart of the weak force. Classical systems experience only our 4D spacetime - they do not probe the two extra timelike directions. However, quantum systems, because they experience the short range weak force, experience the two additional timelike dimensions and hence effectively live in the 6D spacetime, even at low energies.\\

In summary, the six-dimensional theory provides a unified gauge-theoretic and pregeometric origin for both gravitation and the weak interaction. The subsequent subsections clarify the nature of time in the proposed framework, motivate the choice of dimension and signature, and then present the explicit BF-type action that realizes this structure at the classical level.

\subsection{Nature of Time}\label{sec:natureoftimeconnestime}
At a more foundational level, the pregeometric six-dimensional phase prior to symmetry breaking can be characterized spectrally. One may regard the underlying structure as a spectral tuple (\hyperref[sec:algebraautomorphisms]{Appendix~\ref*{sec:algebraautomorphisms}}) adapted to split signature $(3,3)$, formulated in a Krein-space setting to accommodate the indefinite metric. The Dirac operator $\mathcal{D}$ defines the generalized Laplacian $\mathcal{D}^2$ and generates dynamics compatible with the $\mathfrak{so}(3,3)$ symmetry. In such a framework, physical evolution need not be tied to one of the six coordinates; instead, it may be associated with a modular flow, with a Connes-type time parameter emerging from the von Neumann algebra of observables. Therefore, the presence of three timelike directions in the underlying $(3,3)$ signature does not imply the existence of multiple independent ``flows'' of time. At the pregeometric level, the six coordinates—three spacelike and three timelike—enter symmetrically as structural directions of the manifold. They play a static kinematical role analogous to spatial coordinates in ordinary geometry. The notion of dynamical, flowing time is instead associated with Connes-type time parameter, whose origin lies in noncommutative geometry \cite{Singh:2024ven}. \\

In four-dimensional spacetime, one typically makes the tacit identification of Connes time with the single Lorentzian coordinate time, thereby creating the impression that coordinate time itself ``flows.'' In the present framework, these two notions are carefully distinguished. The flowing time is an emergent, algebraically defined modular parameter, whereas coordinate time (whether one or several) remains a geometric label without intrinsic dynamical status. The distinction is analogous to that between absolute time and clock time in classical mechanics, a distinction that was explicitly recognized even in the foundations of Newtonian theory. Within the six-dimensional unified setting, this separation becomes essential for consistently interpreting the split-signature structure.

\subsection{Why 6D, and why signature \texorpdfstring{$(3,3)$}?}
When gravi-weak unification is motivated by the isomorphism $SO(4)\cong SU(2)_R \times SU(2)_L$, one is compelled to treat the two $SU(2)s$ in two different ways. The $SU(2)_R$ is associated with the spacetime geometry of the 4D spacetime and hence with gravity. By contrast, $SU(2)_L$ is not assigned a comparable geometric role, and its curvature is typically set to vanish by assumption. This asymmetric treatment of the two chiral sectors is conceptually unsatisfactory. A natural resolution is to investigate the possibility that each $SU(2)$ factor corresponds to the geometry of a distinct four-dimensional spacetime. In this picture, our observed spacetime carries signature $(3,1)$, while a second four-dimensional spacetime carries the opposite Lorentzian signature $(1,3)$. The two spacetimes intersect along a common two-dimensional interface of signature $(1,1)$ and can be embedded consistently within a six-dimensional manifold of split signature $(3,3)$.The weak interaction is the geometry of this second 4D spacetime, but the two additional timelike dimensions being provided by the second spacetime can also be viewed as internal symmetry directions, from the vantage point of our spacetime.

A strong motivation for 6D spacetime with $SO(3,3)$ symmetry comes from normed division algebras and their relation to the Dirac operator. In essence, a Dirac operator is characterized by the property that its square yields the Laplacian. In the simplest noncommutative division algebra, this being the quaternions, consider a gradient operator $D_3\equiv \hat i \partial/\partial x + \hat j \partial/\partial y + \hat k \partial /\partial z$ made using the three imaginary directions of the quaternion. The square of $D_3$ is minus the 3D Laplacian. However, such a construction  does not work in 4D spacetime with Lorentzian signature. The next admissible case of Laplacian resulting from an algebra-induced gradient Dirac operator is precisely the 6D spacetime with signature $(3,3)$ \cite{Furquan:2025sox}. The split biquaternion \cite{Vaibhav:2021xib} provides six imaginary directions in two pairs of three, having relatively opposite signature. Furthermore, a 4D spacetime is embedded inside of this 6D \cite{Singh2026Preprint}. We therefore investigate the possibility that spacetime is fundamentally 6D, not 4D, and that the gravi-weak unification takes place in 6D. 

The significance of 6D spacetime with signature $(3,3)$ is discussed in some detail in \hyperref[sec:significanceof6Dspacetime]{Appendix~\ref*{sec:significanceof6Dspacetime}}. 

\subsection{6D Action}\label{sec:6dtopologicalbfaction}
We begin by fixing our differential-geometric conventions in $D$ spacetime dimensions. A differential $p$-form $\alpha$ can be expressed in a local orthonormal coframe
$\{e^{i}\}$ as
\begin{align}
    \alpha = \frac{1}{p!} \alpha_{i_{1}\cdots i_{p}} \left( e^{i_{1}} \wedge \cdots \wedge e^{i_{p}} \right)
\end{align}

The Hodge dual operator $\star$ maps a $p$-form into a $(D-p)$-form and is defined
with respect to the spacetime metric $g_{ij}$ and the Levi--Civita tensor
$\epsilon_{i_{1}\cdots i_{D}}$. Acting on the p-form, $\alpha$, the Hodge dual is given by
\begin{align}
    \star \alpha = \frac{1}{p!(D-p)!} \sqrt{g_{ij}} \alpha^{i_{1}\cdots i_{p}} \epsilon_{i_{1}\cdots i_{p}\cdots i_{D}} \left( e^{i_{p+1}} \wedge \cdots \wedge e^{i_{D}} \right)
\end{align}
Specializing now to $D=6$, the Hodge dual maps 4-forms into 2-forms and
2-forms into 4-forms. In particular,
\begin{align}
    &\star \Sigma \hspace{0.2em}\text{is a 2-form}  \hspace{3em} (\because \hspace{0.2em} \Sigma \hspace{0.2em}\text{is a 4-form}) \\
    &\star R \hspace{0.2em}\text{is a 4-form}  \hspace{3em} (\because \hspace{0.2em} R \hspace{0.2em}\text{is a 2-form})
\end{align}
We now consider a general gauge-invariant BF-type action on a six-dimensional
manifold $\mathcal{M}_{6}$. The fundamental fields are a connection $A$ with curvature $R=dA+A\wedge A$, and a Lie-algebra–valued 4-form $\Sigma$. The most
general local action constructed from these fields and polynomial
auxiliary fields takes the form
\begin{align}
    S({\Sigma}, {A}) &= \int_{\mathcal{M}_{6}} \left[ \mathrm{k}_{G} \hspace{0.2em}\epsilon_{IJKLMN} \left( \Sigma^{IJKL}\wedge R^{MN} \right) +  \Phi_{IJKLMN} \left(\Sigma^{IJKL}\wedge \star \Sigma^{MN} \right) \nonumber \right. \\
    &\quad \left. \phantom{{}={}}\phantom{{}={}} + \epsilon_{ABCDEF}\left( \mathrm{\Lambda}_{G} + \mathrm{g}\hspace{0.5em} \Phi^{IJKLMN} \Phi_{IJKLMN}\right) \left( \Sigma^{ABCD}\wedge \star \Sigma^{EF} \right) \right. \nonumber\\
    &\quad \left. \phantom{{}={}}\phantom{{}={}} + \alpha \hspace{0.2em}\epsilon_{IJKLMN} \left( R^{IJ} \wedge \star R^{KLMN}\right) \right] \label{6dgeneralaction}
\end{align}
The first term is the six-dimensional BF term, which couples the 4-form field
$\Sigma$ directly to the curvature $R$ and is topological in nature when considered
alone. The second term introduces an auxiliary tensor field
$\Phi_{IJKLMN}$ as Lagrange multipliers that enforces algebraic constraints on $\Sigma$, analogous to
simplicity constraints in lower-dimensional Plebanski-type formulations.
The third term plays the role of a generalized cosmological and potential term, enabling nontrivial vacuum structure via the self-interaction of
$\Phi_{IJKLMN}$. Finally, the last term is a curvature-squared contribution,
which renders the theory non-topological and introduces propagating degrees of
freedom.

\medskip
\noindent
For compactness, this action can be written in index-free notation as
\begin{align}
    \boxed{S({\Sigma}, {A}) = \int_{\mathcal{M}_{6}} \left[ \mathrm{k}_{G}\left( \Sigma \wedge R \right) +  \Phi \left(\Sigma \wedge \star \Sigma \right) + \left(\mathrm{\Lambda}_{G} + \mathrm{g}\hspace{0.5em}\Phi^{2}\right) \left( \Sigma \wedge \star \Sigma \right) + \alpha \left( R \wedge \star R \right) \right]} \label{6dgeneralaction2}
\end{align}
In the remainder of this
work, we will focus primarily on the first two terms, which already capture the essential geometric structure required for symmetry breaking and dimensional reduction, and the additional terms can be viewed as providing dynamical completion and stability.

\section{MacDowell-Mansouri in SO(3,3)}\label{sec:6dmmso33}
Symmetry breaking via the MacDowell--Mansouri (MM) formalism \cite{macdowell1977unified, wise2010macdowell} provides a way to obtain gravitational dynamics from a gauge theory of a larger group by introducing a controlled reduction to a smaller stabilizer subgroup. In four dimensions this is often phrased as starting from an (A)dS group and projecting the curvature onto the Lorentz subalgebra; therefore, the resulting action is quadratic in the projected curvature and reproduces (Einstein--Hilbert + cosmological + topological) terms after identifying the translational part of the connection with a tetrad (see \cite{stelle1980spontaneously, wilczek1998riemann} for other related approaches).\\

In the proposed six-dimensional framework, the symmetric phase is formulated as an $SO(3,3)$ BF-type theory on a split-signature $(3,3)$ manifold. The key conceptual point is that the symmetry breaking is not implemented via a conventional scalar Higgs potential in six dimensions, but rather via a geometric mechanism intrinsic to first-order/BF formulations: a projector onto a specific subalgebra is introduced and an MM-type term is constructed that depends solely on the projected curvature. Therefore, the order parameter is encoded in how the $SO(3,3)$ curvature is decomposed and projected, rather than in a vacuum expectation value of a fundamental scalar. This section develops this perspective first at the level of general consistency in six dimensions (the boundedness issue), and then via a representative toy breaking $SO(3,3)\to SO(3,1)$ that illustrates the curvature decomposition and the emergence of Einstein--Hilbert-type terms.\\

\subsection{Boundedness problem in 6D}
In six spacetime dimensions, the canonical mass dimensions of fields and couplings substantially constrain the familiar spontaneous symmetry breaking paradigm. For instance, a real scalar field $\phi$ in $D$ dimensions has mass dimension
\begin{align}
    [\phi] = \frac{D-2}{2} \nonumber
\end{align}
Therefore, in $D=6$, $[\phi]=2$. Consequently, the cubic interaction $\phi^3$ has dimension $6$ and is classically renormalisable by power counting, whereas a quartic interaction $\phi^4$ has dimension $8$ and would require a coupling of negative mass dimension, making it non-renormalisable. This fact is in contrast with the familiar $D=4$ case wherein a renormalisable $\phi^4$ potential has mass dimensions of $[\phi]=1$ and naturally corresponds to a Mexican-hat potential.

In $D=6$, a conventional renormalisable symmetry breaking potential takes the form
\begin{align}
    V(\phi) \;=\; \frac{m^2}{2}\,\phi^2 \;+\; \frac{\lambda}{3}\,\phi^3
\end{align}
likely supplemented by higher-order terms that are non-renormalisable in a six-dimensional theory. However, the cubic term $\phi^3$ results in a significant issue: for either sign of $\lambda$, the potential is unbounded from below along one direction in field space. At large $|\phi|$, the $\phi^3$ term dominates and drives $V(\phi)\to -\infty$ for one sign of $\phi$; therefore, the theory lacks a globally stable ground state. Adding a mass term does not address this issue; it may create at best a local metastable extremum, but the energy functional remains unbounded. Therefore, in a purely six-dimensional renormalisable scalar field theory, the standard Higgs mechanism is generically in tension with stability.\\

This observation is particularly relevant for implementing a symmetry breaking that yields physically meaningful lower-dimensional sectors. If a fundamental scalar order parameter is used for breaking $SO(3,3)$ in six dimensions, the power-counting constraints would naturally select a cubic potential, precisely the class that fails to be bounded. \\

In Macdowell--Mansouri formulation, the order parameter is not a scalar potential, but a geometric projection in the gauge algebra, implemented directly at the level of the action. In BF/Plebanski-type formulations, additional algebraic structures are available, such as internal Hodge operators, invariant tensors, and constrained $B$-fields, that enable starting from a topological BF term (symmetric phase), introducing constraints or projectors that select a stabilizer subalgebra, and generating an effective gravitational action from a curvature-squared (or $B\wedge F$ plus constraint) structure, without requiring a scalar potential with a stable minimum.\\

Crucially, the resulting MM-type term is quadratic in the projected curvature and can be arranged to be well-behaved at the classical level, because the instability associated with a fundamental cubic scalar potential is simply absent: the mechanism is rather driven by constraints and geometry.\\

In the remainder of this section we implement this philosophy concretely within the $SO(3,3)$ framework. Before implementing the complete graviweak breaking $SO(3,3)\to SU(2)\times SU(2)$ in later sections, we first present a representative six-dimensional toy model of symmetry breaking from $SO(3,3)\to SO(3,1)$ to exhibit the algebra decomposition, the connection/curvature splitting, the role of the projector, and the emergence of Einstein--Hilbert-type terms from a Macdowell--Mansouri action.

\subsection{Representative symmetry breaking from \texorpdfstring{$\mathrm{SO}(3,3)\to \mathrm{SO}(3,1)$ 
$\linebreak \big(\text{Toy Model}\big)$}
{SO(3,3) to SO(3,1)
(Toy Model)}} \label{subsec:mmconstructionso31}
Herein, we extend the BF reformulation of the MacDowell--Mansouri approach \cite{ wise2010macdowell, durka2012immirzi, ling2001holographic} to the 6D symmetric phase and implement a representative toy model of symmetry breaking from $SO(3,3)$ to $SO(3,1)$. This implementation is intended to illustrate, in the simplest setting in $D=6$, the following key algebraic mechanisms: (i) an enlarged gauge connection splits into stabilizer and coset components, (ii) the stabilizer curvature acquires quadratic corrections from the coset one-forms, and (iii) the MM projection yields an Einstein--Hilbert-type term once an appropriate frame-like combination is identified. The actual graviweak construction is developed in \hyperref[sec:twocopiescurvatureexpansion]{Section~\ref*{sec:twocopiescurvatureexpansion}} and \hyperref[sec:firstandsecondspacetimesectors]{Section~\ref*{sec:firstandsecondspacetimesectors}}.\\

In the original 4D MM construction, an internal vector field $\varphi^{A}$ is introduced in the fundamental of $SO(4,1)$ or $SO(2,3)$ to select a preferred direction in the five-dimensional gauge algebra. In six dimensions, for $SO(3,3)$, the role of the internal vector is played by the projected $\Sigma$-term  via the internal Hodge operator $\star$. \\

To see this, consider the generators of $\mathfrak{so}(3,3)$: $M_{IJ}=-M_{IJ}$.
In breaking $SO(3,3)$ to $SO(3,1)$, i.e.
\begin{align}
    SO(3,3) \rightarrow  SO(3,1) \oplus SO(3,3)/SO(3,1)
\end{align}
the algebra splits as the SO(3,1) Lorentz subalgebra and a coset space, yielding eight vector fields and one scalar field:
\begin{align}
    \mathfrak{so}(3,3) \rightarrow \mathfrak{so}(3,1) \oplus \mathfrak{p}  
\end{align}
where, the SO(3,1) subalgebra is spanned by generators $M_{\mu\nu}$ and the coset space is spanned by $\{M_{\mu4}, M_{\mu5}, M_{45}\}$. The generators $M_{\mu4}$ and $M_{\mu5}$ transform as Lorentz vectors and $M_{45}$ is a Lorentz scalar. This decomposition is the direct six-dimensional analogue of splitting an (A)dS algebra into Lorentz generators plus translational generators, except that in the present case, the coset contains two Lorentz vectors and an additional Lorentz scalar.\\

\noindent A representative BF-type action that implements this idea in 6D is expressed as follows:
\begin{align}
    S_{BF} \sim \int_{\mathcal{M}_{6}} \left[ \left( \Sigma \wedge R \right) \
    -  \left( \Pi(\Sigma) \wedge \Pi(\star \Sigma) \right) 
    - \frac{m^2}{2} \left(
    (\Sigma^{\mu4})^{2} + (\Sigma^{\mu5})^{2} +
    (\Sigma^{45})^{2}
    \right)
    \right] 
\end{align}
where, in index form, $\Pi_{IJ}\mathstrut^{KL} = \delta^{[K}_{I}\delta^{L]}_{J}$ for $SO(3,1)$ indices and zero otherwise.
Here, in the first term, $\Sigma \wedge R$, $\Sigma$ couples to the full curvature form in 6D. This renders the theory topological prior to symmtery breaking. The second term projects out the coset pieces and gives kinetic term to the $SO(3,1)$ part of $\Sigma$. Finally, the third term explicitly breaks the symmetry and gives mass to the eight vector fields and one scalar field. $\Pi$ ensures that only the Lorentz subalgebra drives the symmetry broken action, whereas the coset fields acquire masses and decouple.

\medskip
\noindent
Therefore, the SO(3,3) connection decomposes as follows:
\begin{align}
    A &= \frac{1}{2} A^{IJ}M_{IJ} \\
    &= \underbrace{\frac{1}{2} \omega^{\mu\nu} M_{\mu\nu}}_{\text{Stabilizer}} + \underbrace{A^{\mu4}M_{\mu4} + A^{\mu5}M_{\mu5} + 
    \frac{1}{2} A^{45} M_{45}}_{\text{Coset}}
\end{align}
where $\mu,\nu=0,1,2,3$ are SO(3,1) indices. In matrix form, SO(3,3) gauge connection is decomposed as
\begin{align}
A^{AB} =
\begin{pmatrix}
  \omega^{\mu\nu} & A^{\mu 4} & A^{\mu 5} \\
  -A^{\nu 4} & 0 & A^{45} \\
  -A^{\nu 5} & -A^{45} & 0
\end{pmatrix}
\end{align}
Likewise, the curvature $R = \mathrm{d}A + A \wedge A$,
\begin{align}
    R^{AB} &= dA^{AB} + A^{A}{}_{C} \wedge A^{CB}
\end{align}
The non-zero components of the curvature (categorized as stabilizer, coset-vector, coset-scalar sectors) are calculated as follows.
\paragraph{(i)} The \textbf{stabilizer} sector of the $SO(3,3)$ curvature is given by
\begin{align}
    F^{\mu\nu} &= d\omega^{\mu\nu} + \omega^{\mu}{}_{\rho} \wedge \omega^{\rho \nu} + A^{\mu 4} \wedge A^{\nu 4} + A^{\mu 5} \wedge A^{\nu 5} \\
    &= R^{\mu\nu} + A^{\mu i} \wedge A^{\nu}{}_{i}, \quad (i=4,5)
\end{align}
\paragraph{(ii)} The coset-vector sector of the curvature is given by
\begin{align}
F^{\mu 4} &= d A^{\mu 4} + \omega^{\mu}{}_{\rho} \wedge A^{\rho 4} + A^{\mu 5} \wedge A^{54} \\
&= d A^{\mu 4} + \omega^{\mu}{}_{\rho} \wedge A^{\rho 4} - \varphi \wedge A^{\mu 5}, \quad (A^{54} = -A^{45} = -\varphi)
\end{align}
\begin{align}
F^{\mu 5} &= d A^{\mu 5} + \omega^{\mu}{}_{\rho} \wedge A^{\rho 5} + A^{\mu 4} \wedge A^{45} \\
&= d A^{\mu 5} + \omega^{\mu}{}_{\rho} \wedge A^{\rho 5} + \varphi \wedge A^{\mu 4},  \quad (A^{45} = -A^{54} = \varphi)
\end{align}
We represent both $F^{\mu 4}$ and $F^{\mu 5}$ compactly by employing a covariant derivative as follows:
\begin{align}
    F^{\mu i} &= D A^{\mu i}, \quad D A^{\mu i} = d A^{\mu i} + \omega^{\mu}{}_{\rho} \wedge A^{\rho i} + (a^{i}{}_{j}) \wedge A^{\mu j}
\end{align}
where, $a^i{}_j$ is the $SO(2)$ matrix generated by $\varphi$.
\paragraph{(iii)} The coset-scalar sector is given by
\begin{align}
    F^{45} &= d \varphi + A^{4}{}_{\mu} \wedge A^{\mu 5}
\end{align}
Collecting all the non-vanishing terms, the complete $SO(3,3)$ curvature decomposes as
\begin{align}
F^{AB} =
\begin{pmatrix}
  R^{\mu\nu} + A^{\mu i} \wedge A^{\nu}{}_{i} & D A^{\mu 4} & D A^{\mu 5} \\
  -D A^{\nu 4} & 0 & F^{45} \\
  -D A^{\nu 5} & -F^{45} & 0
\end{pmatrix},
\end{align}
Furthermore, given the idempotent and orthogonal nature of $\Pi$, the curvature splits into stabilizer and coset components,
\begin{align}
    F = \Pi(F) + (1-\Pi)(F) \equiv F_{coset} + F_{Lorentz}
\end{align}
Therefore, the Macdowell--Mansouri Action can be expressed as
\begin{align}
    \int F \wedge\star\Pi(F) = \int \Pi(F)\wedge\star\Pi(F) + \int F_{coset} \wedge \star F_{Lorentz}
\end{align}
However, because $F_{coset} \perp F_{Lorentz}$ under the Killing form, we get 
\begin{align}
    F_{coset} \wedge \star F_{Lorentz} = 0
\end{align}
This enables expressing the MM action as containing the squared stabilizer-projected curvature as follows:
\begin{align}
    \int F \wedge\star\Pi(F) = \int \Pi(F)\wedge\star\Pi(F) \label{mm6d}
\end{align}
Therefore, the symmetry breaking is implemented at the level of the action by discarding (in a gauge-invariant way) the mixed stabilizer--coset pairing, retaining only the squared stabilizer-projected curvature. This is the six-dimensional analogue of the standard MM projection in four dimension.\\

\noindent At this stage, we define two tetrad 1-forms
\begin{align}
    e^{\mu}=A^{\mu4}, \qquad f^{\mu}=A^{\mu5}
\end{align}
These are natural candidates for frame-like fields because they carry one spacetime index and precisely result from the coset components of the enlarged connection. To make contact with the usual gravitational branch, we can introduce the orthogonal linear combinations
\begin{align}
    E^{\mu}=\frac{1}{\sqrt2}\bigl(e^{\mu}+f^{\mu}\bigr), \qquad \Delta^{\mu}=\frac{1}{\sqrt2}\bigl(e^{\mu}-f^{\mu}\bigr)
\end{align}
The combination $E^\mu$ will be interpreted as the effective tetrad on the gravitational branch, whereas $\Delta^\mu$ parameterizes the relative (difference) mode between the two frame-like one-forms. We can regard $\Delta^\mu$ here as an auxiliary/relative mode intrinsic to the present toy model of symmetry breaking, which is likely constrained, gauge-fixed, or rendered heavy by the explicit mass deformation so as not to survive as an independent low-energy degree of freedom.\\

Using these definitions, the stabilizer curvature decomposes as
\begin{align}
    F^{\mu\nu} = R^{\mu\nu} + E^{\mu} \wedge E^{\nu} + 2\Delta^{\mu} \wedge\Delta^{\nu}, \quad (\mu, \nu=0,1,2,3)
\end{align}
Substituting this curvature decomposition into (\ref{mm6d}) and expanding the action yields
\begin{align}
    \int F \wedge\star\Pi(F) &= \int_{\mathcal{M}_{6}} \Pi(F)\wedge\star\Pi(F) \\
    &= \int_{\mathcal{M}_{4}} \Big(R^{ab} + E^{a} \wedge E^{b} + 2\Delta^{a} \wedge\Delta^{b}\Big) \wedge \star\Big(R_{ab} + E_{a} \wedge E_{b} + 2\Delta_{a} \wedge\Delta_{b}\Big) \\
    &=\int_{\mathcal{M}_{4}} \Big[R^{ab} \wedge \star R_{ab} \Big] 
    + \Big[R^{ab} \wedge \star (E_a \wedge E_b) + (E^a \wedge E^b) \wedge \star R_{ab}\Big] \nonumber  \\
    &\quad 
    \hspace{1em}+ \Big[ 2 R^{ab} \wedge \star (\Delta_a \wedge \Delta_b) + 2 (\Delta^a \wedge \Delta^b) \wedge \star R_{ab}\Big] \nonumber  \\
    &\quad 
    \hspace{1em}+ \Big[(E^a \wedge E^b) \wedge \star (E_a \wedge E_b)\Big] \nonumber  \\
    &\quad 
    \hspace{1em}+ \Big[2 (E^a \wedge E^b) \wedge \star (\Delta_a \wedge \Delta_b) + 2 (\Delta^a \wedge \Delta^b) \wedge \star (E_a \wedge E_b)\Big] \nonumber  \\
    &\quad 
    \hspace{1em}+ \Big[4 (\Delta^a \wedge \Delta^b) \wedge \star (\Delta_a \wedge \Delta_b)\Big] \\
    &= \int_{\mathcal{M}_{4}} \underbrace{\Big[R^{ab} \wedge \star R_{ab} \Big]}_{\text{Pontryagin Term}} 
    + \underbrace{\Big[2 R^{ab} \wedge \star (E_a \wedge E_b)\Big]}_{\text{Einstein-Hilbert Term}} 
    + \Big[ 4 R^{ab} \wedge \star (\Delta_a \wedge \Delta_b)\Big] \nonumber  \\
    &\quad 
    \hspace{1em}+ \underbrace{\Big[(E^a \wedge E^b) \wedge \star (E_a \wedge E_b)\Big]}_{\text{Tetrad Terms}} 
    + \Big[4 (E^a \wedge E^b) \wedge \star (\Delta_a \wedge \Delta_b)\Big] \nonumber  \\
    &\quad 
    \hspace{1em}+ \Big[4 (\Delta^a \wedge \Delta^b) \wedge \star (\Delta_a \wedge \Delta_b)\Big]\\
    &= S_{RR} + S_{RE} + S_{R\Delta} + S_{EE} + S_{E\Delta} + S_{\Delta\Delta}
\end{align}

The first term $S_{RR}$ represents the typical curvature-squared topological density in four dimensions (Pontryagin-type, depending on conventions) and does not contribute to local dynamics. The cross-term $S_{RE}$ yields the Einstein--Hilbert term once $E^a$ is identified with the effective tetrad, thereby reproducing standard gravitational dynamics on the branch wherein $E$ is non-degenerate. The $\Delta$-depending terms encode additional interactions associated with the extra coset vector present in this toy model reduction. In the context of the present work, these contributions can be treated as an auxiliary/relative sector, suppressed either by imposing an alignment (gluing) condition between $e \simeq f$ (such that $\Delta^\mu\approx 0$), or by integrating out $\Delta$ when rendered heavy by the explicit mass deformation. In either case, the toy model isolates the essential MM mechanism while preparing for \hyperref[sec:twocopiescurvatureexpansion]{Section~\ref*{sec:twocopiescurvatureexpansion}}, wherein the physically relevant symmetry breaking $SO(3,3)\to SU(2)_L\times SU(2)_R$ reorganizes the coset degrees of freedom without introducing an independent $\Delta$-counterpart.

\section{Symmetry Breaking from \texorpdfstring{$\mathrm{SO}(3,3)\to\mathrm{SU}(2)\times\mathrm{SU}(2)$}
{SO(3,3) to SU(2) x SU(2)}} \label{sec:twocopiescurvatureexpansion}
This section presents the group-theoretic picture underlying the proposed graviweak framework wherein a single six-dimensional gauge symmetry $SO(3,3)$ (appropriate to the split-signature $(3,3)$ phase) undergoes a symmetry-breaking
\[
SO(3,3) \longrightarrow SU(2)_L \times SU(2)_R
\]
Here, the Lie algebra $\mathfrak{so}(3,3)$ decomposes into a $6$-dimensional stabilizer (two commuting $\mathfrak{su}(2)$ terms) plus a $9$-dimensional coset. The stabilizer will furnish two independent chiral sectors, whereas the coset will be interpreted as the ``soldering'' form. In later sections this coset form  will be utilized to construct effective tetrad-like one-forms that generate Plebanski two-forms on the two overlapping 4D spacetime sectors (with opposite Lorentzian signatures), thereby placing gravity and the weak sector on parallel geometric footing.

The Macdowell--Mansouri approach involves constructing an action from a curvature squared term, projecting only the stabilizer part of the curvature. This projection is the symmetry-breaking input: it selects $SU(2)_L \times SU(2)_R$ as the unbroken subgroup and yields an effective action that naturally splits into ``left'' and ``right'' sector contributions plus interaction terms mediated by the coset fields. The present section establishes (i) the stabilizer/coset algebraic structure, (ii) the corresponding decomposition of the $SO(3,3)$ connection and curvature, and (iii) the explicit expansion of the symmetry-broken MacDowell--Mansouri action in this decomposition.
\subsection{Stabilizer and Coset Structure}
\subsubsection{Prerequisite: \texorpdfstring{$\mathrm{SU}(2)\times\mathrm{SU}(2)$}
{SU(2) x SU(2)}}
The group \(\text{SO}(3,3)\) is the indefinite orthogonal group that preserves the quadratic form of split-signature \((3,3)\). It is defined by
\begin{align}
\text{SO}(3,3) = \{M \in \text{GL}(6,\mathbb{R}) : M^T \eta M = \eta, \, \det(M) = 1\}
\end{align}
where the metric signature is given by
\begin{align}
\eta = \text{diag}(+1, +1, +1, -1, -1, -1)
\end{align}
The corresponding Lie algebra \(\mathfrak{so}(3,3)\) consists of all \(6 \times 6\) antisymmetric matrices \(M\) satisfying,
\begin{align}
M^T \eta + \eta M = 0
\end{align}
The Lie algebra of SO(3,3) decomposes under the subgroup action of \(H = \text{SU}(2)_L \times \text{SU}(2)_R\) as follows:
\begin{align}
\mathfrak{so}(3,3) &= \mathfrak{su}(2)_L \oplus \mathfrak{su}(2)_R \oplus \mathcal{K} \\
&= \mathcal{H}_1 \oplus \mathcal{H}_2 \oplus \mathcal{K} \label{decompositionalgebrasu2su2}
\end{align}
where where $H_1 \simeq \mathfrak{su}(2)_L$ and $H_2 \simeq \mathfrak{su}(2)_R$ are the stabilizer subalgebras and $\mathcal{K}$ is the coset subspace (the complement of the stabilizer algebra in \(\mathfrak{so}(3,3)\)). Furthermore, the coset space $\text{SO}(3,3) / (\text{SU}(2)_L \times \text{SU}(2)_R)$ is a homogeneous manifold of dimension 9 and the generators in $\mathcal{K}$ span the tangent space at the identity coset.\\

A convenient way to visualize this splitting is to consider the fundamental representation of $SO(3,3)$, which comprises $6 \times 6$ matrices $X$ preserving a metric of signature $(3,3)$:
\begin{align}
X_{IJ} = \begin{pmatrix}
A_{ab} & B_{ai} \\
-B_{bj} & D_{ij}
\end{pmatrix} \quad\quad (a = 1,2,3 \text{ and } i = 4,5,6) \label{blockformfundrep}
\end{align}
where, indices $I=1-6$; however, to make the decomposition manifest, we use the indices $a = 1,2,3$ for Block A and $i = 4,5,6$ for Block D. Furthermore, $A$ and $D$ are $3 \times 3$ antisymmetric matrices, with 3 generators each; and $B$ is an arbitrary $3 \times 3$ matrix with 9 generators. The off-diagonal block $B$ contains the coset degrees of freedom.\\

\medskip
Furthermore, counting the dimensions validates the decomposition for consistency, 
\begin{align}
    \text{dim}(\mathfrak{su}(2)_L) + \text{dim}(\mathfrak{su}(2)_R) + \text{dim}(\mathcal{K}) &= 3+3+9 \nonumber\\
    &= 15 \nonumber\\
    &= \text{dim}(\mathfrak{so}(3,3)) = \frac{6\cdot5}{2}
\end{align}
Therefore, the stabilizer is $6$-dimensional and the coset is $9$-dimensional, aligning with the block decomposition (\ref{blockformfundrep}).

\subsubsection{Coset Structure}
The 9-dimensional coset generators organize as a \(3 \times 3\) matrix:
\begin{align}
E = \begin{pmatrix} E_{14} & E_{15} & E_{16} \\ E_{24} & E_{25} & E_{26} \\ E_{34} & E_{35} & E_{36} \end{pmatrix}.
\end{align}
Under the action of \((g_R, g_L) \in \text{SU}(2)_R \times \text{SU}(2)_L\), the coset matrix transforms as
\begin{align}
E \mapsto g_R \cdot E \cdot g_L^{-1} \label{cosetleftrighttransformation}
\end{align}
In component form, this transformation is expressed as follows:
\begin{align}
E_{ai} \mapsto (g_R)_{ab} E_{bj} (g_L)^{-1}_{ji} \quad\quad (a = 1,2,3 \text{ and } i = 4,5,6) 
\end{align}
Here, the left index $a$ transforms as a vector under the 3-dimensional adjoint representation of $\text{SU}(2)_R$ and the right index $i$ transforms as a vector under the 3-dimensional adjoint representation of $\text{SU}(2)_L$.

This representation will become important in subsequent constructions: the coset one-forms will behave like tetrads valued in $(\mathbf{3},\mathbf{3})$, and bilinears such as $E\wedge E^T$ and $E^T\wedge E$ will transform purely in the right and left stabilizer factors, respectively. This property enables separation into two sector actions while still permitting coupling via the coset fields.

\subsection{Proposed Construction}
We now present the decomposition of the 15-dimensional Lie algebra of $SO(3,3)$ into two copies of $SU(2)$ using fully explicit generators adapted to the block structure (\ref{blockformfundrep}). The decomposition is expressed as follows:
\begin{align}
\mathfrak{so}(3,3) &= \mathcal{H}_1 \oplus \mathcal{H}_2 \oplus \mathcal{K}
\end{align}
where, the stabilizer and coset terms are generated by the following elements:
\begin{align}
    \mathcal{H}_1 &= \text{span}\{J_1, J_2, J_3\} \\
    \mathcal{H}_2 &= \text{span}\{K_1, K_2, K_3\} \\
    \mathcal{K} &= \text{span}\{E_{14}, E_{15}, E_{16}, E_{24}, E_{25}, E_{26}, E_{34}, E_{35}, E_{36}\}
\end{align}
The stabilizer generators close independently as follows:
\begin{align}
[J_a, J_b] &= \varepsilon_{abc} J_c \quad \text{(SU(2) algebra)} \label{epsilonmap1}\\
[K_i, K_j] &= \tilde{\varepsilon}_{ijk} K_k \quad \text{(SU(2) algebra)} \label{epsilonmap2}
\end{align}
Moreover, because $J_a$ act within Block A and $K_i$ act within Block D in (\ref{blockformfundrep}), they commute:
\begin{align}
[J_a, K_j] = 0 \quad \text{for all } a = 1,2,3 \text{ and } i = 4,5,6.
\end{align}
Therefore, generators in different blocks commute automatically. Furthermore, the coset generators transform nontrivially under both stabilizer factors, and their commutators are consistent with the reductive decompositions:
\begin{align}
[J_a, E_{bj}] &= \varepsilon_{ab}{}^{c} E_{c j}\\
[K_i, E_{aj}] &=  \tilde{\varepsilon}_{ij}{}^{k} E_{a k}\\
[E_{ai}, E_{bj}]
&= \delta_{ij} \varepsilon_{ab}{}^{c} J_c
- \delta_{ab} \tilde{\varepsilon}_{ij}{}^{k} K_k
\end{align}
These relations encode the key geometric mechanism of the later construction: coset--coset commutators feed back into the stabilizer; therefore, bilinears constructed from coset one-forms contribute to the effective stabilizer curvatures. This is precisely the MacDowell--Mansouri pattern: curvature plus (tetrad)$\wedge$(tetrad) appears naturally, and the tetrads themselves originate as coset components of the enlarged connection.

For reference, the proposed construction is summarized in the following table:
\begin{center}
\begin{tabular}{|l|c|l|}
\hline
\textbf{Component} & \textbf{Dimension} & \textbf{Generators} \\
\hline
$\mathcal{H}_1$ & 3 & $\{J_1, J_2, J_3\}$ \\
$\mathcal{H}_2$ & 3 & $\{K_1, K_2, K_3\}$ \\
\hline
Stabilizer $\mathcal{H}_1 \oplus \mathcal{H}_2$ & 6 & $\{J_1, J_2, J_3, K_1, K_2, K_3\}$ \\
Coset $\mathcal{K}$ (Block B) & 9 & $\{E_{14}, E_{15}, E_{16}, E_{24}, E_{25}, E_{26}, E_{34}, E_{35}, E_{36}\}$ \\
\hline
\textbf{Total} & \textbf{15} & All of $SO(3,3)$ \\
\hline
\end{tabular}
\end{center}

\subsubsection{Connection in SO(3,3)}
The starting point is an $SO(3,3)$ connection one-form $A$ on the six-dimensional manifold, valued in $\mathfrak{so}(3,3)$:
\begin{align}
A = A_I dx^I
\end{align}
In breaking $SO(3,3)$ to two copies of $SU(2)$, the SO(3,3) connection decomposes as
\begin{align}
    A &= \frac{1}{2} A^{IJ}M_{IJ} \hspace{9em} (I=1-6)\\
    &= \underbrace{\frac{1}{2} \omega^a J_a}_{SU(2)} + \underbrace{\frac{1}{2}\kappa^i K_i}_{SU(2)} + \frac{1}{\ell}\xi^{ai} E_{ai} \hspace{2em} (a = 1,2,3 \text{ and } i = 4,5,6)
\end{align}
Here $\omega^a$ and $\kappa^i$ denote the $SU(2)_L$ and $SU(2)_R$ connections, respectively, whereas $\xi^{ai}$ is a one-form valued in the coset representation $(\mathbf{3},\mathbf{3})$. The parameter $\ell$ is introduced as a length scale so that $\xi^{ai}$ can be assigned the natural dimension of a soldering form/tetrad-like variable in the subsequent 4D sector interpretations. Although $\omega$ and $\kappa$ constitute internal gauge connections, $\xi$ is the component that will later be identified with geometric frame data.\\
Hereafter, we will use the pure-form notation,
\begin{align}
    A = \underbrace{\frac{1}{2} \omega^a + \frac{1}{2}\kappa^i}_{\text{Stabilizer sector}} + \underbrace{\frac{1}{\ell}\xi^{ai}}_{\text{Coset sector}} \quad (a = 1,2,3 \text{ and } i = 4,5,6)
\end{align}

\subsubsection{Curvature Computation}
The curvature of $A$ is $R = \mathrm{d}A + A \wedge A$,
\begin{align}
    R^{IJ} &= dA^{IJ} + A^{I}{}_{K} \wedge A^{KJ} \quad \quad (I=1-6)
\end{align}
Employing the stabilizer--coset decomposition (\ref{blockformfundrep}), the non-zero components of the curvature $R$ splits into (i) stabilizer-sector curvatures ($F^{ab}, F^{ij}$) spanned by $\mathfrak{su}(2)_R\oplus\mathfrak{su}(2)_L$ and (ii) coset-sector curvatures ($F^{ai}$) that transform covariantly in $(\mathbf{3},\mathbf{3})$.
\paragraph{(i)} The \textbf{stabilizer sector} of the \textbf{curvature} is given by
\begin{align}
    F^{ab} &=  \mathrm{d}A^{ab} + A^{a}_{I} \wedge A^{Ib} \\
            &= \mathrm{d}\omega^{ab} + \omega^{a}_{c} \wedge \omega^{cb} 
            + \frac{1}{\ell^2}\xi^{a}_{i} \wedge \xi^{ib}\\
            &= \left(F^{(1)}_{SU(2)}\right)^{ab} 
            + \frac{1}{\ell^2}\xi^{a}_{i} \wedge \xi^{ib}\\
            &= R_{\mathcal{I}} + \frac{1}{\ell^2} \big(\xi \wedge \xi^{T}\big)
\end{align}
\begin{align}
    F^{ij} &=  \mathrm{d}A^{ij} + A^{i}_{I} \wedge A^{Ij} \\
            &= \mathrm{d}\kappa^{ij} + \kappa^{i}_{k} \wedge \kappa^{kj} 
            + \frac{1}{\ell^2}\xi^{i}_{a} \wedge \xi^{aj}\\
            &= \left(F^{(2)}_{SU(2)}\right)^{ij} + \frac{1}{\ell^2}\xi^{i}_{a} \wedge \xi^{aj} \\
            &= R_{\mathcal{II}} + \frac{1}{\ell^2} \big(\xi^{T} \wedge \xi\big)
\end{align}
For compact notation, we have relabeled the stabilizer terms in the curvature as
\begin{align}
    \left(F^{(1)}_{SU(2)}\right)^{ab} \rightarrow R_{\mathcal{I}} \\
    \left(F^{(2)}_{SU(2)}\right)^{ij} \rightarrow R_{\mathcal{II}}
\end{align}
Likewise, the tetrad bilinears can be expressed compactly as
\begin{align}
    \xi^{a}_{i} \wedge \xi^{ib} = \big(\xi \wedge \xi^{T}\big)^{ab}\\
    \xi^{i}_{a} \wedge \xi^{aj} = \big(\xi^{T} \wedge \xi\big)^{ij}
\end{align}
These bilinears transform in their corresponding stabilizer factors and behave exactly as required by tetrads ($\xi$) transforming as (\textbf{3},\textbf{3}) under the stabilizer. That is, let $\xi \mapsto g_R \xi g_L^{-1}$ as expressed in (\ref{cosetleftrighttransformation}). Then, we can show that
\begin{align}
    \big(\xi \wedge \xi^{T}\big) &\mapsto \big(g_R \xi g_L^{-1} \big)\wedge\big(g_R \xi g_L^{-1} \big)^{T} \nonumber\\
    &= \big(g_R \xi g_L^{-1} \big) \wedge \big(g_L^{-T} \xi g^T_R \big) \nonumber\\
    &= g_R \big(\xi \wedge \xi^{T}\big) g^{T}_R \label{tetradbilinearL}
\end{align}
Likewise,
\begin{align}
    \big(\xi^{T} \wedge \xi\big) &\mapsto g_L \big(\xi^{T} \wedge \xi\big) g^{T}_L \label{tetradbilinearR}
\end{align}
 Therefore, the bilinears $\big(\xi \wedge \xi^{T}\big)^{ab}$ transform purely in the right stabilizer and $\big(\xi^{T} \wedge \xi\big)^{ij}$ transform purely in the left stabilizer; they do not mix. This ensures that once the symmetry is broken to $SU(2)_R\times SU(2)_L$, each sector can be treated independently except through explicit interaction terms.
\paragraph{} Collecting all the non-vanishing terms, the stabilizer sector of the curvature is expressed as follows:
\begin{align}
    F^{(\text{stab})} &= \underbrace{\left(F^{(1)}_{SU(2)}\right)^{ab}}_{\text{First }SU(2)} + \underbrace{\left(F^{(2)}_{SU(2)}\right)^{ij}}_{\text{Second }SU(2)} + \underbrace{\frac{1}{\ell^2} \Big[\xi^{a}_{i} \wedge \xi^{ib} + \xi^{i}_{a} \wedge \xi^{aj}\Big]}_{\text{Tetrads}} \\
    &= \underbrace{R_{\mathcal{I}} + \frac{1}{\ell^2} \big(\xi \wedge \xi^{T}\big)}_{\text{First Sector}} + \underbrace{R_{\mathcal{II}} + \frac{1}{\ell^2} \big(\xi^{T} \wedge \xi\big)}_{\text{Second Sector}}
    \label{so33su2u1stab}
\end{align}
\paragraph{(ii)} The \textbf{coset sector} of the \textbf{curvature} is given by
\begin{align}
     F^{ai} &=  \mathrm{d}A^{ai} + A^{a}_{I} \wedge A^{Ii} \\
            &= \mathrm{d}\xi^{ai} + \omega^{a}_{c} \wedge \xi^{ci} + \xi^{a}_{k} \wedge \kappa^{ki}
\end{align}
This term plays the role analogous to torsion: it measures the compatibility between the stabilizer connections and the coset ``frame'' field.

\subsubsection{Macdowell-Mansouri Symmetry Breaking SO(3,3) to two copies of SU(2)}
We now implement the MacDowell--Mansouri mechanism in the proposed $SO(3,3)$ framework. An action is constructed quadratic in curvature with a projection onto the stabilizer subalgebra; therefore, the unbroken gauge symmetry is $SU(2)_L\times SU(2)_R$ rather than the full $SO(3,3)$. The Macdowell--Mansouri Action as given by,
\begin{align} \label{mmactionsu2u1}
    S_{MM} =  \int F \wedge\star F^{(\text{stab})} &= \int_{\mathcal{M}_{6}} F^{(\text{stab})} \wedge\star F^{(\text{stab})}
\end{align}
The projection to $F^{(\mathrm{stab})}$ is the symmetry-breaking input: it discards the purely coset part from the quadratic invariant, in direct analogy with the toy model construction in \hyperref[subsec:mmconstructionso31]{Section~\ref*{subsec:mmconstructionso31}}.

Substituting the stabilizer sector of the curvature (\ref{so33su2u1stab}) into (\ref{mmactionsu2u1}) and expanding the action yields 
\begin{align}
    S &= \frac{1}{g^2}\int_{\mathcal{M}_{6}} F^{(\text{stab})} \wedge\star F^{(\text{stab})} \\
    &= \frac{1}{g^2}\int \Bigg[R_{I}  + R_{II} + \frac{1}{\ell^2} \Big(\xi^{a}_{i} \wedge \xi^{ib} + \xi^{i}_{a} \wedge \xi^{aj}\Big)\Bigg] \wedge \star \Bigg[ R_{I}  + R_{II} + \frac{1}{\ell^2} \Big(\xi^{a}_{i} \wedge \xi^{ib} + \xi^{i}_{a} \wedge \xi^{aj}\Big)\Bigg]\\
    &= \frac{1}{g^2} \Bigg[\int_{\mathcal{M}_4^R} d^4x \Big\{(R_I \wedge \star R_I) + \frac{1}{\ell^2}R_I \wedge \star \big(\xi^{a}_{i} \wedge \xi^{ib} \big)\Big\} \nonumber \\
    &\quad \hspace{2em} + \int_{\mathcal{M}_4^R} d^4x \Big\{(R_{II} \wedge \star R_{II}) + \frac{1}{\ell^2}R_{II} \wedge \star \big(\xi^{i}_{a} \wedge \xi^{aj} \big)\Big\} \nonumber \\
    &\quad \hspace{2em} + \frac{1}{\ell^4}\int \Big\{ \Big(\xi^{a}_{i} \wedge \xi^{ib} + \xi^{i}_{a} \wedge \xi^{aj}\Big) \wedge \star \Big( \xi^{a}_{i} \wedge \xi^{ib} + \xi^{i}_{a} \wedge \xi^{aj} \Big)\Big\} \Bigg]  \\
    &= \frac{1}{g^2} \Bigg[\int_{\mathcal{M}_4^R} d^4x \Big\{(R_I \wedge \star R_I) + \frac{1}{\ell^2}R_I \wedge \star \big(\xi^{a}_{i} \wedge \xi^{ib} \big) + \frac{1}{\ell^4}\big(\xi^{a}_{i} \wedge \xi^{ib} \big) \wedge \star \big(\xi^{a}_{i} \wedge \xi^{ib} \big) \Big\} \nonumber \\
    &\quad \hspace{2em} + \int_{\mathcal{M}_4^R} d^4x \Big\{(R_{II} \wedge \star R_{II}) + \frac{1}{\ell^2}R_{II} \wedge \star \big(\xi^{i}_{a} \wedge \xi^{aj} \big) + \frac{1}{\ell^4}\big(\xi^{i}_{a} \wedge \xi^{aj} \big) \wedge \star \big(\xi^{i}_{a} \wedge \xi^{aj} \big)\Big\} \Bigg]\\
    &= \frac{1}{g^2} \Bigg[\int_{\mathcal{M}_4^R} d^4x \Big\{(R_I \wedge \star R_I) + \frac{1}{\ell^2}R_I \wedge \star \big(\xi\wedge\xi^T\big) + \frac{1}{\ell^4}\big(\xi\wedge\xi^T\big) \wedge \star \big(\xi\wedge\xi^T\big) \Big\} \nonumber \\
    &\quad \hspace{2em} + \int_{\mathcal{M}_4^R} d^4x \Big\{(R_{II} \wedge \star R_{II}) + \frac{1}{\ell^2}R_{II} \wedge \star \big(\xi^T\wedge\xi \big) + \frac{1}{\ell^4}\big(\xi^T\wedge\xi \big) \wedge \star \big(\xi^T\wedge\xi\big)\Big\} \Bigg]\\
    &= \frac{1}{g^2} \Big(S_{I} + S_{II} \Big)
\end{align}
The terms $S_I$ and $S_{II}$ collect the contributions resulting from $R_I$ with $\xi\wedge\xi^T$ and from $R_{II}$ with $\xi^T\wedge\xi$, respectively. This split is enforced by the transformation properties (\ref{tetradbilinearL}--\ref{tetradbilinearR}): the right bilinear is independent of the left curvature and vice versa. In the subsequent 4D formulation, these terms will be reexpressed in Plebanski form, with $\xi\wedge\xi^T$ and $\xi^T\wedge\xi$ transformed into $SU(2)$-valued two-forms that satisfy simplicity constraints on each spacetime sector. 

The last terms in both $S_I$ and $S_{II}$ are quartic in $\xi$. These terms typically play the role of an effective cosmological constant term (or, more generally, a potential for the symmetry-breaking order parameter), as they involve only  soldering and no stabilizer curvature. In the proposed framework, these terms are also the natural place wherein interface/gluing conditions and additional compatibility constraints will be incorporated once the two-sector picture is implemented and the $\xi$ fields are restricted to overlapping 4D spacetime sectors.

This completes the $SO(3,3)\to SU(2)_R\times SU(2)_L$ breaking at the level of connection and action. In the next section we will reinterpret the two stabilizer sectors as corresponding to two overlapping 4D spacetimes of opposite Lorentz signature and identify the induced Plebanski two-forms that lead to gravity (self-dual) and weak interactions (anti-self-dual) in the appropriate spacetime sectors.

\section{First- and Second-Spacetime Sectors}\label{sec:firstandsecondspacetimesectors}
\noindent\textbf{Why an overlap of the two four-dimensional leaves?}
At this stage the symmetry breaking has produced two chiral sectors to which we assign distinct low-energy
interpretations: one sector corresponds to a Plebanski-type (gravitational-branch) dynamics,
whereas the other corresponds to a Yang--Mills/Higgs (electroweak-branch) dynamics.
These two sectors are treated as coexisting on two separate effective
four-dimensional Lorentzian sectors $M_4^{\mathrm I}$ and $M_4^{\mathrm{II}}$ within the parent $(3,3)$-signature geometry. However, we do not introduce an overlap $S := M_4^{\mathrm I}\cap M_4^{\mathrm{II}}$ merely as a geometric embedding convenience. Rather, the overlap isolates precisely the locus on which the two chiral geometries can be compared and consistently related without forcing them to collapse into a single four-dimensional metric.\\

\noindent
From the BF/Plebanski viewpoint, $S$ functions as a genuine corner/interface. Because the two Lorentzian sectors
carry opposite Lorentz signatures, their canonical structures and internal Lorentz frames are not
a priori compatible.
A consistent variational principle and gauge invariance in the presence of such an interface
requires additional \emph{interface data} (edge modes) and accompanying \emph{gluing constraints}
that match the pullbacks of the chiral fields across $S$.
In this sense, the overlap captures the \emph{only} place where ``gravi--weak cross-talk'' can reside:
any effective coupling between the Lorentzian sectors must be mediated by (or at least consistent with) the interface
edge-mode/gluing algebra, and conversely the same structure controls how strongly these two sectors
can communicate in the low-energy effective theory.\\

\noindent
This also sharpens the program for future work.
Once the reduced interface phase space is derived and quantized (including the corner simplicity
constraints and the edge-mode degrees of freedom), one can in principle compute:
(i) the allowed matching maps between the two leaf geometries (including any conformal
mismatch on $S$), and (ii) the size of any induced mixing or inter-sector effects.
Therefore, the overlap is not an auxiliary assumption but a predictive bottleneck: it localizes the
new degrees of freedom implied by the two-leaf construction and provides a concrete target for
deriving (rather than postulating) any residual gravi--weak couplings.\\

The 4D base manifolds \(M_4^{I}\) and \(M_4^{II}\) are equipped with coordinates \(x^\mu \in \{t, x, y, z\}\) and \(\tilde{x}^\mu \in \{\tilde{x}, \tilde{t}_1,\tilde{t}_2,\tilde{t}_3\}\), respectively, with the corresponding metrics $g^{(I)}_{\mu\nu}$ and $g^{(II)}_{\tilde{\mu}\tilde{\nu}}$. In local frames we use
\begin{align}
\eta_{\mu\nu} = \text{diag}(-1, +1, +1, +1) \\
\tilde{\eta}_{\tilde{\mu}\tilde{\nu}} = \text{diag}(+1, -1, -1, -1)
\end{align}
First, we define the triplet of two-forms as follows:
\begin{align}
    \Sigma^{c} &= \frac{1}{2\ell^2}\varepsilon^{c}\mathstrut_{ab} \xi^{a}_{i} \wedge \xi^{ib} = \frac{1}{2\ell^2}\varepsilon^{c}\mathstrut_{ab} \big(\xi \wedge \xi^{T}\big)^{ab}  \label{sigmafirst}\\
    \tilde{\Sigma}^{k} &= \frac{1}{2\ell^2}\tilde{\varepsilon}^{k}\mathstrut_{ij} \xi^{i}_{a} \wedge \xi^{aj} = \frac{1}{2\ell^2}\tilde{\varepsilon}^{k}\mathstrut_{ij} \big(\xi^{T} \wedge \xi\big)^{ij} \label{sigmasecond}
\end{align}
where, $\varepsilon^{c}\mathstrut_{ab}$ and $\tilde{\varepsilon}^{k}\mathstrut_{ij}$ stem from  (\ref{epsilonmap1}) and (\ref{epsilonmap2}).\\
For the coset one-forms $\xi = \xi_{\mu} dx^{\mu}$ in the first spacetime sector,
\begin{align}
    \big(\xi^{a}_{i} \wedge \xi^{ib}\big)_{\mu\nu} = 2 \xi^{a}\mathstrut_{i[\mu}\xi^{bi}\mathstrut_{\nu]}
\end{align}
Likewise, for the the coset one-forms $\xi = \xi_{\tilde{\mu}} dx^{\tilde{\mu}}$ in the second spacetime sector,
\begin{align}
    \big(\xi^{i}_{a} \wedge \xi^{ja}\big)_{\tilde{\mu}\tilde{\nu}} = 2 \xi^{i}\mathstrut_{a[\tilde{\mu}}\xi^{ja}\mathstrut_{\tilde{\nu}]}
\end{align}
For each of these sectors, we construct the corresponding self-dual and anti-self-dual triple of two-forms using (\ref{sigmafirst}) and (\ref{sigmasecond}) as follows:
\begin{align}
    \Sigma^{(+)c} &= \frac{1}{2}\Big( \mathbbm{1} + \frac{1}{\sqrt{\sigma_{I}}}\star_{I} \Big)\Sigma^{c} \label{tetrad2formsd}\\
    \tilde{\Sigma}^{(-)k} &= \frac{1}{2}\Big( \mathbbm{1} - \frac{1}{\sqrt{\sigma_{II}}}\star_{II} \Big)\tilde{\Sigma}^{k} \label{tetrad2formasd}
\end{align}
where, the Hodge dual operators $\star_{I}$ and $\star_{II}$ correspond to \(M_4^{I}\) and \(M_4^{II}\), respectively.\\

Before specialising to the two chiral Plebanski sectors, it is useful to state explicitly how the
coset–bilinear two-forms $\Sigma^{c}$ and $\tilde\Sigma^{k}$ defined in
\eqref{sigmafirst}--\eqref{sigmasecond} relate to the standard Plebanski 2-forms that will be used
to describe the non-degenerate branches on the individual spacetime sectors.  The 2-forms $\Sigma^{c}$ and
$\tilde\Sigma^{k}$ are constructed as bilinears of the coset (soldering) one-forms $\xi^{a}{}_{i}$
and transform in the adjoint representations of the respective stabilizer $\mathfrak{su}(2)$
factors.  Their Hodge projections $\Sigma^{(+)c}$ and $\tilde\Sigma^{(-)k}$ in
\eqref{tetrad2formsd}--\eqref{tetrad2formasd} isolate the self-dual component with respect to the metric on $M_4^{I}$ and the anti-self-dual component with respect to the metric on
$M_4^{II}$. Therefore, the chiral two-forms employed in the forthcoming Plebanski-type construction acquire a clear algebraic origin: they correspond to the
self/anti-self dual components of coset bilinears constructed from the soldering one-forms resulting in the
$SO(3,3)\to SU(2)_R\times SU(2)_L$ reduction.\\

On the non-degenerate gravitational branches, solving the simplicity constraints ensures the existence of a local
tetrad (coframe) $e^{(\mathcal I)I}$ on $M_4^{I}$ (and $e^{(\mathcal{II})I}$ on $M_4^{II}$). After selecting an appropriate frame and normalization
scale $\ell$, the projected coset two-forms can be identified with the Plebanski self-dual and anti-self-dual 2-forms. This identification canonically fixes the correspondence between the 
adjoint index of the coset $\mathfrak{su}(2)$ factor and the corresponding internal Lorentz index (spatial or temporal, depending on the spacetime sector) employed in the Plebanski-type construction. \\

\noindent In the subsequent sections, the two-forms (\ref{tetrad2formsd}) and (\ref{tetrad2formasd}) will be subject to appropriate simplicity constraints to describe gravity in both \(M_4^{I}\) and \(M_4^{II}\).
\subsection{First D=4 Spacetime Sector (Region \texorpdfstring{$\mathcal{I}$}{I}): Gravity Sector}
\subsubsection{Selfdual two-forms}
In Region $\mathcal{I}$, we employ the self-dual Plebanski 2-forms constructed from the tetrad (coframe) $e^{(\mathcal{I})I}_\mu$ as

\begin{equation}\label{eq:Plebanski_sd}
\Sigma^{(+)i} = e^{(\mathcal{I})0} \wedge e^{(\mathcal{I})i} + \frac{1}{2}\sqrt{\sigma}\,\epsilon^{ijk} e^{(\mathcal{I})}_{j} \wedge e^{(\mathcal{I})}_{k}
\end{equation}
where 
$e^I$ are 1-form coframe fields carrying internal Lorentz indices $I \in \{0,1,2,3\}$. We adopt the convention that $e^{(\mathcal{I})0}$ denotes the temporal 1-form, whereas $e^{(\mathcal{I})i}$ ($i=1,2,3$) denote spatial 1-forms.
The parameter $\sigma = \pm 1$ encodes the orientation ($+1$ corresponds to  right-handed orientation and $-1$ to left-handed orientation). Furthermore,  $\epsilon^{ijk}$ is the Levi-Civita symbol on the spatial indices.\\

The Plebanski 2-forms encode the antisymmetric wedge products of the tetrad in a manner adapted to self-duality. The first term $e^0 \wedge e^i$ captures the components associated with extrinsic geometry, characterizing how spatial hypersurfaces are embedded in spacetime, whereas the second term $\epsilon^{ijk} e_j \wedge e_k$ represents the intrinsic spatial geometry. The relative mixing between these contributions is governed by $\sqrt{\sigma}$, which is closely related to the choice of self-dual or anti-self-dual sector and, in broader formulations, to the Immirzi parameter.

\subsubsection{Simplicity Constraints} We now introduce the simplicity constraints,
which play a central role in BF-type formulations of gravity. Although the unconstrained BF theory is topological and carries no local degrees of freedom, the imposition of simplicity constraints restricts the 
$\Sigma$-field to be “simple,” i.e. a 2-form expressible in terms of a 1-form. This reduction introduces local gravitational degrees of freedom and ensures that the resulting theory is dynamically equivalent to general relativity. In this sense, the simplicity constraints are the mechanism by which geometric spacetime structures, and hence gravity, emerge from an underlying topological field theory.\\

\textbf{Quadratic form}:
The simplicity constraint in its original quadratic form ensures that the $\Sigma$-field has the structure of Plebanski 2-forms. Explicitly, the constraint is expressed as follows:

\begin{equation}\label{eq:simplicity_quad_I}
C^{(+)ij}_{\text{quad}} := \Sigma^{(+)i} \wedge \Sigma^{(+)j} - \frac{1}{3}\delta^{ij} \Sigma^{(+)k} \wedge \Sigma^{(+)k} = 0
\end{equation}

This condition ensures the vanishing of the traceless part of the symmetric $3 \times 3$ matrix of 2-forms $\Sigma^i \wedge \Sigma^j$. Consequently, it constitutes a system of 5 independent equations, corresponding to the five independent components of a traceless symmetric matrix in three dimensions.

In addition, one imposes the degeneracy constraint,
\begin{align}
    V^{(+)ij}_{\text{quad}} :=\frac{1}{3}\delta^{ij} \Sigma^{(+)k} \wedge \Sigma^{(+)k} \neq 0 \label{eq:nondegeneracy}
\end{align}
to ensure that the quadratic constraints do not yield degenerate solutions.\\

\textbf{Linear form}: 
The quadratic form is challenging to implement in quantum theory because products of constraints become complicated. The linear form trades a more restrictive constraint (fewer solutions) for simpler mathematics. Therefore, in recent formulations (particularly important for spin foam quantization programs such as Engle--Pereira--Rovelli--Livine (EPRL) \cite{engle2008flipped, engle2008lqg, livine2008solving} and Freidel-Krasnov (FK) \cite{freidel2008new, livine2007new} models), the simplicity constraint is rewritten in a linear form that is easier to handle mathematically (amenable to quantization). \\

This requires introducing an auxiliary set of 3-forms $n^I$ (one 3-form for each internal Lorentz index) \cite{gielen2010classical}:
\begin{align}
    n^{I}_{\mu\nu\sigma} = \frac{1}{3!}\epsilon^{IJKL}e_\mu^J \wedge e_\nu^K \wedge e_\sigma^L 
\end{align}
where, $\mu, \nu, \sigma \in \{0,1,2,3\}$ are coordinate indices corresponding to $x^\mu \in \{t,x,y,z\}$. The linear simplicity constraints are expressed as
\begin{equation}\label{eq:simplicity_lin_I}
C^{(+)I}_{\text{lin}} := \langle n^I, \Sigma^{(+)IJ} \rangle = \epsilon_{IJKL} n^{I} \cdot \Sigma^{(+)KL}= 0
\end{equation}
Here, $\langle \cdot ,\cdot \rangle$ denotes a contraction of Lorentz indices only. In addition, one must also impose volume constraints:

\begin{equation}\label{eq:volume_constraint_I}
V^{(+)K} := \langle n^I, \Sigma^{(+)IJ} \rangle \wedge e_K = 0
\end{equation}

\subsubsection{Reality Conditions} To obtain a real Lorentzian metric from the chiral variables one must impose
appropriate reality conditions on \(\Sigma^{(+)}\) once the simplicity constraints hold. We assume the standard Plebanski
reality conditions are imposed.
\begin{align}
    \Sigma^{IJ} \wedge \bar{\Sigma}^{KL}=0 \\
    \Sigma^{IJ} \wedge \Sigma^{KL} + \bar{\Sigma}^{IJ} \wedge \bar{\Sigma}^{KL} = 0
\end{align}

\subsubsection{SU(2) Action with constraints} Region $\mathcal{I}$ has signature $(-, +, +, +)$ and is formulated as self-dual BF theory with simplicity constraints. The SU(2) part of the action (using quadratic constraints) is

\begin{equation}\label{eq:S_I}
\boxed{S_\mathcal{I} = \int_{M_\mathcal{I}} \left[ \Sigma^{(+)i} \wedge F^i(A) - \frac{\lambda}{2} \Sigma^{(+)i} \wedge \Sigma^{(+)i} + \frac{1}{2}\varphi^{(+)}_{ij} \Sigma^{(+)i} \wedge \Sigma^{(+)j} \right] d^4x}
\end{equation}

Here, \textbf{$\Sigma^i_{(+)} \wedge F^i(A)$} is the BF kinetic term, $\Sigma^i_{(+)}$ are self-dual 2-forms, $F^i(A) = dA^i + \frac{1}{2}\epsilon^{ijk} A_j \wedge A_k$ is the $\mathfrak{so}(3)$ curvature, and \textbf{$-\frac{\lambda}{2} \Sigma^i_{(+)} \wedge \Sigma^i_{(+)}$} is a quadratic term that, in conjunction with the simplicity constraints, encodes gravity. The constant $\lambda$ is related to the cosmological constant $\Lambda = 3\lambda/\kappa^2$ where $\kappa$ is related to Newton's constant. Furthermore, \textbf{$\varphi^{(+)}_{ij}$} are the Lagrange multiplier fields (traceless, symmetric $3 \times 3$ matrices) that enforce the simplicity constraints.\\

Alternatively, the SU(2) part of the action (using linear constraints) is
\begin{equation}\label{eq:action_I_full}
S_\mathcal{I} = \int_{M_\mathcal{I}} \left[ \Sigma^{(+)i} \wedge F^i(A) + \mu^{(+)}_J \langle n^I, \Sigma^{(+)IJ} \rangle \right]
\end{equation}

\subsection{Second D=4 Spacetime Sector (Region \texorpdfstring{$\mathcal{II}$}{II}): Weak Gauge Sector}
The second four-dimensional sector, denoted as Region $\mathcal{II}$, is the complementary sector in the proposed construction and carries the opposite Lorentz signature relative to Region $\mathcal{I}$. The key structural claim is that the same MacDowell--Mansouri/BF-to-gravity mechanism that yields Einstein dynamics in Region $\mathcal{I}$ admits a second, equally geometric realization on Region $\mathcal{II}$, but now in the opposite chiral sector. This is the point at which the graviweak interpretation becomes available: once the $SO(3,3)$ symmetry is reduced to two chiral $SU(2)$ factors, one $SU(2)$ naturally encodes the gravitational (self-dual) sector on Region $\mathcal{I}$, whereas the other $SU(2)$ governs the anti-self-dual sector on Region $\mathcal{II}$. 

To connect this second $SU(2)$ to the Standard Model, we propose the following identification: the $SU(2)$ connection that appears as the anti-self-dual (chiral) connection on Region $\mathcal{II}$ is interpreted as the weak isospin gauge field $SU(2)_L$. Conceptually, this is motivated by two mutually compatible viewpoints.

\begin{itemize}
    \item \textbf{Geometric viewpoint on Region $\mathcal{II}$.} On the Region $\mathcal{II}$ spacetime sector, the anti-self-dual $SU(2)$ connection is part of a chiral description of a real (flipped-signature) spacetime geometry, in direct analogy with the self-dual Plebanski formulation of gravity. In this perspective, Region $\mathcal{II}$ is genuinely “gravitational,” albeit in the opposite-signature and opposite-chirality sector.
    \item \textbf{Gauge-theoretic viewpoint from Region $\mathcal{I}$.} When the two spacetime sectors are coupled only via the interface degrees of freedom (and appropriate gluing constraints), the Region $\mathcal{II}$ chiral connection is accessed from Region $\mathcal{I}$ as an internal $SU(2)$ gauge field: it is a connection on the internal chiral bundle singled out by the $SO(3,3)\to SU(2)\times SU(2)$ breaking, rather than a connection on the physical Lorentz group of Region $\mathcal{I}$. This matches the role of weak isospin as an internal gauge symmetry acting chirally on matter.
\end{itemize}

In the proposed framework, we remain at the classical level and do not yet introduce a full Standard Model matter sector; nevertheless, the above identification is structurally consistent with the chiral nature of the weak interaction. Herein, because chirality is geometric, it is tied to the anti-self-dual choice on the flipped-signature Region $\mathcal{II}$ sector.

\subsubsection{Antiselfdual two-forms}
In Region $\mathcal{II}$, we employ the anti-self-dual analogue of Plebanski 2-forms constructed from the tetrad (coframe) $e^{(\mathcal{II})I}_\mu$ as
\begin{equation}\label{eq:Plebanski_asd}
\tilde{\Sigma}^{(-)i} = e^{(\mathcal{II})0} \wedge e^{(\mathcal{II})i} - \frac{1}{2}\sqrt{\sigma}\,\epsilon^{ijk} e^{(\mathcal{II})}_{j} \wedge e^{(\mathcal{II})}_{k}
\end{equation}
We adopt the convention that $e^{(\mathcal{II})0}$ denotes the spatial 1-form, whereas $e^{(\mathcal{II})i}$ ($i=1,2,3$) denote temporal 1-forms in the second spacetime with flipped signature.

Given this assignment, \eqref{eq:Plebanski_asd} packages the tetrad wedge products into an internal $SU(2)$ triplet adapted to \emph{anti}-self-duality. In contrast with Region $\mathcal{I}$, the key structural change is the sign in front of the $\epsilon^{ijk}e_j\wedge e_k$ term, which selects the opposite chiral sector under the Hodge dual defined by the Region $\mathcal{II}$ metric.

Geometrically, $\tilde{\Sigma}^{(-)i}$ encodes the same information as the tetrad wedge products, but in a basis compatible with the anti-self-dual decomposition of two-forms in Region $\mathcal{II}$. In the present construction this selection is natural: the flipped signature sector is considered as being governed by the opposite chirality, implying that Region $\mathcal{I}$ and Region $\mathcal{II}$ are related by exchanging self-dual and anti-self-dual sectors while keeping the internal $SU(2)$ structure analogous.

\subsubsection{Simplicity Constraints}
As in Region $\mathcal{I}$, the anti-self-dual theory becomes gravitational only after imposing simplicity constraints, which restrict the admissible two-forms to those resulting from tetrads. In Region $\mathcal{II}$, these constraints are imposed on $\Sigma^{(-)i}$ and select the non-degenerate ``metric'' branch compatible with the flipped-signature geometry.

\textbf{Quadratic form}: The quadratic simplicity constraints enforce that the symmetric matrix $\tilde{\Sigma}^{(-)i}\wedge \tilde{\Sigma}^{(-)j}$ is proportional to $\delta^{ij}$:
\begin{equation}\label{eq:simplicity_quad_II}
C^{(-)ij}_{\text{quad}} := \tilde{\Sigma}^{(-)i} \wedge \tilde{\Sigma}^{(-)j} - \frac{1}{3}\delta^{ij}\,\tilde{\Sigma}^{(-)k} \wedge \tilde{\Sigma}^{(-)k} = 0.
\end{equation}
As before, this removes the traceless symmetric part and yields five independent constraints.

To exclude degenerate configurations and remain on the gravitational branch, we impose the non-degeneracy condition
\begin{align}
    V^{(-)ij}_{\text{quad}} :=\frac{1}{3}\delta^{ij}\,\tilde{\Sigma}^{(-)k} \wedge \tilde{\Sigma}^{(-)k} \neq 0. \label{eq:nondegeneracy_II}
\end{align}

\textbf{Linear form}: A linear form of the simplicity constraints may also be introduced (useful, for instance, in quantization strategies where linear constraints are technically more tractable). Analogous to Region $\mathcal{I}$, we introduce auxiliary 3-forms $n^I$ constructed from the Region $\mathcal{II}$ tetrad:
\begin{align}
    n^{I}_{\tilde{\mu}\tilde{\nu}\tilde{\sigma}}
    = \frac{1}{3!}\epsilon^{IJKL}\,
    e^{(\mathcal{II})J}_{\tilde{\mu}} \wedge e^{(\mathcal{II})K}_{\tilde{\nu}} \wedge e^{(\mathcal{II})L}_{\tilde{\sigma}},
\end{align}
where $\tilde{\mu},\tilde{\nu},\tilde{\sigma}\in\{0,1,2,3\}$ label the coordinates $\tilde{x}^\mu\in\{\tilde{x}, \tilde{t}_1,\tilde{t}_2,\tilde{t}_3\}$ on $M_4^{II}$.
The linear simplicity constraints in Region $\mathcal{II}$ are then expressed as
\begin{equation}\label{eq:simplicity_lin_II}
C^{(-)I}_{\text{lin}} := \langle \tilde{n}^I, \tilde{\Sigma}^{(-)IJ} \rangle
= \epsilon_{IJKL}\, n^{I} \cdot \tilde{\Sigma}^{(-)KL}= 0,
\end{equation}
where $\langle\cdot,\cdot\rangle$ denotes contraction of internal Lorentz indices only.
In addition, one may impose the associated volume-type constraints
\begin{equation}\label{eq:volume_constraint_II}
V^{(-)K} := \langle \tilde{n}^I, \tilde{\Sigma}^{(-)IJ} \rangle \wedge e_K = 0,
\end{equation}
which accompany the linear simplicity conditions in the same spirit as in Region $\mathcal{I}$.

\subsubsection{Reality Conditions}
Although the anti-self-dual variables are naturally complex in a chiral formulation, the spacetime geometry of Region $\mathcal{II}$ is taken to be real (with flipped signature). Therefore, once simplicity holds, one must impose suitable reality conditions on the chiral two-forms. As in Region $\mathcal{I}$, we assume the standard Plebanski reality conditions:
\begin{align}
    \tilde{\Sigma}^{IJ} \wedge \overline{\tilde{\Sigma}}^{KL}=0, \\
    \tilde{\Sigma}^{IJ} \wedge \tilde{\Sigma}^{KL} + \overline{\tilde{\Sigma}}^{IJ} \wedge \overline{\tilde{\Sigma}}^{KL} = 0.
\end{align}
These conditions are understood here to apply to the anti-self-dual sector constructed from $e^{(\mathcal{II})I}$.

\subsubsection{SU(2) Action with constraints}
Region $\mathcal{II}$ has the flipped signature $(+, -, -, -)$ and is formulated as an \emph{anti}-self-dual BF theory supplemented by simplicity constraints. The $SU(2)$ part of the constrained action (quadratic simplicity) is
\begin{equation}\label{eq:S_II}
\boxed{S_{\mathcal{II}} = \int_{M_{\mathcal{II}}} \left[
\tilde{\Sigma}^{(-)i} \wedge F^i(\widetilde{A})
- \frac{\tilde{\lambda}}{2}\,\tilde{\Sigma}^{(-)i} \wedge \tilde{\Sigma}^{(-)i}
+ \frac{1}{2}\varphi^{(-)}_{ij}\,\tilde{\Sigma}^{(-)i} \wedge \tilde{\Sigma}^{(-)j}
\right] d^4\tilde{x} }
\end{equation}
Here $\tilde{\Sigma}^{(-)i}\wedge F^i(\widetilde{A})$ is the BF kinetic term for the anti-self-dual sector, and
\begin{equation}
F^i(\widetilde{A}) = d\widetilde{A}^i + \frac{1}{2}\epsilon^{ijk}\,\widetilde{A}_j \wedge \widetilde{A}_k
\end{equation}
is the $\mathfrak{so}(3)$ curvature of the anti-self-dual $SU(2)$ connection $\widetilde{A}^i$. The term
$-\frac{\tilde{\lambda}}{2}\tilde{\Sigma}^{(-)i}\wedge \tilde{\Sigma}^{(-)i}$ plays the same structural role as in Region $\mathcal{I}$, now for the anti-self-dual sector on $M_4^{II}$; $\tilde{\lambda}$ is related to the effective cosmological constant on Region $\mathcal{II}$ (up to the conventions set by the overall normalization of the action). Finally, $\varphi^{(-)}_{ij}$ are traceless symmetric Lagrange multipliers enforcing the quadratic simplicity constraints \eqref{eq:simplicity_quad_II}.

Alternatively, using linear constraints, the $SU(2)$ part of the Region $\mathcal{II}$ action may be expressed as
\begin{equation}\label{eq:action_II_full}
S_{\mathcal{II}} = \int_{M_{\mathcal{II}}} \left[
\tilde{\Sigma}^{(-)i} \wedge F^i(\widetilde{A})
+ \mu^{(-)}_J \,\langle \tilde{n}^I, \tilde{\Sigma}^{(-)IJ} \rangle
\right],
\end{equation}
where $\mu^{(-)}_J$ are Lagrange multipliers enforcing the linear simplicity constraint \eqref{eq:simplicity_lin_II} (and, if desired, the auxiliary volume-type conditions \eqref{eq:volume_constraint_II}).
On the non-degenerate branch \eqref{eq:nondegeneracy_II}, the constrained theory reproduces the gravitational sector associated with the Region $\mathcal{II}$ tetrad data, now expressed in the anti-self-dual chiral language appropriate to the flipped-signature spacetime.

\subsection{2D Interface (Region \texorpdfstring{$\mathcal{S}$}{S})}
In our framework, the two spacetimes with flipped signatures intersect along a common two-dimensional interface with signature $(-+)$. However, when a gravitational system has a boundary or an interface, the standard variational principle becomes ill-posed. The bulk action contains a  boundary integral that involves both the fields and their normal derivatives, creating over-constraints. This issue is resolved by supplementing the action with a boundary counter-term whose variation exactly cancels the unwanted boundary contributions. This counter-term is identified with the presymplectic potential $\Theta$.

\medskip
\noindent
\textit{Remark:} Although both quadratic and linear constraints are presented for completeness, we will adopt the linear form when analyzing the corner structure.

\subsubsection{Presymplectic Potential:} 
At the 2D interface surface $\mathcal{S}$, the presymplectic structure reduces to 2D. The presymplectic potential is given as follows \cite{freidel2020edge}:
\begin{equation}\label{eq:theta_S}
\Theta^S = \int_S \left[ \delta \tilde{E}_I \wedge \delta n^I - \frac{\beta}{2} \delta \tilde{e}^I \wedge \delta \tilde{e}_I \right]
\end{equation}
where,
$\tilde{E}^I$ denotes the flux variable, canonically conjugate to the internal normal direction and
$n^I$ is the internal normal field orthogonal to S. \\

The first term $\delta \tilde{E}_I \wedge \delta n^I$ represents the extrinsic structure and encodes how S is embedded in the bulk regions. The second term $\delta \tilde{e}^I \wedge \delta \tilde{e}_I$ represents the intrinsic 2D geometry.

The associated presymplectic 2-form, obtained as the exterior derivative of the potential, is given by

\begin{equation}\label{eq:omega_S}
\Omega^S = \delta \Theta^S = \int_S \left[ \delta^2 \tilde{E}_I \wedge \delta n^I - \frac{\beta}{2} \delta^2 \tilde{e}^I \wedge \delta \tilde{e}_I \right]
\end{equation}
where the double variations represent perturbations in the solution space.

The induced 2D metric on the interface S, constructed from the coframe, is given as
\begin{equation}\label{eq:corner_metric}
q_{ab} = \tilde{e}^I_a \tilde{e}^J_b \eta_{IJ}
\end{equation}
where $\eta_{IJ} = \text{diag}(-1, +1)$ on S (signature $(-, +)$).

The corner metric components satisfy a \textbf{non-commutative Poisson algebra}:

\begin{equation}\label{eq:corner_algebra}
\{q_{ab}(x), q_{cd}(y)\} = -\frac{1}{\beta} \left[ q_{ac}\epsilon_{bd} + q_{bc}\epsilon_{ad} + q_{ad}\epsilon_{bc} + q_{bd}\epsilon_{ac} \right] \delta^{(2)}(x,y)
\end{equation}
where $\epsilon_{ab}$ is the Levi-Civita tensor on S with signature $(-, +)$.\\

The corner Poisson brackets close under a larger algebra referred to as the \textbf{corner symmetry algebra}:

\begin{equation}
\mathfrak{g}^S = \text{Diff}(S) \ltimes [\mathfrak{sl}(2, \mathbb{C})^S \oplus \mathfrak{sl}(2, \mathbb{R})_\parallel^S]
\end{equation}

This is a semidirect product of:
$\text{Diff}(S)$, the diffeomorphisms (coordinate reparametrizations) of the 2D interface;
$\mathfrak{sl}(2, \mathbb{C})^S$, the complex Lorentz algebra generated by $\tilde{E}^I \wedge n^I$, corresponding to boosts perpendicular to S; and
$\mathfrak{sl}(2, \mathbb{R})_\parallel^S$, the real Lorentz algebra generated by $\tilde{e}^I \wedge \tilde{e}_I$, corresponding to rotations within S.\\

The semidirect product notation reflects the action of diffeomorphisms on the internal symmetry generators.

\subsubsection{Corner constraints:} 
When the simplicity constraint is pulled back to the 2D interface S, the constraint decomposes into first-class and second-class components.

The corner simplicity constraint (from the bulk) takes the following form \cite{freidel2021edge}:
\begin{equation}\label{eq:constr_simp_corner}
\mathcal{C}^S_a = B^I_a - \frac{1}{\beta} S^I_a \approx 0
\end{equation}
where, $B^I_a$ denotes the boost component of the flux and
$S^I_a$ represents the spin component of the frame fields.
Both are pulled back to the 2D surface.

Furthermore, we introduce the 2D Hodge star $\star_S$, which allows for a decomposition of the constraint into holomorphic and anti-holomorphic components:

\begin{equation}\label{eq:holomorphic_decomp}
C^+_a := \frac{1}{2}(C_a + i \star_S C_a) \quad \text{(holomorphic/second-class)}
\end{equation}

\begin{equation}\label{eq:antiholomorphic_decomp}
C^-_a := \frac{1}{2}(C_a - i \star_S C_a) \quad \text{(anti-holomorphic/first-class)}
\end{equation}

The $i$ and Hodge star $\star_S$ (the 2D Hodge dual on signature $(-, +)$ surface) create a special algebraic structure.

 In two dimensions with Lorentzian signature, the Hodge star naturally separates self-dual and anti-self-dual components, closely mirroring the decomposition encountered in the bulk theory.

The holomorphic component $C^+_a$ cannot be eliminated by any gauge transformation and therefore constitutes a set of second-class constraints. By contrast, the anti-holomorphic component $C^-_a$ generates gauge transformations and is thus first-class.

The second-class constraint $C^+_a = 0$ generates the algebra $\mathfrak{sl}(2,\mathbb{R})^S_{\parallel}$, which coincides precisely with the non-commutative metric algebra discussed earlier. This constraint cannot be gauge-fixed away; it fundamentally limits the corner phase space.

Consistently, the Poisson brackets among the second-class constraints close on themselves:
\begin{equation}
\{C^+_a(x), C^+_b(y)\} \propto (\text{corner metric and flux terms}) \delta^{(2)}(x,y)
\end{equation}

In the bulk, both self-dual and anti-self-dual sectors of the Lorentz algebra are available to generate gauge transformations, and the simplicity constraints, being first-class, can be used to eliminate redundancy.

However, at the mixed-signature interface, the distinction between temporal and spatial directions becomes ambiguous. This ambiguity obstructs the full gauge interpretation of the simplicity constraint: one sector (the holomorphic component) becomes second-class and acquires the status of a genuine geometric constraint.

Consequently, the corner geometry emerges as a physical degree of freedom exhibiting an intrinsically non-commutative structure.

\subsubsection{Gluing constraints:} 
The overlap of the two spacetimes results in incompatible canonical structures at the interface. Specifically,

In \textbf{Region I}, the canonical Poisson bracket algebra is:

\begin{equation}
\{e^I_\mu(x), \omega^{JK}_\nu(y)\}_I \propto \delta_\mu^\nu \delta^I_J \delta^{(3)}(x-y)
\end{equation}

In \textbf{Region II}, the canonical algebra is:

\begin{equation}
\{e^I_\mu(x), \omega^{JK}_\nu(y)\}_{II} \propto \delta_\mu^\nu \delta^I_J \delta^{(3)}(x-y)
\end{equation}

Although these algebras appear formally identical, they are defined on spacetimes with opposite signatures. Consequently, the generators $\omega^{JK}$ acquire distinct physical interpretations:
\begin{itemize}
\item in Region $\mathcal{I}$, $\omega^{JK}$ generates boosts and rotations in a spacetime of signature $(-,+,+,+)$;
\item in Region $\mathcal{II}$, $\omega^{JK}$ generates Lorentz transformations in a spacetime of signature $(+,+,+,-)$.
\end{itemize}

Therefore, we introduce edge modes $\phi(x)$, which are auxiliary dynamical fields present exclusively on the 2D interface S. The edge mode $\phi$ acts as a frame-matching field, specifying how to translate the reference frame (internal Lorentz labeling) from Region II's convention into Region I's convention.

These edge modes take values in the group
\begin{equation}\label{eq:edge_mode_field}
\phi(x) \in SL(2, \mathbb{C})^S
\end{equation}
a six-dimensional (real) Lie group parametrizing 3 spatial rotations and 3 Lorentz boosts.

These edge mode enforces the following first-class constraint (the gluing condition):

\begin{equation}\label{eq:constr_glue}
\mathcal{G}[\phi] = \Sigma^i_{(+)}|_S - U_\phi \cdot \Sigma^i_{(-)}|_S \approx 0
\end{equation}
where $U_\phi \in SL(2, \mathbb{C})^S$ is the edge mode matrix, acting on the internal Lorentz indices.

More explicitly, if $U_\phi = (U_\phi)^i_j$, the gluing reads

\begin{equation}\label{eq:gluing_explicit}
\Sigma^i_{(+)}|_S = (U_\phi)^i_j \Sigma^j_{(-)}|_S
\end{equation}

Therefore, dressing the $\Sigma$-field from Region II by the edge-mode transformation yields precisely the $\Sigma$-field as observed from Region I.

The constraint $\mathcal{G}[\phi] = 0$ is first-class because of the following:
\begin{enumerate}
\item It commutes with all other constraints (with appropriate correction terms)
\item It generates a gauge symmetry: if $\phi(x) \to \phi(x) g(x)$ for any $g(x) \in SL(2, \mathbb{C})^S$, the gluing condition remains satisfied (the $g$ factors cancel).
\item For any solution satisfying the gluing condition, you can arbitrarily transform $\phi \to \phi g$, and you get an equivalent solution (same physical state).
\end{enumerate}

First-class constraints generate gauge redundancy rather than physical constraints. Accordingly, the edge mode $\phi$ does not represent a physical degree of freedom, but rather encodes the relative choice of internal Lorentz frames across the interface.

To implement the gluing condition in the action, we introduce the boundary term
\begin{equation}\label{eq:gluing_action}
\boxed{S_{\text{glue}} = \int_S \mu^i(x) \left[ \Sigma^i_{(+)}|_S - U_\phi(x) \cdot \Sigma^i_{(-)}|_S \right]}
\end{equation}
where $\mu^i(x)$ are Lagrange multiplier fields (3-component vectors on S).

Variation with respect to $\mu^i$ enforces the gluing condition $\mathcal{G}[\phi] = 0$, and variation with respect to $\phi$ yields equations that determine $\phi$ in terms of the bulk $\Sigma$-fields from the two spacetimes, rendering $\phi$ dynamical.

\section{Complete action incorporating coupling constants and the \texorpdfstring{$\mathrm{SU}(2)\times \mathrm{U}(1)$}{SU(2) x U(1)} sector}
\label{sec:couplngconstantactionandu1}
In this section we consolidate the Plebanski-type actions in the two spacetime regions $M_{\mathcal I}$ and $M_{\mathcal{II}}$ in conjunction with gluing constraints formulated in the previous section and explain how the bare parameters $\lambda$, $\tilde\lambda$ and $\varphi^{(\pm)}_{ij}$ are related to the physical constants: Gravitational constant $G_{N}$, the cosmological constants $\Lambda_{\mathcal I}$ and $\Lambda_{\mathcal{II}}$, the weak gauge coupling $g_{\text{YM}}$ and the Fermi constant $G_{F}$. A single dimensionless ``master coupling'' $g$ is associated with the full action and controls both gravitational and weak interactions after the appropriate phase structure is imposed. Finally, we provide the extension to $SU(2) \times U(1)$ on both spacetime regions $M_{\mathcal I}$ and $M_{\mathcal{II}}$.

We start from a chiral Plebanski formulation split into two spacetime regions, $M_{\mathcal I}$ and $M_{\mathcal{II}}$, separated by a three-dimensional hypersurface $S$. The complete action is given by,
\begin{align}
    \boxed{S = \frac{1}{g^2} \Big( S_{\mathcal{I}} + S_{\mathcal{II}} + S_{\text{glue}} \Big)}
\end{align}
where $g$ is a dimensionless master coupling. The three contributions are defined as follows.

\paragraph{Region $\mathcal I$: Self-dual (gravity) sector.}
In the first region $M_{\mathcal I}$ with signature $(-,+,+,+)$, we employ a self-dual SU(2) triplet of 2-forms $\Sigma^{(+)i}$ constructed from a coframe $e^{I}_{(\mathcal I)}$.
The action is as follows.
\begin{equation}
  S_{\mathcal I}
  =
  \int_{M_{\mathcal I}}
  \left[
    \Sigma^{(+)i} \wedge F^{i}(A)
    - \frac{\lambda}{2}\,\Sigma^{(+)i} \wedge \Sigma^{(+)i}
    + \frac{1}{2}\,\varphi^{(+)}_{ij}\,
      \Sigma^{(+)i} \wedge \Sigma^{(+)j}
  \right]
  \label{eq:SI-bare}
\end{equation}
Here $A^{i}$ is an $\mathfrak{su}(2)$ connection with curvature $F^{i}(A)$, $\lambda$ is a dimensionful parameter, and $\varphi^{(+)}_{ij}$ are Lagrange multipliers enforcing the simplicity constraints which ensure that $\Sigma^{(+)i}$ comes from a tetrad and encodes a metric on $M_{\mathcal I}$.

\paragraph{Region $\mathcal{II}$: Anti-self-dual (weak) sector.}
In the second region $M_{\mathcal{II}}$ with flipped signature $(+,-,-,-)$, we use an anti-self-dual SU(2) triplet of 2-forms $\tilde\Sigma^{(-)i}$ constructed from a coframe $e^{I}_{(\mathcal{II})}$. The action is
\begin{equation}
  S_{\mathcal{II}}
  =
  \int_{M_{\mathcal{II}}}
  \left[
    \tilde\Sigma^{(-)i} \wedge F^{i}(\tilde A)
    - \frac{\tilde\lambda}{2}\,\tilde\Sigma^{(-)i} \wedge \tilde\Sigma^{(-)i}
    + \frac{1}{2}\,\varphi^{(-)}_{ij}\,
      \tilde\Sigma^{(-)i} \wedge \tilde\Sigma^{(-)j}
  \right]
  \label{eq:SII-bare}
\end{equation}
where $\tilde A^{i}$ an $\mathfrak{su}(2)$ connection on $M_{\mathcal{II}}$ and $\tilde\lambda, \varphi^{(-)}_{ij}$ are the analogous parameters and Lagrange multipliers in this region.

\paragraph{Gluing term.}
Across the hypersurface $S$, the two chiral sectors are glued by imposing a generalized matching condition on the 2-forms:
\begin{equation}
  S_{\text{glue}}
  =
  \int_{S} \mu^{i}(x)\,
  \Big[
    \Sigma^{(+)i}\big|_{S}
    - U_{\phi}(x) \cdot \tilde\Sigma^{(-)i}\big|_{S}
  \Big],
  \label{eq:Sglue}
\end{equation}
where $\mu^{i}(x)$ are Lagrange multipliers and $U_{\phi}(x)\in SU(2)$ is a local gauge transformation relating the two SU(2) frames on the interface. This term enforces continuity (up to gauge rotation) of the chiral 2-forms across $S$ and guarantees a consistent variational principle for the combined system.

\subsection{
Region \texorpdfstring{$\mathcal{I}$}{I}: 
Identification of \texorpdfstring{$G_{N}$}{GN} and \texorpdfstring{$\Lambda_{\mathcal{I}}$}{LambdaI}}

We now show how the parameters $g$ and $\lambda$ in \eqref{eq:SI-bare} are associated with Newton's constant $G_{N}$ and the cosmological constant $\Lambda_{\mathcal I}$ in the usual Einstein--Hilbert description.

Varying $S_{\mathcal I}$ with respect to $\Sigma^{(+)i}$ yields
\begin{equation}
  F^{i}(A)
  - \lambda\,\Sigma^{(+)i}
  + \varphi^{(+)}_{ij}\,\Sigma^{(+)j}
  = 0,
\end{equation}
or equivalently
\begin{equation}
  F^{i}(A)
  = \lambda\,\Sigma^{(+)i}
    - \varphi^{(+)}_{ij}\,\Sigma^{(+)j}.
  \label{eq:F-Sigma-EOM}
\end{equation}
We decompose the symmetric multipliers into a traceless part and a trace part:
\begin{equation}
  \varphi^{(+)}_{ij}
  = \Psi^{(+)}_{ij} + \lambda\,\delta_{ij},
  \qquad
  \Psi^{(+)}_{i}{}^{i} = 0.
  \label{eq:phi-decomp}
\end{equation}
Substituting \eqref{eq:phi-decomp} into \eqref{eq:F-Sigma-EOM} and defining $\Psi^{(+)i}{}_{j}:= -\Psi^{(+)}_{ij}$ gives the standard chiral GR form
\begin{equation}
  F^{i}(A)
  = \Psi^{(+)i}{}_{j}\,\Sigma^{(+)j}
    + \lambda\,\Sigma^{(+)i}.
  \label{eq:F-decomp-lambda}
\end{equation}
Comparison with the usual decomposition of the self-dual curvature,
\begin{equation}
  F^{i}(A)
  = \Psi^{(+)i}{}_{j}\,\Sigma^{(+)j}
    + \frac{\Lambda_{\mathcal I}}{3}\,\Sigma^{(+)i},
  \label{eq:F-decomp-L}
\end{equation}
shows that, at the level of the local curvature equation, the coefficient of $\Sigma^{(+)i}$ is $\lambda = \Lambda_{\mathcal I}/3$. We now verify this by reducing the action to its metric form.

The simplicity constraints enforced by $\varphi^{(+)}_{ij}$ imply that the $\Sigma^{(+)i}$ can be written as the Plebanski self-dual 2-forms built from a tetrad $e^{I}_{(\mathcal I)}$. One then has the
well-known identities
\begin{align}
  \int_{M_{\mathcal I}} \Sigma^{(+)i}\wedge F^{i}(A)
  &= \int_{M_{\mathcal I}} d^{4}x\,\sqrt{-g_{(\mathcal I)}}\,
     R\big(g_{(\mathcal I)}\big),
  \label{eq:SigmaF-to-R}
  \\
  \int_{M_{\mathcal I}} \Sigma^{(+)i}\wedge\Sigma^{(+)i}
  &= 6 \int_{M_{\mathcal I}} d^{4}x\,\sqrt{-g_{(\mathcal I)}}.
  \label{eq:SigmaSigma-to-vol}
\end{align}
Inserting \eqref{eq:SigmaF-to-R}--\eqref{eq:SigmaSigma-to-vol} into \eqref{eq:SI-bare} and restoring the overall $1/g^{2}$ gives
\begin{equation}
  S_{\mathcal I}
  = \frac{1}{g^{2}}
    \int_{M_{\mathcal I}} d^{4}x\,\sqrt{-g_{(\mathcal I)}}\,
    \left[
      R\big(g_{(\mathcal I)}\big)
      - 6\lambda
      + \ldots
    \right],
  \label{eq:SI-metric}
\end{equation}
where the terms encoded by $\Psi^{(+)}_{ij}$ (i.e.\ the Weyl curvature squared contributions in the chiral
language) have been ignored.

The Einstein--Hilbert action with cosmological constant in
$M_{\mathcal I}$ is
\begin{equation}
  S_{\text{EH}+\Lambda}^{(\mathcal I)}
  = \frac{1}{16\pi G_{N}}
    \int_{M_{\mathcal I}} d^{4}x\,\sqrt{-g_{(\mathcal I)}}\,
    \big[ R(g_{(\mathcal I)}) - 2\Lambda_{\mathcal I} \big].
  \label{eq:SEH-metric}
\end{equation}
Matching the coefficients of the Ricci scalar and of the volume term between \eqref{eq:SI-metric} and \eqref{eq:SEH-metric} yields
\begin{equation}
  \frac{1}{g^{2}} = \frac{1}{16\pi G_{N}},
  \qquad
  6\lambda = 2\Lambda_{\mathcal I}
  \;\Longrightarrow\;
  \Lambda_{\mathcal I} = 3\lambda.
  \label{eq:GN-Lambda-Id}
\end{equation}
Therefore, the master coupling $g$ fixes the gravitational constant via $G_{N} = g^{2}/(16\pi)$, whereas $\lambda$ determines the cosmological constant in Region $\mathcal I$ as $\Lambda_{\mathcal I} = 3\lambda$.\\

\noindent The $SU(2)_{R}$ action for the first spacetime (along with the coupling constants) is expressed as
\begin{align}
    \boxed{S_{\mathcal{I}} = \frac{1}{16\pi G_{N}}\int \left( \Sigma^{(+)} \wedge R^{(+)} - \frac{1}{2}\left(\varphi^{(+)} + \frac{\Lambda}{3} \mathbbm{I} \right) \Sigma^{(+)} \wedge \Sigma^{(+)} \right)}\label{eq:SI-coupling}
\end{align}
The corresponding equations of motion are as  follows:
\begin{align}
    \frac{\delta S_{\mathcal{I}}}{\delta A^{(+)}} = 0 &\implies \mathrm{D}\Sigma^{(+)} \equiv \mathrm{d}\Sigma^{(+)} + A^{(+)} \wedge A^{(+)} = 0\hspace{2em} \\
    \frac{\delta S_{\mathcal{I}}}{\delta \Sigma^{(+)}} = 0 &\implies \boxed{R^{(+)} + \left( \varphi^{(+)} + \frac{\Lambda}{3}\mathbbm{I}\right)\Sigma^{(+)} = 0} \hspace{2.5em} (\text{vacuum EFE})\\
    \frac{\delta S_{\mathcal{I}}}{\delta \varphi^{(+)}} = 0 &\implies \Sigma^{(+)}\wedge \Sigma^{(+)} - \frac{\Lambda}{3}\mathbbm{I} \left(\Sigma^{(+)}\wedge \Sigma^{(+)}\right) = 0
\end{align}

\subsection{Region \texorpdfstring{$\mathcal{II}$}{II}: 
Weak \texorpdfstring{$\mathrm{SU}(2)$}{SU(2)} coupling and Fermi constant}
In contrast with conventional graviweak scenarios where the anti-self-dual sector is driven into a purely Yang--Mills phase, in our construction the region $M_{\mathcal{II}}$ is itself described by a bona fide Plebanski gravity theory. The bare action \eqref{eq:SII-bare} is structurally identical to \eqref{eq:SI-bare}, but constructed from an anti-self-dual triplet
$\tilde\Sigma^{(-)i}$ and an independent connection $\tilde A^{i}$.

Variation with respect to $\tilde\Sigma^{(-)i}$ yields
\begin{equation}
  F^{i}(\tilde A)
  = \tilde\lambda\,\tilde\Sigma^{(-)i}
    - \varphi^{(-)}_{ij}\,\tilde\Sigma^{(-)j}.
  \label{eq:FII-EOM}
\end{equation}
Decomposing the multipliers into traceless and trace parts analogous to \eqref{eq:phi-decomp},
\begin{equation}
  \varphi^{(-)}_{ij}
  = \Psi^{(-)}_{ij} + \tilde\lambda\,\delta_{ij},
  \qquad
  \Psi^{(-)}_{i}{}^{i} = 0,
  \label{eq:phi-minus-decomp}
\end{equation}
Imposing the anti-self-dual simplicity constraints in $M_{\mathcal{II}}$ expresses $\tilde\Sigma^{(-)i}$ as Plebanski two-forms built from a tetrad $e^{I}_{(\mathcal{II})}$ with flipped signature $(+---)$, and the bulk action reduces to an Einstein--Hilbert form. 

In this sense, Region $\mathcal{II}$ is described by an anti-self-dual gravity theory with a gravitational constant 
$G_{W} = g^{2}/(16\pi)$ 
(set by the overall prefactor \(1/g^{2}\)) and its own cosmological constant $\Lambda_{\mathcal{II}} = 3\tilde\lambda$; however, its connection appears as an internal $SU(2)_{L}$ gauge field to observers in Region $\mathcal I$.

To make contact with the Standard Model, we now explain how the anti-self-dual connection $\tilde A^{i}$ in Region $\mathcal{II}$ gives rise to the internal weak isospin gauge field seen in Region $\mathcal I$ and how its coupling is related to the parameters of the Plebanski action.

As constructed in the previous section, let $S$ be the common codimension-2 interface, with induced 2D Lorentzian metric $h_{ab}$ and volume form $\epsilon_{S}$. The gluing constraint \eqref{eq:Sglue} implies, in particular, a matching of the induced
connections on $S$ up to a local $SU(2)$ rotation $U_{\phi}(x)$:
\begin{equation}
  A^{i}\big|_{S}
  \simeq U_{\phi}(x)\cdot \tilde A^{i}\big|_{S}\cdot U_{\phi}(x)^{-1}
         + U_{\phi}(x)\,dU_{\phi}(x)^{-1}.
  \label{eq:connection-matching}
\end{equation}
We define the ``boundary weak isospin connection" by
\begin{equation}
  A^{i}_{w}
  := \tilde A^{i}\big|_{S},
  \label{eq:Aw-def}
\end{equation}
which is an $\mathfrak{su}(2)$-valued one-form on $S$. 

From the perspective of Region $\mathcal I$, $A^{i}_{w}$ furnishes the restriction to $S$ of an $SU(2)_{L}$ gauge field $A^{i}_{\mu}(x)$ defined on $M_{\mathcal I}$ and enters covariant derivative of a left-handed Weyl spinor $\psi_{L}^{\alpha i}$ in Region $\mathcal I$ as
\begin{equation}
  D_{\mu}\psi_{L}^{\alpha i}
  = \partial_{\mu}\psi_{L}^{\alpha i}
    + \omega_{\mu}{}^{\alpha}{}_{\beta}\,\psi_{L}^{\beta i}
    + i g_{w} A_{\mu}{}^{j} (T_{j})^{i}{}_{k}\,\psi_{L}^{\alpha k},
  \label{eq:fermion-cov-deriv}
\end{equation}
where $\omega_{\mu}{}^{\alpha}{}_{\beta}$ is the spin connection in Region $\mathcal I$, $T_{j}$ are the generators of $SU(2)$ in the appropriate representation, and $g_{w}$ is the weak isospin coupling.
Right-handed spinors are singlets under this internal $SU(2)$, aligning with the Standard Model:
\begin{equation}
  D_{\mu}\psi_{R}^{\dot\alpha}
  = \partial_{\mu}\psi_{R}^{\dot\alpha}
    + \omega_{\mu}{}^{\dot\alpha}{}_{\dot\beta}\,\psi_{R}^{\dot\beta}.
\end{equation}

At the level of the gauge-field dynamics, the effective weak-sector action in Region $\mathcal I$ is of the Yang--Mills form
\begin{equation}
  S_{\text{weak}}^{(\mathcal I)}
  = -\frac{1}{4g_{w}^{2}}
    \int_{M_{\mathcal I}} d^{4}x\,\sqrt{-g_{(\mathcal I)}}\,
      \mathrm{Tr}\big( F_{\mu\nu}(A_{w}) F^{\mu\nu}(A_{w}) \big)
  + S_{\text{Higgs}} + S_{\text{ferm}},
  \label{eq:S-weak-YM}
\end{equation}

\noindent\textit{Remark (first-order BF--YM form)}.\footnote{Yang-Mills theory can be formulated as a deformed (topological) BF theory \cite{baez1996four, cattaneo1998four}.}
Although we have written the weak gauge-field dynamics directly in the second-order Yang--Mills form
\eqref{eq:S-weak-YM}, it is useful---particularly when comparing with BF-type formulations---to rewrite
the kinetic term in first-order form by introducing an auxiliary $\mathfrak{su}(2)$-valued two-form
$B_{w}$ on $M_{\mathcal I}$. Equivalently, one may take
\begin{equation*}
  S_{\text{YM}}^{(\mathcal I)}[A_{w},B_{w}]
  = \frac{1}{g_{w}^{2}}
    \int_{M_{\mathcal I}}
      \mathrm{Tr}\!\left(
        B_{w}\wedge F(A_{w})
        - \frac{1}{2}\,B_{w}\wedge \star_{(\mathcal I)} B_{w}
      \right),
\end{equation*}
where $F(A_{w})=\tfrac12 F_{\mu\nu}(A_{w})\,dx^{\mu}\wedge dx^{\nu}$ and $\star_{(\mathcal I)}$ is the Hodge
dual defined by the Region $\mathcal I$ metric $g_{(\mathcal I)}$. The $B_{w}$ equation of motion is
algebraic, $F(A_{w})=\star_{(\mathcal I)}B_{w}$ (equivalently $B_{w}=-\star_{(\mathcal I)}F(A_{w})$), and
substituting it back reproduces the $F_{\mu\nu}F^{\mu\nu}$ term in \eqref{eq:S-weak-YM}. In particular,
dropping the quadratic ``$B_{w}^{2}$'' term would return a topological BF theory with no propagating
Yang--Mills degrees of freedom; thus the appearance of this term is precisely what yields the
dynamical electroweak gauge sector once the Region $\mathcal I$ metric is available.

In Eqn. (\ref{eq:S-weak-YM})  $F_{\mu\nu}(A_{w})$ is the field strength of the internal $SU(2)_{L}$ connection extending from the boundary data \eqref{eq:Aw-def}. The value of $g_{w}$ is determined by matching this effective description to the gravitational normalization inherited from Region $\mathcal{II}$ through the corner.

To see this, consider a ``thin-slab'' neighbourhood of $S$ in $M_{\mathcal{II}}$, with local coordinates $(\ell_{1},\ell_{2},y^{a})$ such that $y^{a}$ parametrize $S$ and $(\ell_{1},\ell_{2})$ are normal
coordinates in the two independent orthogonal directions. The Einstein--Hilbert form can then be expressed schematically as
\begin{equation}
  S_{\mathcal{II}}
  \;\sim\;
  \frac{1}{16\pi G_{W}}
  \int d\ell_{1}\,d\ell_{2}
  \int_{S} d^{2}y\,\sqrt{-h}\, R\big(g_{(\mathcal{II})}\big),
\end{equation}
and, upon expanding $R(g_{(\mathcal{II})})$ in terms of the intrinsic curvature of $S$ and the extrinsic data of the normals, we obtain boundary kinetic terms for the induced connection $A^{i}_{w}$. Retaining
only the leading, curvature-squared contribution on $S$ (and integrating over the small normal directions) yields an effective action of the form
\begin{equation}
  S_{\text{weak},S}
  = -\frac{1}{4 g_{w}^{2}}
    \int_{S} d^{2}y\,\sqrt{-h}\,
      \mathrm{Tr}\big( F_{ab}(A_{w}) F^{ab}(A_{w}) \big)
  + \cdots,
  \label{eq:S-weak-S}
\end{equation}
with the identification
\begin{equation}
  \frac{1}{g_{w}^{2}}
  = c_{w}\,\frac{1}{16\pi G_{W}}
  = c_{w}\,\frac{1}{g^{2}},
  \label{eq:gw-vs-g}
\end{equation}
where $c_{w}$ is a (dimensionful) constant determined by the geometry and thickness of the neighbourhood of $S$ in $M_{\mathcal{II}}$. Equation \eqref{eq:gw-vs-g} expresses the weak isospin coupling $g_{w}$ in terms of the same master coupling $g$ that sets the gravitational scale in Region $\mathcal{II}$.

Finally, we connect the above geometric identification to the physical Fermi constant $G_{F}$. After making the identification \eqref{eq:Aw-def} in Region $\mathcal I$ and introducing the conventional Higgs sector that breaks $SU(2)_{L}\times U(1)_{Y}$\footnote{The $U(1)$ sector in our construction is explained in \hyperref[sec:su(2)xu(1)extension]{Section~\ref*{sec:su(2)xu(1)extension}}} to $U(1)_{\text{em}}$, the charged weak bosons acquire a mass
\begin{equation}
  M_{W}^{2}
  = \frac{1}{4} g_{w}^{2} v^{2},
\end{equation}
where $v$ is the electroweak vacuum expectation value. Integrating out the $W$ bosons at energies $E \ll M_{W}$ yields the four-fermion Fermi interaction
\begin{equation}
  \mathcal{L}_{\text{Fermi}}
  = -\frac{G_{F}}{\sqrt{2}}
    \bigl(
      \bar\psi\gamma^{\mu}P_{L}\psi
    \bigr)
    \bigl(
      \bar\psi\gamma_{\mu}P_{L}\psi
    \bigr),
\end{equation}
where
\begin{equation}
  \frac{G_{F}}{\sqrt{2}}
  = \frac{g_{w}^{2}}{8 M_{W}^{2}}
  = \frac{1}{2 v^{2}}.
  \label{eq:Gf-standard}
\end{equation}
Using \eqref{eq:gw-vs-g}, we can express $G_{F}$ in terms of the master coupling and the gravitational parameters of Region $\mathcal{II}$:
\begin{equation}
  G_{F}
  = \frac{g_{w}^{2}}{4\sqrt{2} M_{W}^{2}}
  = \frac{1}{4\sqrt{2} M_{W}^{2}}\frac{g^{2}}{c_{w}}
  = \frac{1}{4\sqrt{2} M_{W}^{2}}\frac{16\pi G_{W}}{c_{w}}
  \label{eq:GF-GW-Id}
\end{equation}

\noindent The $SU(2)_{L}$ action for the second spacetime is
\begin{align}
    \boxed{S_{\mathcal{II}} = \frac{1}{16\pi G_{W}}\int \left( \tilde{\Sigma}^{(-)} \wedge \tilde{R}^{(-)} - \frac{1}{2}\left(\varphi^{(-)}  + \frac{\tilde{\Lambda}}{3} \mathbbm{I} \right) \tilde{\Sigma}^{(-)} \wedge \tilde{\Sigma}^{(-)} \right)}\label{eq:SII-coupling}
\end{align}

\noindent The equations of motion are as  follows:
\begin{align}
    \frac{\delta S_{\mathcal{II}}}{\delta \tilde{A}^{(-)}} = 0 &\implies \mathrm{D}\tilde{\Sigma}^{(-)} \equiv \mathrm{d}\tilde{\Sigma}^{(-)} + \tilde{A}^{(-)} \wedge \tilde{A}^{(-)} = 0\hspace{2em} \\
    \frac{\delta S_{\mathcal{II}}}{\delta \tilde{\Sigma}^{(-)}} = 0 &\implies \boxed{\tilde{R}^{(-)} + \left( \varphi^{(-)} + \frac{\tilde{\Lambda}}{3}\mathbbm{I}\right)\tilde{\Sigma}^{(-)} = 0} \hspace{2.5em} (\text{Analogous vacuum EFE})\\
    \frac{\delta S_{\mathcal{II}}}{\delta \varphi^{(-)}} = 0 &\implies \tilde{\Sigma}^{(-)}\wedge \tilde{\Sigma}^{(-)} - \frac{\tilde{\Lambda}}{3}\mathbbm{I} \left(\tilde{\Sigma}^{(-)}\wedge \tilde{\Sigma}^{(-)}\right) = 0
\end{align}

\noindent\textit{Remark.} Using \eqref{eq:GN-Lambda-Id} and \eqref{eq:GF-GW-Id}, we observe that the gravitational constants for both the spacetime sectors are equal, $G_{N}=G_{W}$.

\subsection{Extension to \texorpdfstring{$\mathrm{SU}(2)\times \mathrm{U}(1)$}{SU(2) x U(1)}}\label{sec:su(2)xu(1)extension}
A perspective central to the proposed framework is that the two spacetime sectors Region $\mathcal{I}$ and Region $\mathcal{II}$ admit a gauge-theoretic description in terms of a corresponding $SU(2)$ theory, which ultimately results from a spontaneously broken phase of a larger chiral symmetry $SU(2) \times U(1)$ for both sectors.\\

As stated in the Introduction section, the present study is situated within a broader $E_8 \times \omega E_8$ unification program \cite{Kaushik, Singh:2024ven, Singh:2025xxv}, wherein each $E_8$ sector branches to $SU(3)\times E_6$. The two $E_6$ sectors encode the matter and field content associated with the standard model and general relativity, whereas the two additional $SU(3)$s are related to the 6D spacetime and to the internal symmetry space \cite{Singh2026Preprint}. 

Within this setting, the proposed trinification of $E_6$ provides the following interpretation of the branching of the two $E_8$, which is discussed in detail in \cite{Singh:2025xxv}:
\begin{equation}
\begin{split}
E_{8L}  & \rightarrow SU(3)^{geom}_{L} \times E_{6L} \xrightarrow{E_{6L}} SU(3)_c \times SU(3)_{F,L} \times SU(3)_L \xrightarrow{SU(3)_L} SU(2)_L\times U(1)_Y  \\
E_{8R} & \rightarrow SU(3)^{geom}_{R} \times E_{6R} \xrightarrow {E_{6R}}  SU(3)_{c'} \times SU(3)_{F,R} \times SU(3)_R \xrightarrow{SU(3)_R} SU(2)_R\times U(1)_{Ydem} \\
\end{split}
\end{equation}
and
\begin{equation}
{\bf 248}  = (\bf 78, \bf 1) \oplus (\bf 1, \bf 8) \oplus (\bf 27, \bf 3) \oplus  (\overline{\bf 27}, \overline{\bf 3})
\end{equation} 
Of the three $SU(3)$s resulting from the branching of $E_{6L}$, the $SU(3)_c$ implements the color gauge symmetry of QCD. Furthermore, $SU(3)_{F,L}$ is the non-gauged global flavor symmetry which is responsible for three left-handed fermion generations \cite{Singh:2025xxv} described by the exceptional Jordan algebra $J_3({\mathbb O}_c)$. $SU(3)_L$ branches as $SU(2)_L \times U(1)_Y$ giving rise to the electroweak sector, and the $SU(2)_L$ acts only on left-handed particles. 

Regarding the branching of the second $E_6$, which we label $E_{6R}$, the  $SU(3)_{c'}$ symmetry is preferably not gauged (so as to be consistent with phenomenology) - it is global and explicitly broken at the electroweak scale. Its role is to rotate the so-called Jordan frame which arises in the Peirce decomposition of the exceptional Jordan algebra matrices \cite{Singh:2025xxv}. The $\omega SU(3)_{F,R}$ is the global flavor symmetry which describes three generations of right-handed fermions. The $SU(3)_R$ branches as $SU(2)_R \times U(1)_{Ydem}$. Here, the spontaneously broken $SU(2)_R$ gives rise to general relativity, and the unbroken $U(1)_{dem}$ resulting from $U(1)_{Ydem}\rightarrow U(1)_{dem}$ is a new force, dubbed dark electromagnetism. It is sourced by the square-root of mass and can be made cosmologically and phenomenologically safe. The emergence of general relativity in this manner, and the preceding gravi-weak unification, was briefly discussed in \cite{Kaushik} and its detailed analysis is the subject of the present paper.

Instead of working with the $SU(3)_L \times SU(3)_R$ Yang-Mills theory on 6D, we have worked with a BF theory on 6D. This is justified because in our theory, there is no dynamical metric prior to the electroweak symmetry breaking, and a topological theory is preferred before the breaking. The equivalence between an $SU(3)_L \times SU(3)_R$ YM and the present treatment using a BF theory is as follows. 

\paragraph{Yang--Mills as a BF theory (the ``Yang--Mills branch'').}
On any spacetime where a metric (hence a Hodge operator $\star$) is available, an ordinary
Yang--Mills theory admits a first-order BF-type rewriting with an auxiliary field.
For a gauge group $G$ with connection $A$ and curvature $F=dA+A\wedge A$, the standard
second-order action
\begin{equation}
S_{\rm YM}[A]
= -\frac{1}{2g^2}\int \Tr\!\big(F\wedge \star F\big)
\label{eq:YM_second_order}
\end{equation}
is classically equivalent to the first-order ``BF--YM'' action
\begin{equation}
S_{\rm BF\text{-}YM}[A,B]
= \int \Tr\!\big(B\wedge F\big)
-\frac{g^2}{2}\int \Tr\!\big(B\wedge \star B\big),
\label{eq:BFYM_first_order}
\end{equation}
where $B$ is an adjoint-valued $(D\!-\!2)$-form (so in $D=6$ one has $B\in\Omega^4(M_6,\mathfrak g)$,
while on an effective 4D leaf one has $B\in\Omega^2(M_4,\mathfrak g)$).
Varying \eqref{eq:BFYM_first_order} with respect to $B$ gives
\begin{equation}
F - g^2\,\star B = 0
\qquad\Longrightarrow\qquad
B = g^{-2}\,\star^{-1}F,
\label{eq:B_on_shell}
\end{equation}
and eliminating $B$ reproduces \eqref{eq:YM_second_order} (with the usual overall sign convention
fixed by the spacetime signature and the definition of $\star$).
In particular, for $G=SU(3)_L\times SU(3)_R$ one may take
$A=A_L\oplus A_R$, $B=B_L\oplus B_R$ and a sum of two copies of \eqref{eq:BFYM_first_order} with
couplings $g_L,g_R$.

This clarifies the status of using a \emph{pure} BF term in the pre-transition phase:
prior to electroweak symmetry breaking (in the premetric/topological regime) there is no
dynamical $\star$ available, so only the topological term $\int\Tr(B\wedge F)$ is natural.
After symmetry breaking, the emergence/localisation of 4D leaf geometries supplies the required
Hodge operators on the leaves, and the gauge sectors can then be placed on the \emph{Yang--Mills
branch} by including the quadratic $B\wedge\star B$ deformation, thereby recovering the familiar
$F_{\mu\nu}F^{\mu\nu}$ kinetic terms for the electroweak (and gravidem) gauge fields.

\paragraph{Where the $U(1)$ factors enter, and why $GL(2,\mathbb C)$ is a natural package.}
The $SO(3,3)\rightarrow SU(2)_R\times SU(2)_L$ breaking intrinsic to the gravi--weak BF sector
produces only the two $SU(2)$ stabilizer factors; it does \emph{not} generate additional abelian
generators. The required $U(1)$'s therefore have to come from the trinification branchings
\begin{equation}
SU(3) \ \longrightarrow\ SU(2)\times U(1),
\qquad
\mathfrak{su}(3)=\mathfrak{su}(2)\oplus \mathfrak{u}(1)\oplus(\mathbf 2_{+1}\oplus \mathbf 2_{-1}),
\qquad
Y \propto \mathrm{diag}(1,1,-2),
\label{eq:su3_to_su2_u1_decomp}
\end{equation}
(and similarly for the right-sector $U(1)_{Y_{\rm dem}}$).
On the unbroken $SU(2)$ doublet block, the $U(1)$ generator acts as a multiple of the identity,
so it is convenient to regard
\begin{equation}
SU(2)\times U(1) \simeq U(2)\quad(\text{up to a } \mathbb Z_2 \text{ quotient}),
\label{eq:su2u1_as_u2}
\end{equation}
and, in a chiral/complex formulation, to work with its complexification
\begin{equation}
U(2)^{\mathbb C}\ \simeq\ GL(2,\mathbb C),
\qquad
\mathfrak{gl}(2,\mathbb C)=\mathfrak{sl}(2,\mathbb C)\oplus \mathbb C\,\mathbf 1,
\qquad
\mathcal A = A^i \tau_i + A^0\,\mathbf 1.
\label{eq:gl2c_split}
\end{equation}
Here, the traceless part $A^i$ corresponds to the $SU(2)$ connection already present in the
$SO(3,3)$ reduction, while the trace part $A^0$ provides the abelian $U(1)$ connection (identified
with $U(1)_Y$ on the electroweak leaf and with $U(1)_{Y_{\rm dem}}$ on the gravidem leaf).
The physically compact $SU(2)\times U(1)$ is then recovered by the appropriate reality conditions
(selecting $U(2)\subset GL(2,\mathbb C)$) in conjunction with the charge normalisation inherited from the
parent $SU(3)$ embedding \eqref{eq:su3_to_su2_u1_decomp}.

Equivalently (and sometimes more transparently), one may keep the gravi--weak $SO(3,3)$ BF sector
unchanged and append commuting abelian factors at the level of the total gauge algebra,
$\mathfrak g_{\rm tot}=\mathfrak{so}(3,3)\oplus\mathfrak u(1)_Y\oplus\mathfrak u(1)_{Y_{\rm dem}}$,
with the $U(1)$'s understood as the unbroken remnants of $SU(3)_{L,R}\rightarrow SU(2)_{L,R}\times U(1)$.\\

\paragraph{Extension to SU(2)$\times$U(1).} To incorporate $SU(2) \times U(1)$ on both spacetime sectors, we can extend the gauge groups from $SL(2,\mathbbm{C})$ to $GL(2,\mathbbm{C})$ \cite{robinson1994formulation}. The action (\ref{eq:SI-coupling}) becomes
\begin{align}
    S_{I} &= \int \left( \Sigma^{(+)}\mathstrut_{SU(2)_{R}} \wedge R^{(+)}\mathstrut_{SU(2)_{R}} - \frac{1}{2}\left(\varphi^{(+)} + \frac{\Lambda}{3} \mathbbm{I} \right) \Sigma^{(+)}\mathstrut_{SU(2)_{R}} \wedge \Sigma^{(+)}\mathstrut_{SU(2)_{R}} \right) \nonumber \\
    &\quad \phantom{{}={}} + \left( \Sigma^{0}\mathstrut_{U(1)_{Ydem}} \wedge R^{0}\mathstrut_{U(1)_{Ydem}} - \frac{1}{2} k \Sigma^{0}\mathstrut_{U(1)_{Ydem}} \wedge \Sigma^{0}\mathstrut_{U(1)_{Ydem}}\right) - \varphi^{0} \Sigma^{(+)}\mathstrut_{SU(2)_{R}} \wedge \Sigma^{0}\mathstrut_{U(1)_{Ydem}} \\
    &= \int \left[ \left(\Sigma^{(+)}\mathstrut \wedge R^{(+)}\right)_{SU(2)_{R} \times U(1)_{Ydem}} - \frac{1}{2}
    \begin{pmatrix}
    \varphi^{(+)}  + \frac{\Lambda}{3} \mathbbm{I}  & \varphi^{0} \\
    \varphi^{0} & k 
    \end{pmatrix} \left( \Sigma^{(+)} \wedge \Sigma^{(+)} \right)_{SU(2)_{R} \times U(1)_{Ydem}}\right]
\end{align}
The field equations corresponding to the $U(1)_{Ydem}$ sector are as follows:
\begin{align}
    \frac{\delta S_{I}}{\delta \varphi} &= 0 \implies \Sigma^{(+)}\mathstrut_{SU(2)_{R}} \wedge \Sigma^{0}\mathstrut_{U(1)_{Ydem}} = 0 \\
    \frac{\delta S_{I}}{\delta \Sigma^{0}\mathstrut_{U(1)_{Ydem}}} &= 0 \implies k\Sigma^{0}\mathstrut_{U(1)_{Ydem}} = F^{2}_{Ydem}
\end{align}
See \cite{torres2011unified} for a more general constraint.\\

\noindent Likewise, the action (\ref{eq:SII-coupling}) becomes
\begin{align}
    S_{II} &= \int \left( \tilde{\Sigma}^{(-)}\mathstrut_{SU(2)_{L}} \wedge \tilde{R}^{(-)}\mathstrut_{SU(2)_{L}} - \frac{1}{2}\left(\varphi^{(-)}  + \frac{\tilde{\Lambda}}{3} \mathbbm{I} \right) \tilde{\Sigma}^{(-)}\mathstrut_{SU(2)_{L}} \wedge \tilde{\Sigma}^{(-)}\mathstrut_{SU(2)_{L}} \right) \nonumber \\
    &\quad \phantom{{}={}} + \left( \tilde{\Sigma}^{0}\mathstrut_{U(1)_{Y}} \wedge \tilde{R}^{0}\mathstrut_{U(1)_{Y}} - \frac{1}{2} \tilde{k} \tilde{\Sigma}^{0}\mathstrut_{U(1)_{Y}} \wedge \tilde{\Sigma}^{0}\mathstrut_{U(1)_{Y}}\right) - \tilde{\varphi}^{0} \tilde{\Sigma}^{(-)}\mathstrut_{SU(2)_{L}} \wedge \tilde{\Sigma}^{0}\mathstrut_{U(1)_{Y}} \\
    &= \int \left[ \left(\tilde{\Sigma}^{(-)}\mathstrut \wedge \tilde{R}^{(-)}\right)_{SU(2)_{L} \times U(1)_{Y}} - \frac{1}{2}
    \begin{pmatrix}
    \varphi^{(-)} + \frac{\tilde{\Lambda}}{3} \mathbbm{I}  & \tilde{\varphi}^{0} \\
    \tilde{\varphi}^{0} & \tilde{k} 
    \end{pmatrix} \left( \tilde{\Sigma}^{(-)} \wedge \tilde{\Sigma}^{(-)} \right)_{SU(2)_{L} \times U(1)_{Y}} \right]
\end{align}
The corresponding field equations involving the $U(1)_{Y}$ sector are as follows:
\begin{align}
    \frac{\delta S_{II}}{\delta \tilde{\varphi}^{0}} &= 0 \implies \Sigma^{(+)}\mathstrut_{SU(2)_{L}} \wedge \tilde{\Sigma}^{0}\mathstrut_{U_{Y}(1)} = 0 \\
    \frac{\delta S_{II}}{\delta \tilde{\Sigma}^{0}\mathstrut_{U_{Y}(1)}} &= 0 \implies \tilde{k}\tilde{\Sigma}^{0}\mathstrut_{U_{Y}(1)} = F^{2}_{Y}
\end{align}

\section{Discussion}\label{sec:discussionsection}
\subsection{Why the weak interaction can be both geometry and an internal gauge symmetry}
\label{subsec:weak_geometry_vs_internal_gauge}

A potential point of confusion in gravi--weak unification is the following.
In conventional QFT the electroweak interaction is described by a Yang--Mills
connection on a principal bundle over our $4$D spacetime; hence, it is an \emph{internal} symmetry.
By contrast, in our framework the weak interaction is also interpreted as the
(curved) geometry of a \emph{second} embedded $4$D spacetime of opposite signature,
overlapping with our spacetime in a neutral $(1,1)$ interface.
\footnote{See the kinematic two-leaf construction and overlap discussion in the accompanying note \cite{Singh2026Preprint}.}
The goal of this subsection is to state the precise mathematical sense in which these two viewpoints
are the \emph{same statement} in different guises.

\paragraph{(A) Leaf selection as a reduction of the $6$D frame bundle.}
Let $(M_6,g)$ be a $6$D pseudo-Riemannian manifold of signature $(3,3)$ and let $P_{\mathrm{SO}(3,3)}\to M_6$
be its orthonormal frame bundle. A choice of an embedded (or localized) $4$D ``leaf''
$\Sigma\subset M_6$ is equivalent, pointwise, to specifying a rank-$2$ oriented negative plane
$N_x\subset T_xM_6$ (the ``discarded'' directions) and its orthogonal complement
$W_x:=N_x^\perp\subset T_xM_6$ (the ``tangent-to-the-leaf'' directions). In this setting one has the stabilizer
\begin{equation}
H \;=\; \mathrm{Stab}(N)\;\simeq\; \mathrm{SO}(3,1)\times \mathrm{SO}(2),
\end{equation}
and the corresponding Lie-algebra splitting
\begin{equation}
\mathfrak{so}(3,3)\;=\;\mathfrak{so}(W)\;\oplus\;\mathfrak{so}(N)\;\oplus\;(W\wedge N),
\label{eq:so33_split}
\end{equation}
where $W\wedge N$ are precisely the generators that mix the $4$D leaf directions with the two
discarded directions.

\paragraph{(B) The same split turns a $6$D spin connection into ``gravity + gauge + Higgs''.}
Let $\Omega$ be an $\mathfrak{so}(3,3)$-valued connection one-form on $P_{\mathrm{SO}(3,3)}$ (e.g.\ the $6$D spin
connection, or the unified gauge connection in the BF/MM formulation). Under the reduction to $H$,
$\Omega$ decomposes uniquely as
\begin{equation}
\Omega \;=\; \omega \;+\; A \;+\; \Phi,
\qquad
\omega\in\Omega^1(\Sigma,\mathfrak{so}(W)),\;
A\in\Omega^1(\Sigma,\mathfrak{so}(N)),\;
\Phi\in\Omega^1(\Sigma, W\wedge N).
\label{eq:connection_decomp}
\end{equation}
The crucial observation is:
\begin{itemize}
\item $\omega$ is the ordinary $4$D Lorentz/spin connection on the leaf (gravity sector).
\item $A$ is a connection for the \emph{normal-frame rotations} on $N$, i.e.\ a genuine
      Yang--Mills gauge field from the $4$D viewpoint (for codimension-$2$ this is locally $\mathrm{SO}(2)\simeq \mathrm{U}(1)$).
\item $\Phi$ transforms as a tensor under $H$ (it is not a connection); it is the natural ``coset'' field.
      When the leaf selection is dynamical (Lorentz--Higgs mechanism), these coset components are the
      ones that are eaten / become massive, precisely because they would rotate the leaf directions into the
      discarded directions and hence do not preserve the leaf constraint.
\end{itemize}

\paragraph{(C) Why this is \emph{exactly} the standard notion of an internal gauge symmetry.}
A $4$D observer confined to $\Sigma$ can only use coordinates and tangent vectors in $W$.
Therefore, local transformations that rotate only the discarded directions $N$ at fixed base-point $x\in\Sigma$
act \emph{vertically} on the reduced bundle: they do not move $x$ and are invisible as spacetime transformations.
Equivalently, for any matter field $\psi$ which is a section of a vector bundle associated to the reduced
principal bundle $P_H|_\Sigma$, a local normal rotation $h(x)\in \mathrm{SO}(N)$ acts as
\begin{equation}
\psi(x)\;\mapsto\;\rho\!\big(h(x)\big)\,\psi(x),
\qquad x\ \text{fixed},
\end{equation}
which is precisely what ``internal'' gauge symmetry means in QFT. The corresponding gauge potential is $A$ in
\eqref{eq:connection_decomp}, and its curvature is the $\mathfrak{so}(N)$ component of the $6$D curvature
$F(\Omega)$.

\paragraph{(D) Where the \texorpdfstring{$\mathrm{SU}(2)$}{SU(2)} fits, and why ``two extra timelike directions'' can still match it.}
At the purely kinematic level of selecting a $4$D leaf in a $(3,3)$ bulk, the normal rotation group of a rank-$2$
plane is $\mathrm{SO}(2)\simeq \mathrm{U}(1)$, so one should not expect an \emph{unbroken} $\mathrm{SU}(2)$ to arise
from codimension-$2$ normal rotations alone. The appearance of $\mathrm{SU}(2)$ in the gravi--weak program instead
uses additional structure present in the $(3,3)$ bulk (e.g.\ the compact $\mathrm{SU}(2)_L\times \mathrm{SU}(2)_R$
acting isometrically on the two $\mathrm{Im}\,\mathbb{H}$ factors in the $(3+3)$ split), in conjunction with the specific
symmetry breaking pattern that yields $\mathrm{SU}(2)\times \mathrm{U}(1)$ sectors on the two leaves.
From the $4$D perspective, the two directions exclusive to the ``weak leaf'' are not macroscopic propagation
directions; they parameterize an internal orbit space (often described in our present article as a $2$-dimensional
$\mathrm{SU}(2)$ symmetry space such as $S^2\simeq \mathbb{CP}^1$), while the full $\mathrm{SU}(2)$ acts as the local
structure group on the associated bundle.\footnote{This is the precise sense in which two extra directions can
correspond to an $\mathrm{SU}(2)$ internal structure: the directions correspond to a coset/orbit (dimension $2$),
while the gauge group acting on the fiber is $\mathrm{SU}(2)$ (dimension $3$).}

\paragraph{(E) Physical interpretation.}
In summary: the ``weak interaction as geometry of a second $4$D spacetime'' is the statement that the same unified
$6$D connection $\Omega$ induces a genuine spacetime spin connection on the second leaf, whereas the ``weak
interaction as an internal gauge symmetry'' is the statement that, upon restricting to our leaf, the components
of $\Omega$ that rotate (or encode the orientation of) the directions transverse to our leaf act vertically and
therefore appear as internal gauge fields. The overlap/interface ensures that both leaves can be compared
covariantly, while the leaf constraint (implemented kinematically via projectors or dynamically via Lorentz--Higgs
fields) ensures that the extra directions are not freely propagating macroscopic dimensions.

\subsection{MacDowell--Mansouri breaking, Lorentz--Higgs order parameters, and the electroweak Higgs mechanism}

It is useful to separate three distinct notions of ``symmetry breaking'' that appear in gravi--weak unification.

\paragraph{(i) MacDowell--Mansouri (MM) type breaking in the BF formulation.}
In the present work, the starting point is an $SO(3,3)$ BF-type theory in $D=6$.
The MM mechanism is implemented by splitting the curvature into stabilizer and coset pieces via an
idempotent projector $\Pi$ onto the stabilizer subalgebra, so that the effective action depends on
$\Pi(F)$ rather than on the full $F$.

Upon restriction to an appropriate $4$-dimensional leaf and the gravitational (simplicity) branch,
this yields a first-order Plebanski form of gravity.
This MM-type breaking is therefore a geometric reduction and branch selection in a gauge-gravity
(BF) system; it is not the particle-physics Higgs mechanism.

\paragraph{(ii) ``Lorentz--Higgs'' order parameters and localization of the two $4$D leaves.}
In the broader $D=6$ geometric picture, one may also introduce Higgs-like order parameters whose role is
to select (or localize) two overlapping $4$-dimensional leaves inside $M_6$, thereby reducing the
$SO(3,3)$ symmetry to the corresponding Lorentz subgroup on each leaf.
This use of the word ``Higgs'' refers to Lorentz/group reduction in spacetime geometry, and should not be
confused with electroweak symmetry breaking in internal gauge space.

\paragraph{(iii) Electroweak Higgs breaking on the electroweak leaf.}
Independently of (i)--(ii), a realistic electroweak sector requires spontaneous symmetry breaking
\begin{equation}
SU(2)_L\times U(1)_Y \longrightarrow U(1)_{\rm em},
\end{equation}
generated by a Higgs vacuum expectation value. In particular, the physical photon--$Z$ mixing is produced
by the Higgs VEV, rather than by any ``kinetic mixing'' originating from the overlap of generators in the
unified $SO(3,3)$ connection (the latter can at most induce portal-like couplings between sectors).

\subsection{Do we need an \texorpdfstring{$\mathrm{SU}(2)_R\times \mathrm{U}(1)\to \mathrm{U}(1)$}{SU(2)R x U(1) -> U(1)} Higgs step to recover GR?}

In the present BF/MM framework, general relativity is recovered by selecting the gravitational branch
(simplicity constraints) of the chiral $SU(2)$ sector, i.e.\ the Plebanski/Palatini form of gravity.
This does not require Higgs-breaking the gravitational chiral $SU(2)$ down to $U(1)$.
If an additional right-sector breaking pattern such as
$SU(2)_R\times U(1)_{Y_{\rm dem}}\to U(1)_{\rm dem}$ is invoked for dark electromagnetism or mass-ratio
phenomenology \cite{Singh:2025xxv}, then it constitutes an additional dynamical assumption and must be accompanied by a clear
specification of the corresponding order parameter, its scale, and the decoupling of any heavy vector
modes, while preserving the gravitational simplicity branch.

We would also like to make the important observation that $SU(2)_R \times U(1)_{Ydem}$ is a perturbatively renormisable quantum field theory, like the electroweak theory,  and unlike GR and unlike the Fermi four-point interaction. In this manner, by bringing in $U(1)_{Ydem}$, we circumvent the vexing problem of quantizing general  relativity. Instead, GR is seen as an emergent classical theory which is not to be quantised per se, just as the Fermi interaction is not to be quantised.

\subsection{Ultra-soft breaking in the right-handed gravidem sector:
\texorpdfstring{$\linebreak \mathrm{SU}(2)_R \times \mathrm{U}(1)_{Y_{\rm dem}} \rightarrow \mathrm{U}(1)_{\rm dem}$}
{SU(2)R x U(1){Ydem} -> U(1)dem}}
\label{sec:discussion_ultrasoft_RH}

A central viewpoint adopted in this work is that (on our leaf) the gravitational sector
admits a gauge-theoretic description in terms of an $SU(2)_R$ theory which is ultimately
to be understood as a spontaneously broken phase of a larger right-handed symmetry
$SU(2)_R \times U(1)_{Y_{\rm dem}}$. In parallel, an unbroken $U(1)_{\rm dem}$ subgroup
is singled out, which we associate with dark electromagnetism and the mass--charge
dictionary used in the square-root mass programme \cite{Singh:2025xxv}.

\paragraph{Electroweak-analogue Higgs breaking on the right.}
To mimic electroweak symmetry breaking in the right-handed (``gravidem'') sector, we posit
a Higgs field $\Phi_R$ transforming non-trivially under
\begin{equation}
G_{\rm dem} \;:=\; SU(2)_R \times U(1)_{Y_{\rm dem}}
\qquad\longrightarrow\qquad
U(1)_{\rm dem}\,,
\end{equation}
with a vacuum expectation value that selects the unbroken generator
\begin{equation}
Q_{\rm dem} \;\equiv\; T^3_R + \frac{1}{2}\,Y_{\rm dem}\,,
\qquad\text{so that}\qquad
Q_{\rm dem}\,\langle \Phi_R\rangle = 0\,.
\end{equation}
Writing the $SU(2)_R$ and $U(1)_{Y_{\rm dem}}$ gauge fields as $W^i_{R}$ and $B_{\rm dem}$,
with couplings $g_R$ and $g'_{\rm dem}$, the standard diagonalisation defines
\begin{align}
W_R^\pm &:= \frac{1}{\sqrt{2}}\left(W_R^1 \mp i W_R^2\right),\\[2pt]
A_{\rm dem} &:= \sin\theta_R\, W_R^3 + \cos\theta_R\, B_{\rm dem},\\
Z_R &:= \cos\theta_R\, W_R^3 - \sin\theta_R\, B_{\rm dem},\\
\tan\theta_R &:= \frac{g'_{\rm dem}}{g_R}\,,
\end{align}
with the mass pattern (for a minimal doublet-like breaking)
\begin{equation}
m_{W_R} \;=\; \frac{1}{2}g_R v_R,\qquad
m_{Z_R} \;=\; \frac{1}{2}v_R\sqrt{g_R^2+(g'_{\rm dem})^2},\qquad
m_{A_{\rm dem}} \;=\; 0\,.
\end{equation}

\paragraph{Ultra-soft (practically unbroken) regime on observable scales.}
We focus on the phenomenological corner in which the right-handed massive vectors are
\emph{ultralight},
\begin{equation}
m_{W_R},\,m_{Z_R} \;\ll\; H_0\,,
\end{equation}
so that their Compton lengths $\lambda_{W_R,Z_R}\sim 1/m_{W_R,Z_R}$ exceed the Hubble radius.
Equivalently, for any experimentally accessible distance $r$ one has $e^{-m r}\simeq 1$.
In this regime, the right-handed sector is observationally close to the \emph{unbroken}
$SU(2)_R\times U(1)_{Y_{\rm dem}}$ phase as far as range-suppression effects are concerned,
while still possessing a well-defined \emph{selected} unbroken subgroup $U(1)_{\rm dem}$
(which is crucial for the dark charge assignment and the associated mass-ratio framework).

A technically natural way to realize the ultra-soft regime is to keep the breaking scale
$v_R$ high while taking $g_R$ and $g'_{\rm dem}$ sufficiently small so that
$m_{W_R},m_{Z_R}\sim g\,v_R$ are tiny, thereby avoiding a late-time thermal phase transition
in the visible cosmological history.

\paragraph{Phenomenological caveat (stated as an assumption).}
Because $U(1)_{\rm dem}$ is long-range, consistency with equivalence-principle and fifth-force
constraints requires that the effective couplings/portals of $A_{\rm dem}$ (and of the
ultralight $W_R^\pm,Z_R$) to macroscopic visible-sector matter are extremely suppressed,
or that an appropriate sequestering/screening mechanism is operative.
In the present paper we treat this as a model assumption that defines the viable parameter
regime, and defer a dedicated phenomenological bound analysis to future work.

\subsubsection{Speculative relation between force ranges and cosmological scales on the two leaves}

A recurring structural feature of the two-leaf picture is that each leaf carries its own Lorentzian metric, while the leaves intersect on a mixed-signature two-dimensional interface. This raises a subtle but important point: any statement comparing a length measured on the gravitational leaf to a length measured on the electroweak leaf is only meaningful once a matching prescription (``gluing'') for the two metric structures is specified on the overlap.

Motivated by the fact that, in BF/Plebanski-type formulations, the fundamental two-form data naturally determine a conformal geometry and additional input is required to fix an overall scale, we record here a phenomenological possibility: on the overlap/interface we assume that the induced metrics agree only up to a (possibly dynamical) conformal factor,
\begin{equation}
g^{(L)}_{ab}\big|_{\mathcal S} \;=\; \Omega^2\, g^{(R)}_{ab}\big|_{\mathcal S}\,,
\qquad
\text{equivalently}\qquad
e^{(L)I}_{a}\big|_{\mathcal S} \;=\; \Omega\, e^{(R)I}_{a}\big|_{\mathcal S}\,,
\label{eq:conformal-gluing}
\end{equation}
where $\mathcal S$ denotes the interface and $\Omega$ is a dimensionless hierarchy parameter (or
a slowly varying field on $\mathcal S$).

With this prescription, proper lengths along directions that can be compared across the overlap
scale as $\ell_L = \Omega\, \ell_R$. One may then contemplate an ``IR/UV correspondence'' between
the two sectors in which the natural correlation scale on each leaf is set by the range of the
corresponding massive gauge sector:
\begin{equation}
R_H^{(R)} \sim \lambda_{\rm soft} \equiv m_{\rm soft}^{-1}\,,
\qquad
R_H^{(L)} \sim \lambda_W \equiv m_W^{-1}\,,
\label{eq:hubble-compton}
\end{equation}
where $R_H^{(R)}$ is the Hubble (or more conservatively: IR curvature/correlation) scale inferred
on the gravitational leaf, while $\lambda_W$ is the weak correlation length set by the electroweak
mass gap on the electroweak leaf.  Combining \eqref{eq:conformal-gluing}--\eqref{eq:hubble-compton}
suggests the hierarchy estimate
\begin{equation}
\Omega \;\sim\; \frac{R_H^{(L)}}{R_H^{(R)}} \;\sim\; \frac{H_0}{m_W} \;\sim\; 10^{-44}\,,
\label{eq:omega-estimate}
\end{equation}
i.e.\ an enormous relative normalization between the two metric structures.

We emphasize that \eqref{eq:conformal-gluing}--\eqref{eq:omega-estimate} are not derived here; they
are a speculative phenomenological ansatz whose role would be to sharpen the question of how
macroscopic gravitational distances can appear as microscopic correlation lengths from the
electroweak-leaf viewpoint (and conversely).  Making this proposal consistent requires:
(i) a concrete dynamical origin for $\Omega$ (e.g.\ an order parameter associated with leaf
localization or corner data), and
(ii) compatibility with the corner/gluing constraint structure on the interface, including the
matching of canonical data and Poisson brackets across the overlap.
\medskip

A related (and more conservative) possibility already discussed in this work is that the right-handed
gravidem sector sits in an ultra-soft regime in which the associated massive vectors are so light
that their Compton lengths exceed the Hubble scale on observable distances, rendering the sector
practically indistinguishable from an unbroken phase as far as range-suppression effects are
concerned.

\subsection{Experimental signatures and observational constraints}
\label{subsec:exp_signatures}

The construction proposed here has two immediate phenomenological pressure points:
(i) the appearance of an unbroken, long-range $U(1)_{\rm dem}$ gauge field $A_{\rm dem}$, in conjunction with an
ultra-soft $SU(2)_R$ sector (light $W_R^\pm$ and $Z_R$ in the regime discussed in \hyperref[sec:discussion_ultrasoft_RH]{Section~\ref{sec:discussion_ultrasoft_RH}});
and (ii) the possibility of chirality-selective couplings of the effective ``gravitational'' interactions.
Both features generically imply new long-range forces and/or violations of the universality of free fall unless
all portals between the $SU(2)_R\times U(1)_{\rm dem}$ sector and visible matter are extremely suppressed.
(Indeed, the need for such suppression is already noted explicitly in the ultra-soft discussion:
``Consistency with equivalence-principle and fifth-force constraints requires ... extremely suppressed couplings ...'';
see \hyperref[sec:discussion_ultrasoft_RH]{Section~\ref{sec:discussion_ultrasoft_RH}}.)

\paragraph{1. Fifth-force phenomenology from $U(1)_{\rm dem}$ and ultra-soft vectors.}
Any additional light boson that couples (even feebly) to Standard-Model matter induces a macroscopic force.
A convenient experimental parameterization (used widely in the fifth-force literature) is a Yukawa modification
of Newton's law between neutral, unpolarized bodies~\cite{Adelberger2009TorsionBalanceReview}:
\begin{equation}
V(r)\;=\;-\frac{G_N m_1 m_2}{r}\left[1+\alpha\,e^{-r/\lambda}\right],
\label{eq:yukawa_param}
\end{equation}
with range $\lambda = m_{\rm new}^{-1}$ and relative strength $\alpha$.
Short-range torsion-balance experiments test precisely this structure down to sub-millimeter separations and place
stringent limits on gravitational-strength Yukawa interactions at $\lambda\sim 10^{-4}\,{\rm m}$ scales
(e.g.\ E\"ot-Wash has tested the inverse-square law down to $\sim 50~\mu{\rm m}$)~\cite{Lee2020ISL52um,Adelberger2009TorsionBalanceReview}.
If $A_{\rm dem}$ is exactly massless and couples composition-dependently (as would be natural for a new $U(1)$ charge),
then the force is \emph{not} of the universal $m_1 m_2$ form and instead becomes composition dependent:
equivalence-principle tests dominate the constraints~\cite{Touboul2022MICROSCOPE}.

\paragraph{2. Equivalence principle (UFF/WEP) tests as a primary discriminator.}
The MICROSCOPE satellite has tested the weak equivalence principle (WEP) by comparing the free-fall accelerations
of different materials (Ti and Pt) and found no violation at the $\eta\sim 10^{-15}$ level~\cite{Touboul2022MICROSCOPE}.
Any non-universal coupling of the $U(1)_{\rm dem}$ or ultra-soft $SU(2)_R$ vectors to ordinary matter is therefore
forced to be extremely small (or screened) over orbital-to-laboratory scales.
This is the sharpest and cleanest near-term experimental handle on the scenario if any portal to visible matter exists.

\paragraph{3. Chirality- and spin-dependent gravity from right-handed couplings.}
A particularly sharp qualitative prediction would arise if ``only right-handed fermions gravitate'' in the literal,
low-energy sense. Even if macroscopic matter is unpolarized, such a coupling generically induces
parity-odd and spin-dependent forces once one considers polarized samples or relativistic fermions.
Existing torsion-pendulum experiments with polarized electrons already strongly constrain new macroscopic-range
spin-dependent interactions~\cite{Terrano2015SpinElectrons,Adelberger2009TorsionBalanceReview}, and classic polarized-mass
equivalence-principle tests constrain anomalous spin-gravity couplings~\cite{Ritter1990PolarizedMasses}.
In practice, therefore, a viable implementation of right-handed gravitational participation must explain why the
effective low-energy gravitational coupling to ordinary (mostly nonrelativistic) matter appears universal to extremely
high accuracy, while still retaining a right-handed gauge/gravity interpretation in the ultraviolet.

\paragraph{\it Cosmological and GW probes of gravitational chirality.}
Even if chirality-selective couplings of matter to the effective gravitational sector are screened in ordinary
laboratory settings, a generic expectation in parity-odd (or chiral) extensions of gravity is birefringence
between left- and right-circular tensor modes. This opens complementary tests that do not rely on polarized
bulk matter.
On cosmological scales, net tensor chirality can be searched for via parity-odd CMB polarization correlators
(TB and EB) and related signatures \cite{McDonoughAlexander2018ChiralGWString,AlexanderMartin2005BirefringentGW,AlexanderPeskinSheikhJabbari2006LeptogenesisGW,AlexanderYunes2009dCSReview}.
In addition, Alexander and McDonough emphasized that circularly polarized primordial tensor modes can source a
(nonstandard) Stokes-$V$ component of the CMB, providing another parity-sensitive channel (though typically
highly suppressed in minimal inflationary settings) \cite{AlexanderMcDonough2019CMBVmode}.
On astrophysical scales, parity-violating generation/propagation effects lead to amplitude birefringence and
phase modifications in compact-binary waveforms, which can be constrained with LIGO/Virgo/KAGRA and future
detectors, and can be sharpened with multimessenger counterparts \cite{YunesEtAl2010GRBParity,AlexanderYunes2018GWParity}.

\paragraph{\it Parity-violating gravity and (dynamical) Chern--Simons terms.}
A widely used effective-field-theory parameterization of parity violation in gravity is
\emph{(dynamical) Chern--Simons} (dCS) modified general relativity, reviewed in detail by
Alexander and Yunes~\cite{AlexanderYunes2009dCSReview} (see also the original 4D proposal of
Jackiw and Pi~\cite{JackiwPi2003CS}).
In its standard form, the Einstein--Hilbert action is augmented by a pseudoscalar coupling to the
gravitational Pontryagin density,
\begin{equation}
S_{\rm dCS}
\;=\;
\frac{1}{16\pi G}\int d^4x\,\sqrt{-g}\,
\bigg[
R
+\frac{\alpha_{\rm CS}}{4}\,\vartheta\,
{}^{\ast}\!RR
\bigg]
\;+\; S_{\vartheta}\,,
\qquad
{}^{\ast}\!RR \equiv \frac{1}{2}\,\epsilon^{\mu\nu\rho\sigma}
R_{\mu\nu\alpha\beta}R_{\rho\sigma}{}^{\alpha\beta}\,,
\label{eq:dCS_action}
\end{equation}
where $\vartheta$ is a pseudoscalar (axion-like) field and $S_{\vartheta}$ denotes its kinetic/potential sector.
The key structural point is that ${}^{\ast}\!RR$ is a total derivative in 4D; therefore, parity-violating
\emph{dynamical} effects arise only when the coefficient is effectively spacetime dependent
(e.g.\ via $\nabla_\mu\vartheta\neq 0$)~\cite{AlexanderYunes2009dCSReview}.

This bears a suggestive relation to the present BF/MacDowell--Mansouri reduction: in the toy MM expansion,
a curvature-squared topological density already appears as the ``Pontryagin-type'' term
$R^{ab}\wedge\star R_{ab}$ (cf.\ the discussion around $S_{RR}$), and it drops out of local dynamics when its
coefficient is constant.
Accordingly, an effective dCS-like parity-violating modification on our 4D leaf would require an additional
ingredient that promotes the would-be topological sector to a dynamical interaction---for example, a
pseudoscalar order parameter or overlap/interface degree of freedom whose low-energy remnant couples as
$\vartheta\,{}^{\ast}\!RR$.
Deriving (or ruling out) such an induced coupling from the 6D theory---e.g.\ via symmetry breaking data,
corner/interface terms, or anomaly-induced contributions once chiral matter is included---is an interesting
direction for future work.
If present, the resulting parity violation would generically imply birefringent propagation of
gravitational waves and parity-odd correlators in cosmological observables (as reviewed in
Ref.~\cite{AlexanderYunes2009dCSReview}), providing a clean observational handle on chirality-dependent
gravitational dynamics.

\paragraph{4. Additional probes and model-dependent handles.}
Depending on the detailed portal structure and cosmological history, further constraints and/or signatures may arise from:
(i) dark-radiation bounds if $U(1)_{\rm dem}$ remains thermally populated;
(ii) precision electroweak fits if there is any mixing with hypercharge or $SU(2)_L$;
(iii) parity-violating gravitational propagation effects, which can be bounded with gravitational-wave polarimetry,
if the gravitational sector itself violates parity at an observable level.
These probes are highly model-dependent and require specifying the portal operators and cosmological initial conditions.

\subsection{Interface edge-mode algebra and its relation to EPR,  and locality in 6D}
\label{subsec:interface-nonlocality}

\paragraph{Motivation.}
The 6D EPR-resolution proposal of Furquan--Singh--Wesley \cite{Furquan:2025sox} treats electroweak symmetry breaking as a
\emph{quantum-to-classical} transition in which a $(3,3)$-signature parent theory yields two
overlapping $4$D ``leaves'' with one $(1,1)$ plane in common.\footnote{In the notation of
\cite{Furquan:2025sox}, the overlap can be taken to be $(t_1,x_1)$.}
In this picture, (classical) detectors live only on the familiar (gravitational) $4$D leaf, while
quantum systems can probe the full 6D structure, including the second leaf at the weak length/time
scale \cite{Furquan:2025sox}.

Independently, our $SO(3,3)$ BF/Plebanski construction also produces two chiral sectors (and, after
symmetry breaking, two effective low-energy branches) glued on an interface $\mathcal{S}$.
This subsection explains how the \emph{same interface degrees of freedom} that are required for
a well-posed variational principle and canonical gluing in the BF theory provide a natural home for
the ``$H\oplus H'$ mixing'' operators invoked in the Tsirelson-violation discussion of
\cite{Furquan:2025sox}. The key point is that such mixing operators are \emph{not} an arbitrary ansatz
once the interface edge-mode phase space is quantized.

\paragraph{(i) The $H\oplus H'$ structure and cross-terms in CHSH.}
Furquan--Singh--Wesley emphasize that the relevant state space for an EPR pair is the
\emph{direct sum} of Hilbert spaces associated with the two $4$D leaves,
\begin{equation}
\mathcal{H}_{\rm complete} \;=\; \mathcal{H}\;\oplus\;\mathcal{H}' \,,
\end{equation}
and that a generic observable on $\mathcal{H}_{\rm complete}$ may have off-block-diagonal terms.
These off-block-diagonal pieces are precisely what generate the ``cross-terms''
(e.g.\ $r_{AB}$ below) in the CHSH correlator and can, in principle, lead to violations of the
Tsirelson bound \cite{Furquan:2025sox}:
\begin{align}
r_{AB} &= \alpha^* \beta \,\langle \psi_1 | AB | \psi_2\rangle
       + \beta^* \alpha \,\langle \psi_2 | AB | \psi_1\rangle \,,
\label{eq:rAB}\\[2mm]
F &= E(A,B) + E(A',B) + E(A,B') - E(A',B') \nonumber\\
  &= |\alpha|^2 F_1 + |\beta|^2 F_2 + r \,,
\qquad
r=r_{AB}+r_{A'B}+r_{AB'}-r_{A'B'} \,,
\label{eq:CHSH-cross}
\end{align}
with the conclusion that sufficiently nonzero $r$ can push $F$ beyond $2\sqrt2$
(and even up to the PR value $4$ for specific $r$) \cite{Furquan:2025sox}.

\medskip
\noindent
\emph{Interpretive point:} In the BF setting, a nonzero $r$ should be viewed as a \emph{computable
matrix element} of interface operators that map states between the two branches, rather than as an
ad hoc parameter.

\paragraph{(ii) Interface edge modes and gluing constraints in the BF theory.}
On the BF/Plebanski side, the overlap/interface $\mathcal{S}$ requires additional data beyond the
bulk fields in order to (a) restore gauge invariance in the presence of boundaries/corners and
(b) implement consistent gluing of the two leaves' internal frames.
Concretely, one introduces edge modes
\begin{equation}
\varphi(x) \;\in\; SL(2,\mathbb{C})_{\mathcal{S}}
\qquad (x\in\mathcal{S})
\end{equation}
encoding the map between the internal frames of each leaf.
The basic (kinematic) gluing condition on $\mathcal{S}$ takes the form
\begin{equation}
\Sigma^{(+)}\big|_{\mathcal{S}}
\;=\;
U_{\varphi}\,\cdot\, \Sigma^{(-)}\big|_{\mathcal{S}} \,,
\label{eq:gluing-sigma}
\end{equation}
where $U_{\varphi}\in SO(3,1)$ is the Lorentz transformation induced by $\varphi$ on the relevant
tensor representation.
A stronger version noted in the BF analysis includes complex conjugation (encoding the
chiral/anti-chiral matching across the two leaves),
\begin{equation}
\Sigma^{(+)}\big|_{\mathcal{S}}
\;=\;
U_{\varphi}\,\cdot\, \overline{\Sigma^{(-)}}\big|_{\mathcal{S}} \,,
\label{eq:gluing-sigma-conj}
\end{equation}
which is the structure one should keep in mind when comparing to the ``two-leaf'' picture of
\cite{Furquan:2025sox}.

In addition to gluing, the interface supports corner simplicity constraints.
In our notation, these constraints can be expressed as
\begin{equation}
C_a \;=\; B_a^{I} - \frac{1}{\beta}\, S_a^{I}\;\approx\;0 \,,
\qquad a=1,2 \;\;{\rm (tangent\;to\;}\mathcal{S}{\rm )}\,,
\label{eq:corner-simplicity}
\end{equation}
with a decomposition into holomorphic/anti-holomorphic parts (using the induced Hodge dual on
$\mathcal{S}$),
\begin{equation}
C_a^{\pm} \;=\; \tfrac12\left(C_a \pm i \star_{\mathcal{S}} C_a\right)\,.
\label{eq:corner-holo}
\end{equation}
The BF Hamiltonian analysis implies that $C_a^{+}$ is second class (and generates an
$\mathfrak{sl}(2,\mathbb{R})_{\mathcal{S}\parallel}$ structure), whereas $C_a^{-}$ is first class
(and generates $\mathfrak{so}(2)_{\mathcal{S}}$ rotations).
The second-class character manifests through nontrivial Poisson brackets of the form
$\{C_a^{+}(x),C_b^{+}(y)\}\propto \delta^2(x-y)$ with coefficients depending on corner flux/metric
data, and leads to a \emph{non-commutative corner metric algebra} after
Dirac reduction.

Finally, on each leaf the basic canonical pair is (schematically) a chiral connection and its
conjugate momentum, with Poisson bracket (Ashtekar-type variables)
\begin{equation}
\{A^i_a(x),\widetilde{E}^{\,b}_j(y)\}
\;=\;
i\,\delta^b_a\,\delta^i_j\,\delta^{(3)}(x,y)\,,
\label{eq:bulk-PB}
\end{equation}
as recorded in our BF analysis. Restricting \eqref{eq:bulk-PB} to the
interface and performing the Dirac reduction with \eqref{eq:corner-simplicity}--\eqref{eq:corner-holo}
is the minimal step needed to make the interface algebra explicit.

\paragraph{(iii) Minimal interface algebra to quantize (so mixing becomes derived).}
The preceding structures imply that the \emph{minimal} interface phase space that must be quantized
to obtain a derived mixing operator consists of:
\begin{enumerate}
\item the pullbacks to $\mathcal{S}$ of the bulk holonomy--flux variables (or equivalently
$(A,\widetilde{E})$ restricted to $\mathcal{S}$) on each leaf, with brackets descending from
\eqref{eq:bulk-PB};
\item the edge mode $\varphi(x)\in SL(2,\mathbb{C})_{\mathcal{S}}$ implementing the frame map;
\item the first-class gluing constraint \eqref{eq:gluing-sigma} (or \eqref{eq:gluing-sigma-conj}),
and the corner simplicity constraints \eqref{eq:corner-simplicity}--\eqref{eq:corner-holo},
including the Dirac bracket induced by the second-class subset.
\end{enumerate}
Quantizing this reduced algebra produces a boundary/corner Hilbert space carrying an
$SL(2,\mathbb{C})$ representation at the interface, in conjunction with a constrained (non-commutative)
corner geometry sector.

\medskip
\noindent
\emph{Key identification:}
Upon quantization, the edge mode becomes an operator that \emph{intertwines} the two leaf Hilbert
spaces. Abstractly one obtains an operator
\begin{equation}
\widehat{\mathcal{U}}_{\varphi}:\;\mathcal{H}' \rightarrow \mathcal{H}
\end{equation}
(and its adjoint), which provides the canonical candidate for the off-block-diagonal entries in an
observable on $\mathcal{H}\oplus\mathcal{H}'$:
\begin{equation}
\widehat{\mathcal{O}}
\;\sim\;
\begin{pmatrix}
\widehat{\mathcal{O}}_{\mathcal{H}} & \widehat{\mathcal{M}}\\[1mm]
\widehat{\mathcal{M}}^{\dagger} & \widehat{\mathcal{O}}_{\mathcal{H}'}
\end{pmatrix},
\qquad
\widehat{\mathcal{M}} \;\equiv\; \widehat{\mathcal{U}}_{\varphi}\;\;(\text{optionally combined with }\widehat{\rm complex\;conjugation})\,.
\label{eq:mixing-operator}
\end{equation}
In this reading, the CHSH cross-terms \eqref{eq:rAB}--\eqref{eq:CHSH-cross} are expectation values
of products of such interface intertwiners between leaf states.

\paragraph{(iv) Why Tsirelson violations are not (yet) seen.}
The nonlocality paper stresses that detectors are classical and confined to our $4$D spacetime
branch, so experimentally accessible observables ``live'' in $\mathcal{H}$ and their eigenstates span
only $\mathcal{H}$ \cite{Furquan:2025sox}. In the BF language this corresponds to restricting to
approximately block-diagonal operators in \eqref{eq:mixing-operator}, i.e.\ suppressing
$\widehat{\mathcal{M}}$.
A concrete BF-theoretic mechanism for this suppression would be that exciting interface/corner
degrees of freedom requires access to the weak-scale geometry of the second leaf, which is
operationally inaccessible to macroscopic detectors \cite{Furquan:2025sox}.

\paragraph{(v) Phenomenological links and what remains to be done.}
Our present BF paper notes that matching the induced metrics on $\mathcal{S}$ may only be possible up to a
conformal factor $\Omega$, and that combining correlation scales suggests $\Omega\sim H_0/m_W\sim
10^{-44}$; crucially, this is explicitly flagged as an ansatz whose consistency requires (i) a
dynamical origin for $\Omega$ and (ii) compatibility with the interface corner/gluing constraint
structure, including matching of canonical data and Poisson brackets across the overlap. This dovetails with the nonlocality picture, where the ``short distance'' in the
second leaf is a central ingredient in the apparent 4D nonlocality and in the extreme difficulty of
testing finite collapse-propagation times (e.g.\ $t\sim 10^{-26}\,$s) in Bell experiments
\cite{Furquan:2025sox}.

A second clear point of contact is the long-range $U(1)_{\rm dem}$ sector: the nonlocality paper
invokes a ``dark photon'' propagating on the second leaf as a candidate mediator of collapse
information \cite{Furquan:2025sox}, while the BF construction highlights that an unbroken long-range
$U(1)_{\rm dem}$ in conjunction with an ultra-soft $SU(2)_R$ sector generically implies new long-range
forces and/or violations of the universality of free fall unless portals to visible matter are
extremely suppressed. Thus, existing fifth-force and equivalence-principle bounds
already provide indirect constraints on the plausibility of any collapse-mediation mechanism based on
such sectors.

\medskip
\noindent
\textbf{Theory tasks (minimal, internal to the BF framework):}
(i) derive the full presymplectic potential including the interface/corner terms that introduce
$\varphi$ (rather than postulating it), and compute the reduced Dirac bracket algebra on
$\mathcal{S}$;
(ii) quantize the resulting corner/edge algebra and solve the gluing constraints
\eqref{eq:gluing-sigma} or \eqref{eq:gluing-sigma-conj} to obtain the physical Hilbert space;
(iii) compute the induced magnitude of off-diagonal matrix elements $\widehat{\mathcal{M}}$ in
\eqref{eq:mixing-operator} and assess whether they are parametrically suppressed by $\Omega$ or by a
mass gap.

\medskip
\noindent
\textbf{Experimental handles (indirect but concrete):}
(i) improved bounds on long-range forces or dark photons (and chirality-selective couplings) constrain
any viable ``collapse mediator'' sector; (ii) high-precision Bell tests designed to bound (even if
not resolve) a finite collapse-propagation time scale can be reinterpreted as constraints on the
effective interface-mediated mixing amplitude; and (iii) any laboratory-accessible portal that
couples visible matter to $U(1)_{\rm dem}$ would generically correlate with deviations from standard
quantum predictions, including (in principle) small nonzero cross-term effects in CHSH-type
experiments.

\bigskip

\paragraph{Summary of this Section.}
Absent an explicit portal construction and a quantitative computation of the induced couplings,
the most robust experimental statement is a \emph{constraint}: any appreciable coupling of $A_{\rm dem}$ and/or the
ultra-soft $SU(2)_R$ vectors to visible matter would have already produced detectable equivalence-principle violation
or fifth-force signals~\cite{Touboul2022MICROSCOPE,Lee2020ISL52um,Adelberger2009TorsionBalanceReview}, so such couplings must be
extremely suppressed or screened.
Conversely, if a controlled portal is introduced, the scenario becomes testable via a correlated pattern of:
(1) composition-dependent forces (WEP violation)~\cite{Touboul2022MICROSCOPE}, (2) deviations from the inverse-square law at short range~\cite{Lee2020ISL52um},
and (3) spin- and parity-dependent gravitational effects~\cite{Terrano2015SpinElectrons,Ritter1990PolarizedMasses}.

\section{Outlook and conclusions}
\label{sec:outlook-conclusions}

\paragraph{Summary and main message.}
This study develops and analyzes a gravi--weak construction in which an $SO(3,3)$ BF-type gauge theory on a six-dimensional manifold of split signature, after an algebraic MacDowell--Mansouri-type projection, yields two overlapping four-dimensional Lorentzian sectors. The primary objective was to provide a symmetric, geometric origin for both general relativity and the electroweak $\mathrm{SU}(2)$ sector within a single gauge framework and to identify the algebraic and kinematic ingredients required to realise that objective. Consequently, we implemented an explicit symmetry reduction $\mathrm{SO}(3,3)\to \mathrm{SU}(2)\times\mathrm{SU}(2)$, computed the stabilizer--coset decomposition of the connection and curvature, introduced effective tetrads, and recast each chiral sector in a constrained BF/Plebanski form such that the typical simplicity and reality conditions can be imposed.\\

\paragraph{Primary Findings} The key results can be summarized as follows.
\begin{enumerate}
    \item The MacDowell--Mansouri-type symmetry breaking of the six-dimensional curvature produces two $SU(2)$ chiral sectors. The selection of the non-degenerate branch by the simplicity constraints yields a gravitational sector governed by first-order Plebanski-type dynamics and a second sector whose connection realizes an $\mathrm{SU}(2)$ gauge theory interpretable as weak isospin.
    \item The construction produces two four-dimensional sectors of opposite Lorentz signature that overlap on a mixed-signature interface; the interface supports edge modes and gluing constraints that furnish a presymplectic/canonical structure for matching the chiral data.
    \item Extending the chiral structure to include an additional $U(1)$ factor permits an electroweak embedding and the standard Higgs mechanism on the second spacetime sector while enabling an ultra-soft right-handed sector with a surviving long-range $U(1)_{\mathrm{dem}}$ on the first spacetime sector.
\end{enumerate}

\noindent The study's stated objectives have been met: we presented an explicit algebraic realisation of the symmetry breaking, demonstrated how effective tetrads and Plebanski-type actions result from the projected connection, and identified the interface degrees of freedom necessary for consistent gluing. These results support the central claim that gravitation (as expressed in the Plebanski formulation) and weak isospin can be regarded as distinct low-energy phases resulting from a common six-dimensional gauge structure.

\paragraph{Future Research Directions.}
Although the present study establishes the symmetry-breaking architecture and the BF/Plebanski
reformulation of the left/right sector actions, the following key tasks remain:

\begin{enumerate}
\item \textbf{Systematic analysis of symmetry-breaking order parameters \ (MacDowell--Mansouri vs.\ Higgs).}
MacDowell--Mansouri-type breaking connstitutes an action-level reduction of gauge symmetry via a
selected stabilizer/embedding (equivalently, an ``order parameter'' selecting a subgroup in the
BF language). By contrast, the electroweak Higgs mechanism is a dynamical spontaneous symmetry
breaking within a Yang--Mills--Higgs sector. A natural next step is to clearly
distinguish and relate these mechanisms: identifying which symmetry reductions are structural
(leaf selection and subgroup embedding), which are dynamical (Higgs breaking), and which degrees
of freedom remain physical in each resulting phase.

\item \textbf{Matter content, chirality, and anomalies.}
Given the intrinsically chiral character of the proposal, a complete implementation should include:
(i) explicit coupling of chiral fermions to the appropriate sector in a manner consistent with the BF/Plebanski structure,
(ii) a clear map between ``geometric'' and ``internal'' viewpoints of the weak interaction in
the presence of matter fields, and
(iii) a check of anomaly cancellation within each sector, including the possible role of the interface
in any anomaly inflow-type mechanism.
\end{enumerate}

\paragraph{Concluding remark.}
The two-sector construction developed in this study provides a concrete framework in which the weak interaction admits a dual interpretation: as a geometric structure associated with a second Lorentzian spacetime sector, and as an internal gauge symmetry when viewed from within a single four-dimensional sector. These perspectives are not mutually exclusive but result as complementary descriptions of a common underlying gauge structure. By incorporating both gravitational dynamics and weak isospin within a common $\mathrm{SO}(3,3)$ BF framework, the analysis clarifies how distinctions between spacetime and internal symmetries can emerge as sector-dependent features of a unified structure. In this sense, the proposal reframes the conceptual boundary between geometry and gauge symmetry, and establishes a technically explicit platform from which both its mathematical consistency and phenomenological viability can be further assessed.

\bigskip

\bigskip

\noindent{\bf Acknowledgements}: For helpful discussions and collaboration during the early stages of this work, it is our pleasure to  thank Torsten Asselmeyer-Maluga, Felix Finster, Niels Gresnigt, Antonino Marciano, Claudio Paganini, and Emanuele Zappala.

\pagebreak
\bigskip

\noindent{\bf APPENDIX}
\appendix

\section{Motivation for selfdual and anti-selfdual formulations}
\label{sec:motivationforsdasd}
The adoption of chiral self-dual and anti-selfdual variables in 4D gravity is motivated by algebraic, variational, and topological simplications that are unique to four dimensions and that influence both classical and quantum formulations. At the algebraic level, the space of two-forms admits an intrinsic two-fold splitting into chiral subspaces, which reduces the effective number of independent components and reveals the spinorial content of the theory. From the variational perspective, chiral actions are polynomial in the basic fields and admit a transparent set of constraints whose resolution yields the spacetime metric as a derived object. Topologically, the use of self-dual and antiself-dual variables aligns the gravitational field with well-studied nonperturbative sectors of gauge theory such as instantons (discussed in \hyperref[sec:motivationforsu2gravity]{Appendix~\ref*{sec:motivationforsu2gravity}}), thereby connecting geometric and gauge-theoretic pictures of four-dimensional field configurations. 

A property specific to 4 dimensions is that the Hodge star operator $\star$ maps 2-forms to 2-forms, $\star:\Lambda^{2} \rightarrow \Lambda^{2}$. Because the operator $\star$ squares to the identity $\mathbbm{1}$ on the space of 2-forms $\Lambda^{2}$, we have
\begin{align}
    \omega = \star^{2} \omega = \lambda^{2} \omega \label{starsquare}
\end{align}
where, $\omega\in\Lambda^{2}$. Hence, $\star$ has two eigenvalues: $\lambda=\pm 1$ (Euclidean \& split signatures) or $\lambda=\pm \mathrm{i}$ (Lorentzian signature). The presence of only two eigenvalues results in a decomposition of the space of $\Lambda^{2}$ into self-dual $\Lambda^{+}$ and anti-self-dual $\Lambda^{-}$ eigenspaces: $\Lambda^{2} = \Lambda^{+} \oplus \Lambda^{-}$. The corresponding selfdual and antiselfdual projectors onto these eigenspaces can be expressed as
\begin{align}
    \mathcal{P}_{\pm} = \frac{1}{2} \left( \mathbbm{1} \pm \frac{1}{\lambda} \star \right) \label{selfantiselfproj}
\end{align}
Defining the curvature $F$ as an operator on $\Lambda^{2}$, we can decompose it w.r.t $\star$ using the selfdual and antiselfdual projectors (\ref{selfantiselfproj}) as follows,
\begin{align}
    F = (\mathcal{P}_{+}F\mathcal{P}_{+} + \mathcal{P}_{+}F\mathcal{P}_{-}) + (\mathcal{P}_{-}F\mathcal{P}_{+} + \mathcal{P}_{-}F\mathcal{P}_{-}) \label{curvdecomp}
\end{align}
This decomposition is typically represented as a block matrix \cite{atiyah1978self}\cite{krasnov2020formulations}
\begin{align}
    \mathscr{R} = \begin{pmatrix}
                    A & B \\
                    B^{*} & C
                \end{pmatrix}
\end{align}
where,
\begin{align}
    A &= \mathcal{P}_{+}F\mathcal{P}_{+} \\
    B &= \mathcal{P}_{+}F\mathcal{P}_{-} = (\mathcal{P}_{-}F\mathcal{P}_{+})^{\dagger} \\
    C &= \mathcal{P}_{-}F\mathcal{P}_{-}
\end{align}
This decomposition is useful because the blocks A, B, and C yield specific geometric interpretations: A and C are related to the self-dual or antiself-dual Weyl curvature, whereas the off-diagonal block B encodes the Ricci part of the curvature. Specifically, the vanishing of $B$ is equivalent to the following two statements: (i) the Einstein condition ($R_{\mu\nu} \sim g_{\mu \nu}$) (i.e. the Ricci Tensor contains only the trace parts),
and (ii) the curvature $F$ commutes with the Hodge star operator $\star$ \cite{brans1974complex},
\begin{align}
    \mathcal{P}_{+}F\mathcal{P}_{-} = (\mathcal{P}_{-}F\mathcal{P}_{+})^{\dagger}=0 \Leftrightarrow R_{\mu\nu} \sim g_{\mu \nu} \Leftrightarrow \star F = F \star \label{curvcomstar}
\end{align}
Therefore, Einstein metrics select curvature tensors that act diagonally with respect to the selfdual or antiself-dual splitting. Therefore, gravitational dynamics can be encoded naturally by conditions imposed purely on one chiral block of the curvature.\\ 

\noindent From (\ref{starsquare}) and (\ref{curvcomstar}), it follows that,
\begin{align}
    F = \frac{1}{\lambda^{2}} \star F \star
\end{align}
The selfdual sector is given as,
\begin{align}
    F_{+} &= \mathcal{P}_{+}F\mathcal{P}_{+} + \mathcal{P}_{+}F\mathcal{P}_{-} \\
    &= \frac{1}{2} \left( \mathbbm{1} + \frac{1}{\lambda} \star \right) F \frac{1}{2} \left( \mathbbm{1} + \frac{1}{\lambda} \star \right) + \frac{1}{2} \left( \mathbbm{1} + \frac{1}{\lambda} \star \right) F \frac{1}{2} \left( \mathbbm{1} - \frac{1}{\lambda} \star \right) \\
    &= \frac{1}{4} \left( F + \frac{1}{\lambda} \star F + \frac{1}{\lambda} F \star + \frac{1}{\lambda^{2}} \star F \star \right) + \frac{1}{4} \left( F + \frac{1}{\lambda} \star F - \frac{1}{\lambda} F \star - \frac{1}{\lambda^{2}} \star F \star \right)\\
    &= \frac{1}{4} \left( \mathbbm{1} + \frac{1}{\lambda} \star \right) \left( F + \frac{1}{\lambda^{2}} \star F \star\right) + 0  \\
    &= \frac{1}{2} \mathcal{P}_{+} \left( 2F\right) \\
    &= \mathcal{P}_{+} F \label{selfdualcurv}
\end{align}
where, (\ref{selfdualcurv}) indicates that $F_{+}$ is the selfdual projection of curvature $F$. \\
Likewise, we can show that the antiselfdual sector is given as,
\begin{align}
    F_{-} &= \mathcal{P}_{-}F\mathcal{P}_{+} + \mathcal{P}_{-}F\mathcal{P}_{-} \\
    &= \mathcal{P}_{-} F \label{antiselfdualcurv}
\end{align}
It follows from (\ref{curvdecomp}), (\ref{selfdualcurv}), and (\ref{antiselfdualcurv}) that the curvature $F$ decomposes as,
\begin{align}
    F = F_{+} + F_{-}
\end{align}
Moreover, the selfdual and antiselfdual decompositions are directly associated with the representation theory of Lorentz group. The Lie algebra of the Lorentz group in the complex domain splits as: $\mathfrak{su}(2) \oplus \mathfrak{su}(2)$ (more discussion in \hyperref[sec:chirality-sl2c-twoleaves]{Appendix~\ref*{sec:chirality-sl2c-twoleaves}}). The real compact form $\mathfrak{su}(2)$ arises as the self-dual (or anti-self-dual) subalgebra on spatial slices or after performing a chiral reduction. Furthermore, the $SU(2)$-valued connection $A$ decomposes as $SL(2,\mathbbm{C})\oplus \overline{SL(2,\mathbbm{C}}$,
\begin{align}
    A = A_{+}+A_{-}
\end{align}
where,
\begin{align}
    (A_{\mu}^{IJ})_{+} = \frac{1}{2}\left(A_{\mu}^{IJ} - i\star A_{\mu}^{IJ} \right) \\
    (A_{\mu}^{IJ})_{-} = \frac{1}{2}\left(A_{\mu}^{IJ} + i\star A_{\mu}^{IJ} \right)
\end{align}
The (anti-) self-dual connection $A_{\pm}$ can be proven to yield (anti-) self-dual curvature $F_{\pm}$. That is,
\begin{align}
    F_{\pm}(A) = F(A_{\pm})
\end{align}
The reduction in algebraic complexity owing to the chiral split is reflected in component counting. A general two-form in four dimensions has six independent components; after splitting into $\Lambda^{+}\oplus\Lambda^{-}$ these become two triplets. When one uses self-dual variables (connections and two-forms valued in an $\mathfrak{su}(2)$ algebra), the number of independent connection components on a spacetime point effectively reduces from 24 (for a general Lorentz connection) to 12 (for the chiral connection), whereas the tetrad retains its 16 components; therefore, the resulting first-order chiral action is more economical in field content relative to nonchiral Palatini or Einstein–Cartan formulations.\\

\noindent We can write the first-order chiral action for gravity in the selfdual sector as,
\begin{align}
    S[e,A^{(+)}] &= \frac{1}{8\pi G} \int \epsilon_{IJKL} \left( (e^{I} \wedge e^{J})\mathcal{P}_{+} \wedge R^{(+)}\mathstrut^{KL} - \frac{\Lambda}{4!} (e^{I} \wedge e^{J} \mathcal{P}_{+} \wedge e^{K} \wedge e^{L}) \right) \\
    &=\frac{1}{8\pi G} \int \epsilon_{IJKL} \left( (e^{I} \wedge e^{J})^{(+)} \wedge R^{(+)}\mathstrut^{KL} - \frac{\Lambda}{4!} (e^{I} \wedge e^{J} \wedge e^{K} \wedge e^{L})^{(+)} \right) \label{firstchiralgr}
\end{align}
which is the first order formulation of General Relativity, polynomial in the fields, comprising only $16 + 12$ components (16 tetrad and 12 connection components). This is significantly economical than other formalisms such as the Palatini and Einstein-Cartan cases that depend on $16+24$ and $10+40$ components, respectively.\\

Furthermore, by implementing a $1+3$ decomposition of spacetime in the action (\ref{firstchiralgr}), the Hamiltonian formulation of Ashtekar can be derived \cite{jacobson1987left}\cite{jacobson1988covariant}, which is an alternative to Ashtekar's approach of starting from the canonical transformation of the phase space of general relativity \cite{ashtekar1987new}. The Hamiltonian analysis of (\ref{firstchiralgr}) demonstrates a subtle advantage of the selfdual formulation over other formalisms.  For instance, in the Palatini formalism, although all the constraints and evolution equations are polynomial in fields, two of the constraints are of second class. To address this, reduced canonical variables are typically introduced. However, the first class constraints no longer retain the polynomial form in terms of these new variables. Conversely, in the selfdual formulation (\ref{firstchiralgr}), the second class constraints can be resolved by a different set of canonical variables such that the first class constraints continue to retain their polynomial form \cite{ashtekar1989cp}. This significantly simplifies the constraint equations and provides a strong motivation for using the self dual formulation.

\pagebreak
\section{Motivation for Gravity as an \texorpdfstring{$\mathrm{SU}(2)$}{SU(2)} Gauge Theory}\label{sec:motivationforsu2gravity}
Gauge theories are typically formulated using the Lagrangian, which is invariant under local transformations of a symmetry group, such as $U(1)$ for electromagnetism, $SU(2)$ for the weak force, and $SU(3)$ for the strong force. In each case, the fundamental dynamical variable is a connection on a principal bundle, and the field strength arises as its curvature. The dynamics are governed by gauge-invariant functionals constructed from this curvature.

Prior studies have demonstrated that General Relativity admits an analogous reformulation, wherein the fundamental variable is not the spacetime metric but an $SU(2)$ connection. By exploiting the spin structure of the Lorentz group and the decomposition of $SO(1,3)$ into self-dual and anti-self-dual sectors, one may replace the metric description by a chiral gauge-theoretic formulation in which gravitational dynamics are encoded in an $SU(2)$ connection and associated two-forms, thereby connecting the principal $SU(2)$-bundle formalism with the Riemannian geometric formulation of gravity. 

The relevance of $SU(2)$ is not accidental. As the double cover of $SO(3)$, it captures the rotational subgroup of the Lorentz group $SO(1,3)$. In canonical formulations of gravity, most notably in the Ashtekar variables, the phase space of gravity is coordinatized by an $SU(2)$ connection on spatial slices and its conjugate densitized triad. The resulting constraint algebra exhibits the structure of a non-Abelian gauge theory supplemented by diffeomorphism constraints. This makes gravity resemble standard gauge theory, making it more amenable to quantization.

\subsection{Motivation 1: Hamiltonian Analysis}
A primary motivation for employing an $SU(2)$ gauge-theoretic formulation of gravity emerges from its canonical structure. Implementing a 3+1 spacetime foliation of the selfdual formulation (\ref{firstchiralgr}) results in the following selfdual action \cite{romano1993geometrodynamics, jacobson1988covariant}: 
\begin{align}
    S = \frac{1}{i} \int d^{4}x \left( \tilde{E}^{a}_{i} \dot{A}^{i}_{a} + \frac{i}{2} \undertilde{N} \epsilon_{ijk} \tilde{E}^{a}_{i}\tilde{E}^{b}_{j}F^{k}_{ab} - N^{a}\tilde{E}^{b}_{i}F^{i}_{ab} + t^{a}A^{i}_{a}\mathcal{D}_{a}(\tilde{E}^{a}_{i}) \right) \label{selfdualashtekar}
\end{align}
Here, $A^{i}_{a}$ is an $SU(2)$ connection on the spatial hypersurface, $\tilde{E}^{a}_{i}$ is the densitized triad, $F^{i}_{ab}$ is the curvature of $A^{i}_{a}$, and $\undertilde{N}$, $N^{a}$, and $t^{i}$ are Lagrange multipliers enforcing the constraints. The action (\ref{selfdualashtekar})  is first-order in time derivatives and makes explicit that $\tilde{E}^{a}_{i}$ and $A^{i}_{a}$ are conjugate variables. \\

\noindent The Poisson brackets in terms of these conjugate variables $\tilde{E}^{a}_{i}$ and $A^{i}_{a}$ are as follows:
\begin{align}
    \{\tilde{E}^{a}_{i}(x),\tilde{E}^{b}_{j}(y)\} &= 0 \\
    \{A^{a}_{i}(x),A^{b}_{j}(y)\} &= 0 \\
    \{A^{a}_{i}(x),\tilde{E}^{b}_{j}(y)\} &= i \delta^{i}_{j} \delta^{b}_{a} \delta^{3}(x-y)
\end{align}
This structure is formally identical to that of a non-Abelian Yang--Mills theory. \\

\noindent The Hamiltonian readily follows from (\ref{selfdualashtekar}) and is given by,
\begin{align}
    H = \frac{1}{i} \int d^{3}x \left( -\frac{i}{2} \undertilde{N} \epsilon_{ijk} \tilde{E}^{a}_{i}\tilde{E}^{b}_{j}F^{k}_{ab} + N^{a}\tilde{E}^{b}_{i}F^{i}_{ab} - t^{a}A^{i}_{a}\mathcal{D}_{a}(\tilde{E}^{a}_{i}) \right) 
\end{align}
The constraints are given by,
\begin{align}
    \mathcal{G}_{i} &= \mathcal{D}_{a}(\tilde{E}^{a}_{i}) = 0 \\
    \mathcal{V}_{a} &= \tilde{E}^{b}_{i}F^{i}_{ab} = 0 \\
    \mathcal{S} &= \epsilon_{ijk}\tilde{E}^{a}_{i}\tilde{E}^{b}_{j}F^{k}_{ab} = 0
\end{align}
Evaluating the Poisson bracket of the Gauss constraint $\mathcal{G}_{i}$ with the fields $\tilde{E}^{a}_{i}$ and $ A^{i}_{a}$  yields 
\begin{align}
    \{\mathcal{G}_{a}(x),\tilde{E}^{b}_{i}(y)\} &= \epsilon_{ijk} \tilde{E}^{a}_{k}(x)\delta^{3}(x-y) \\
    \{\mathcal{G}_{a}(x),\tilde{A}^{b}_{i}(y)\} &= \epsilon_{ijk}\tilde{A}^{a}_{k}(x)\delta^{3}(x-y)
\end{align}
We observe that these generate the infinitesimal $SU(2)$ gauge transformations. From the perspective of canonical quantization, this reformulation is decisive. The basic variables are holonomies of an $SU(2)$ connection and fluxes of $\tilde{E}^{a}_{i}$, precisely paralleling non-Abelian gauge theory. Therefore, the gravitational degrees of freedom are encoded in a connection on a principal $SU(2)$ bundle over the spatial slice, rather than in the spatial metric itself.

This canonical equivalence provides a strong foundational justification for adopting $SU(2)$ as the fundamental gauge group underlying gravitational dynamics.

\subsection{Motivation 2: SU(2) Instantons}
A deeper structural motivation for formulating gravity as an $SU(2)$ gauge theory results from the nontrivial correspondence between self-dual $SU(2)$ Yang--Mills connections and Riemannian geometry. This correspondence is not merely heuristic: in specific settings, self-dual gauge configurations can be shown to coincide with the self-dual part of the Levi--Civita connection of an appropriate metric\cite{krasnov2012gauge}. Such results strongly suggest that the $SU(2)$ gauge structure captures intrinsic geometric content of gravity.

\paragraph{Self-dual Yang--Mills configurations.}
Let $A^i$ be an $SU(2)$ connection with curvature
\begin{align}
    F^i = dA^i + \frac{1}{2}\epsilon^{i}{}_{jk} A^j \wedge A^k.
\end{align}
In Euclidean signature, instanton solutions are defined by the (anti-)self-duality condition
\begin{align}
    F^i = \pm \star F^i,
\end{align}
where $\star$ denotes the Hodge dual with respect to a background metric. These configurations extremize the Yang--Mills action within fixed topological sectors characterized by the second Chern number
\begin{align}
    k = \frac{1}{8\pi^2}\int \mathrm{tr}(F \wedge F).
\end{align}
Therefore, self-dual connections represent finite-action, topologically nontrivial field configurations that minimize the action in each sector.

\paragraph{Equivalence with gravitational connections.}
A remarkable theorem establishes that, the index-1 self-dual $SU(2)$ instanton is gauge-equivalent to the self-dual part of the Levi--Civita connection of a metric conformally related to the standard metric on $S^4$ \cite{atiyah1978self}. This result was employed in the work of Krasnov and collaborators \cite{krasnov2012gauge} and was later emphasized in gauge-theoretic approaches to gravity, demonstrating that certain Yang--Mills instantons are gravitational instantons. In other words, in the self-dual sector, solutions of the Yang--Mills equations coincide with geometric data resulting from Riemannian metrics.

This correspondence is conceptually significant: it indicates that the configuration space of self-dual $SU(2)$ connections overlaps nontrivially with the moduli space of conformal gravitational structures. Consequently, the $SU(2)$ gauge field does not merely mimic gravitational degrees of freedom; it can encode them exactly in appropriate sectors.

\paragraph{Metric reconstruction from $SU(2)$ two-forms.}
The relation between $SU(2)$ gauge data and spacetime geometry becomes even more explicit in the Plebanski formalism. Given a triple of $su(2)$-valued two-forms $\Sigma^i$ satisfying suitable nondegeneracy and simplicity constraints, one may reconstruct a spacetime metric via the Urbantke formula,
\begin{align}
    \sqrt{|g|}\, g_{\mu\nu} \propto 
    \epsilon^{\alpha\beta\gamma\delta}\,
    \epsilon_{ijk}\,
    \Sigma^i_{\mu\alpha}
    \Sigma^j_{\nu\beta}
    \Sigma^k_{\gamma\delta}.
\end{align}
When the $\Sigma^i$ result from tetrads, this metric coincides (up to conformal factor and orientation) with the metric whose self-dual spin connection has curvature $F^i$. Therefore, the spacetime geometry is algebraically reconstructible from $SU(2)$-valued two-form data. The metric is not considered fundamental; rather, it emerges from gauge-theoretic variables.

\paragraph{Implications for a gauge-theoretic description of gravity.}
These results collectively provide a strong rationale for adopting $SU(2)$ as the gauge group underlying gravitational dynamics:

\begin{enumerate}
    \item In the self-dual formulation, the gravitational phase space is naturally expressed in terms of an $SU(2)$ connection and its conjugate momentum, rendering gravity structurally analogous to Yang--Mills theory.
    \item Nonperturbative sectors of $SU(2)$ gauge theory (instantons) possess direct geometric interpretations as gravitational configurations in appropriate contexts.
    \item The fundamental spacetime metric can be reconstructed from $SU(2)$ two-form data, supporting the perspective that geometry can be derived from gauge-theoretic variables.
\end{enumerate}

Within the framework developed in this study, wherein $SO(3,3)$ symmetry breaks to $SU(2)_R \times SU(2)_L$, this correspondence acquires additional significance. One chiral $SU(2)$ factor encodes self-dual gravitational dynamics in the $(3,1)$ sector, whereas the opposite chirality governs the complementary sector. Therefore, the instanton--gravitational correspondence substantiates the choice of $SU(2)$ not as a formal convenience, but as a structure intrinsically capable of carrying gravitational degrees of freedom. For further discussion of the gauge-theoretic perspective on gravity, see for example \cite{atiyah1978self, krasnov2012gauge} and references therein.

\subsection{Advantages}
Formulating gravity as an $SU(2)$ gauge theory yields several conceptual and technical advantages relative to the purely metric description. These advantages are particularly relevant in the context of a unified gauge-theoretic framework.
\begin{itemize}
    \item \textbf{Existence of controlled deformations of General Relativity.}  
    In the chiral $SU(2)$ formulation, the Einstein--Hilbert action is replaced by a gauge-invariant functional of $SU(2)$-valued two-forms and a connection. This approach permits an infinite-parameter family of modified gravity theories, all propagating two local degrees of freedom similar to that in GR, but comprising additional topological coupling constants. This indicates that GR is not the only consistent theory of massless spin-2 fields; rather, it occupies a distinguished point within a broader gauge-theoretic landscape. 

    \item \textbf{Polynomial and compact action functional.}  
    As explained in \hyperref[sec:motivationforsdasd]{Appendix~\ref*{sec:motivationforsdasd}}, in the $SU(2)$ (Plebanski or self-dual) formulation, the gravitational action is polynomial in the fundamental fields. By contrast, the metric Einstein--Hilbert action is non-polynomial owing to the presence of the inverse metric and the square root of the determinant. Therefore, the gauge-theoretic action possesses a structurally simpler Lagrangian density, which is advantageous both conceptually and technically, particularly in perturbative expansions and in the analysis of constraint algebras.

    \item \textbf{Canonical structure analogous to Yang--Mills theory.}  As observed in the Hamiltonian analysis, the phase space is coordinatized by an $SU(2)$ connection $A^i_a$ and its conjugate densitized triad $\tilde{E}^a_i$, with Poisson brackets mirroring those of standard Yang--Mills theory. The Gauss constraint generates internal $SU(2)$ gauge transformations, while the remaining constraints encode diffeomorphism invariance. This parallel with non-Abelian gauge theory renders the gravitational constraint algebra more transparent and facilitates the application of quantization techniques developed for Yang--Mills systems.
\end{itemize}

Collectively, these features indicate that the $SU(2)$ gauge formulation is not merely an alternative rewriting of General Relativity, but a structurally robust framework that uncovers hidden symmetries, admits controlled deformations, and aligns gravity with the broader language of non-Abelian gauge theory. This makes it particularly suitable for the unified graviweak construction developed in the present study.

\pagebreak
\section{Significance of 6D spacetime with signature \texorpdfstring{$(3,3)$}{(3,3)}}\label{sec:significanceof6Dspacetime}

In this Appendix we explain why a 6D spacetime with signature (3,3) is a plausible description of spacetime structure, because of various theoretical reasons. These reasons, as discussed below, are independent of such a spacetime also being the appropriate arena for gravi-weak unification.

A convenient entry point is the standard 4D Minkowski-space Dirac operator. Dirac (1928) sought an operator whose square reproduces the Klein--Gordon dynamics for a relativistic scalar,
\begin{equation}
\hbar^2 \left(-\frac{1}{c^2}\frac{\partial^2\ }{\partial t^2} +\frac{\partial^2\ }{\partial x^2} + \frac{\partial^2\ }{\partial y^2} +\frac{\partial^2\ }{\partial z^2}  \right)\psi = m^2 c^2 \psi
\label{3kg}
\end{equation}
and he proposed a first-order (linear) equation of the form
\begin{equation}
D\psi \equiv i\hbar \left ( \gamma_0 \frac{1}{c} \frac{\partial\ }{\partial t} +\gamma_1 \frac{\partial\ }{\partial x_1} +\gamma_2\frac{\partial\ }{\partial x_2} +\gamma_3 \frac{\partial\ }{\partial x_3}\right)\psi = mc\psi 
\label{3de}
\end{equation}
with the requirement that $D^2\psi=m^2c^2\psi$, so that $D^2$ matches the differential operator appearing in (\ref{3kg}). This forces the objects $\gamma_\mu$ to satisfy the familiar anti-commutation relations
\begin{eqnarray}
\gamma_0^2 = I, \quad \gamma_1^2 = \gamma_2^2 = \gamma_3^2 = - I, \quad
\gamma_\mu\gamma_\nu = - \gamma_\nu \gamma_\mu \quad {\rm for}\quad  \mu\neq \nu
\end{eqnarray}
which immediately shows that the $\gamma_\mu$ cannot be ordinary commuting numbers; matrices provide a concrete realisation. One standard $4\times4$ choice, 
in the Pauli--Dirac representation, is conveniently written using $2\times2$ Pauli matrices:
\begin{equation}
\gamma_0=\gamma^0 = \begin{pmatrix} I\; & 0\\ 0\; & -I\;\end{pmatrix}, \qquad
\gamma_k=-\gamma^k = \begin{pmatrix} 0\; & -\sigma_k\\ \sigma_k\; & 0\;\end{pmatrix}
\end{equation}
With $x_0=ct$ and $\partial_\mu=\frac{\partial\ }{\partial x^\mu}$, one may compress (\ref{3de}) further into
\begin{equation}
D\psi\equiv i\hbar \gamma^\mu\partial_\mu \psi = mc\psi
\end{equation}
and, in the presence of an electromagnetic potential $A^\mu$, the usual minimal coupling prescription
$i\hbar\partial^\mu \rightarrow i\hbar\partial^\mu - eA^\mu$ yields
\begin{equation}
\gamma_\mu \left( i\hbar\partial^\mu - eA^\mu\right)\psi = mc\psi
\end{equation}
Here, $\gamma_\mu\partial^\mu=\gamma^\mu\partial_\mu$ and $\partial^\mu=\frac{\partial\ }{\partial x_\mu}$. The Dirac wavefunction is a four-component spinor,
\begin{equation}
\psi = \begin{pmatrix} \psi_1 \\ \psi_2\\ \psi_3\\ \psi_4\\\end{pmatrix}
\end{equation}
whose entries satisfy the Klein--Gordon equation. The point to emphasise for our purposes is structural: the $\gamma_\mu$ generate a Clifford algebra (anti-commutation and squaring to $\pm I$), specifically $Cl(1,3)$. Since Clifford algebras are intimately tied to quaternionic structures, it is natural to ask whether the Dirac operator admits a quaternionic interpretation. It does, and this is precisely where six dimensions enter.

In a lecture, Michael Atiyah \cite{rf3Atiyah} remarks that the Dirac operator is already implicit in Hamilton's discovery of quaternions. Concretely, using the quaternionic imaginary units $\hat i,\hat j,\hat k$, one can form the 3D ``Dirac'' operator as the spatial gradient
\begin{equation}
D_3 = \hat i \frac{\partial\ }{\partial x_1} + \hat j \frac{\partial\ }{\partial x_2} + \hat k \frac{\partial\ }{\partial x_3} 
\label{3d3}
\end{equation}
whose square gives $D_3^2=-\nabla^2$. Recall the quaternion algebra
\begin{eqnarray}
& q=a_0 + a_1 \hat i + a_2 \hat j + a_3 \hat k \; , \quad \hat i^2 = \hat j^2 = \hat k^2 = -1\; , \quad \hat i \hat j \hat k = -1\nonumber\\
& \hat i \hat j = - \hat j \hat i = \hat k, \quad \hat j \hat k = - \hat k \hat j =\hat  i\; , \quad \hat k \hat i = - \hat i \hat k = \hat j
\end{eqnarray}
with real coefficients $a_\mu$ and ${\mathbb H}$ denoting the quaternion division algebra. Allowing complex coefficients produces biquaternions ${\mathbb C}\otimes{\mathbb H}$. From the Clifford-algebra viewpoint, quaternions correspond to $Cl(0,2)$ and biquaternions to $Cl(2)$. The automorphism group of ${\mathbb H}$ is $SU(2)$, which geometrically encodes spatial rotations; biquaternions, in turn, generate the Lorentz algebra $SL(2,\mathbb C)$, the double cover of $SO(1,3)$ on 4D Minkowski spacetime. (For context, $SL(2, {\mathbb H})$ is the double cover of $SO(1,5)$; we return to related remarks below.) With the quaternionic conjugate $\tilde q=a_0-a_1\hat i-a_2\hat j-a_3\hat k$, the norm is $q\tilde q=a_0^2+a_1^2+a_2^2+a_3^2$, and for purely imaginary quaternions ($a_0=0$) this reduces to the standard 3D Euclidean quadratic form. The quaternion product also reproduces the familiar dot and cross products: writing $q=a_0+{\bf a}$ and $q'=b_0+{\bf b}$ one has
\begin{equation}
qq'  = (a_0 + {\bf a})(b_0 + {\bf b}) = a_0b_0 + a_0 {\bf b} + b_0 {\bf a} - {\bf a}.{\bf b} + {\bf a} \times {\bf b}
\end{equation}
where ${\bf a}.{\bf b}$ and ${\bf a}\times{\bf b}$ are the standard scalar and vector products in 3D. Historically, quaternion algebra thus contains vector analysis as a special re-packaging.

Returning to the 4D relativistic setting, one might try to imitate (\ref{3d3}) by writing a quaternionic 4D gradient operator
\begin{equation}
D_4 =  i \frac{\partial\ }{\partial x_0} + \hat i \frac{\partial\ }{\partial x_1} + \hat j \frac{\partial\ }{\partial x_2} + \hat k \frac{\partial\ }{\partial x_3} 
\end{equation}
but the obstruction is immediate: a Klein--Gordon factorisation needs four mutually non-commuting directions, whereas the commuting scalar imaginary $i=\sqrt{-1}$ does not supply a fourth quaternionic unit. This mismatch is the key clue. There exists, however, a natural quaternionic (more precisely, split-biquaternionic) gradient operator that plays the role of a Dirac operator---but it lives most naturally in six dimensions with split signature $(3,3)$. In that sense, the usual 4D Dirac operator $D$ should be viewed as a reduction/specialisation of a more primitive $D_6$.

Insisting (in Atiyah's spirit) that the Dirac operator is a genuine quaternionic-type gradient operator strongly suggests that the underlying arena is 6D rather than 4D, with the extra two directions being time-like. In our proposal, this 6D spacetime undergoes symmetry breaking and effectively contains two overlapping 4D Lorentzian submanifolds: an ordinary 4D Minkowski spacetime and a second 4D ``anti-Minkowski'' spacetime with flipped signature, sharing one time and one space direction. Geometry on our 4D spacetime is sourced by gravitation, while the geometry on the second 4D spacetime is sourced by the weak interaction. From the 4D viewpoint this second geometry appears as an ``internal'' gauge symmetry, but the claim is that it is, in origin, a spacetime symmetry in disguise. Before symmetry breaking, both sectors belong to a unified 6D gravi-weak phase. The split-signature biquaternionic structure is precisely what naturally supplies the needed $D_6$.\footnote{A related approach by Koivisto et al. employs a biquaternionic extension and a Higgs-like Cartan ``khronon" field that dynamically selects an imaginary time direction, yielding an emergent chiral gravitational sector from an underlying Spin(4) gauge formulation of gravity \cite{koivisto2025spin}.}

This also clarifies why ordinary (real) quaternions are not well adapted to 4D special relativity. The quaternion norm $q\tilde q=\tilde q q=a_0^2+a_1^2+a_2^2+a_3^2$ is positive definite: it is preserved by $SO(4)$ rotations, not by Lorentz transformations preserving $x_0^2-x_1^2-x_2^2-x_3^2$. Introducing biquaternions resolves this. A Hermitian biquaternion is characterised by $\tilde q=q^*$ (complex conjugate $q^*$): it has a real scalar part and an imaginary vector part,
$q_h=a_0+a_1 i\hat i+a_2 i\hat j+a_3 i\hat k$, and its norm becomes Lorentzian:
$q_h\tilde q_h=\tilde q_h q=a_0-a_1^2-a_2^2-a_3^2$. Likewise an anti-Hermitian biquaternion
$q_{ah}=a_oi+a_1\hat i+a_2\hat j+a_3\hat k$ satisfies ${\tilde q}_{ah}=-q^*$ and yields
$q_{ah}\tilde q_{ah}=-a_0^2+a_1^2+a_2^2+a_3^2$, i.e.\ the Lorentzian quadratic form with flipped signature.

Accordingly, one may represent a 4D Minkowski event by the Hermitian biquaternion
\begin{equation}
x = x_0 + i{\bf x}, \qquad {\bf x} = x_1 \hat i + x_2 \hat j + x_3 \hat k
\end{equation}
and realise Lorentz transformations by quaternionic conjugation: $x\mapsto qx\tilde q^{*}$ with $q=u+iv$ (quaternions $u,v$) and $q\tilde q=1$; rotations correspond to $q^*=q$ while boosts correspond to $q^*=\tilde q$. For a detailed quaternionic account of special relativity we refer to Lambek \cite{rf3lambek1, rf3lambek2, rf3lambek3}.

In the same spirit, the Hermitian quaternionic gradient operator associated with Minkowski space is
\begin{equation}
D_{4h} = \frac{\partial \ }{\partial x_0} - i\nabla \equiv  \frac{\partial \ }{\partial x_0} - i\hat i \frac{\partial\ }{\partial x_1} - i\hat j \frac{\partial\ }{\partial x_2} - i\hat k \frac{\partial\ }{\partial x_3} 
\end{equation}
which transforms under Lorentz transformations exactly like $x$: $D_{4h}\rightarrow qD_{4h}\tilde q^{*}$. As is well known, this operator packages Maxwell theory succinctly:
\begin{equation} 
D_{4h}F + J =0
\end{equation}
with $F={\bf B}+i{\bf J}$ and $J=\rho+i{\bf J}$ transforming as $F\rightarrow qF\tilde q^*$ and $J\rightarrow qJ\tilde q^*$.

However, while $D_{4h}$ is well-suited for relativistic electromagnetism, it does not by itself provide a clean quaternionic rewriting of the Dirac equation. That problem motivated a long line of work attempting to build a Dirac theory directly from quaternionic differential operators (Dirac \cite{rf3dirac1945}, Conway \cite{rf3conway1948}, Lambek \cite{rf3lambek2}, Morita \cite{rf3morita1986}). Following Morita's analysis \cite{rf3morita1986}, one uses a special subset of $SL(2,{\mathbb H})$ (realising the 4D Lorentz group) in which elements are naturally described by a unit quaternion (rotation data) in conjunction with a pure quaternion (boost data); here ``pure'' means vanishing scalar part, i.e.\ $\tilde q=-q$.

Morita considers $2\times2$ quaternionic matrices $A\in SL(2,{\mathbb H})$ obeying $A\sigma_1\tilde A=\lambda\sigma_1$ and $A\sigma_3\tilde A=\mu\sigma_3$ with $\lambda=\mu=1$, which may be written as
\begin{equation}
A=\begin{pmatrix} Q & -P\\P & Q\\ \end{pmatrix}, \quad |Q|^2 - |P|^2 =1, \quad \tilde P Q + \tilde Q P =1, \quad R\equiv {Q^{-1}P }= - \tilde{R}
\end{equation}
and act on $X=x^i\Gamma_i$ by $X\mapsto X'=AX\tilde A^{T}$, thereby inducing $x\mapsto x'=\Lambda x$. The relevant $\Gamma$-matrices are
\begin{equation}
\Gamma^0 = \begin{pmatrix} 1 \ & . \\ . \ & 1\end{pmatrix}, \quad \Gamma^1 = \begin{pmatrix} . \ & -\hat i\\ \hat i \ & .\end{pmatrix}, \quad \Gamma^2 = \begin{pmatrix} . \ & -\hat j\\ \hat j \ & .\end{pmatrix}, \quad \Gamma^3 = \begin{pmatrix} . \ & -k \\ k \ & .\end{pmatrix}
\end{equation}  
where the vector part ${\bf \Gamma}$ generates boosts, and rotations are generated by
\begin{equation}
\Lambda^1 = \begin{pmatrix} i\  & . \\ .\  & i\end{pmatrix}, \quad \Lambda^2 = \begin{pmatrix} j \ & . \\ .\  & j\end{pmatrix}, \quad \Lambda^3 = \begin{pmatrix} k & . \\ . & k\end{pmatrix}
\end{equation}
One parametrisation writes $A$ in terms of pure quaternions $p,q$ and also in terms of a unit quaternion $a$ and a pure quaternion $\mu$:
\begin{eqnarray}
& A=\exp B(p,q), \quad B(q,p) = \begin{pmatrix} q\ & -p\\ p \  & q\end{pmatrix},\nonumber\\ & A=R(a)B(\mu), \quad |a|=1=Q/|Q|, \quad B(\mu) = \frac{1}{\sqrt{1+\mu^2}}\begin{pmatrix} 1\ & \mu \\ -\mu\ & 1\\ \end{pmatrix}
\end{eqnarray}
Here $R(a)$ (built from the $\Lambda$'s) encodes rotations while $B(\mu)$ (built from the pure $\Gamma$'s) encodes boosts. With $\tilde\Gamma^0=\Gamma^0$ and $\tilde\Gamma^r=-\Gamma^r$, one verifies the Clifford relation
$\Gamma^i\tilde\Gamma^j+\Gamma^j\tilde\Gamma^i=-2\eta^{ij}$, paralleling the usual Dirac algebra
$\gamma^i\gamma^j+\gamma^j\gamma^i=2\eta^{ij}$. Using a two-component spinor $\phi$ with quaternionic entries, Morita writes the Dirac equation as
\begin{equation} {\tilde{\partial}}\phi = m\sigma_3\phi\hat k, \qquad \tilde{\partial}\equiv \tilde{\Gamma}^i\partial_i  \end{equation}
corresponding to $(\gamma^\mu\partial_\mu+m)\phi=0$ and its complex conjugate. While this exhibits the geometric rôle of quaternions in the gradient operator, it is aesthetically less direct than the 3D operator (\ref{3d3}); moreover, the appearance of $\sigma_3$ and $\hat k$ reflects an essentially arbitrary choice. One could just as well have used $\hat i$ and $\sigma_1$, or $\hat j$ and $\sigma_2$. This arbitrariness is a strong hint that the truly natural setting is larger than 4D.

To arrive at a more intrinsic quaternionic Dirac operator, consider instead the Clifford algebra $Cl(0,3)$ with generators $e_1,e_2,e_3$ satisfying $e_1^2=e_2^2=e_3^2=-1$. The eight basis elements
$(1,e_1,e_2,e_3,e_1e_2,e_2e_3,e_3e_1,e_1e_2e_3)$ split into two quaternionic quartets:
$(1,e_1,e_2,e_1e_2)$ and $(e_1e_2e_3,e_2e_3,e_3e_1,e_3)$. Writing $\omega\equiv e_1e_2e_3$, one has $\tilde\omega=-\omega$ and $\omega^2=1$, so the second quartet is $\omega(1,-e_1,-e_2,-e_1e_2)$. Thus $Cl(0,3)$ is the algebra of split biquaternions $(\mathbb H\oplus\omega\mathbb H)$, also expressible as ${\mathbb C}'\otimes\mathbb H$ with ${\mathbb C}'=(1,\omega)$. Complexifying gives $\mathbb C\otimes(\mathbb H+\omega\mathbb H)$, denoted $Cl(3)$. The two quaternion copies together provide six distinct imaginary units: $(e_1,e_2,e_1e_2)$ and $(-\omega e_1,-\omega e_2,-\omega e_3)$. We denote these as $(\hat i,\hat j,\hat k)$ and $(\hat l,\hat m,\hat n)$, respectively.

The key algebraic feature is that within each triple the units anticommute, while elements from different triples commute. Interpreting these as two commuting 3D vector structures suggests a natural 6D event variable
\begin{equation}
x_6=t_1\hat{i}+t_2\hat{j}+t_3\hat{k}+\omega(x_1\hat{l}+x_2\hat{m}+x_3\hat{n})
\label{3ve6}
\end{equation}
with invariant quadratic form
\begin{equation}
x_6\tilde{x}_6=t_1^2+t_2^2+t_3^2-x_1^2-x_2^2-x_3^2
\label{3ve6sq}
\end{equation}
displaying precisely the split signature $(3,3)$ associated with $SO(3,3)$. Note that $x_6$ is not Hermitian. Writing the components as $(01,02,03,1,2,3)$, the corresponding 6D Dirac (gradient) operator is
\begin{equation}
D_6=\hat{i}\partial_{01}+\hat{j}\partial_{02}+\hat{k}\partial_{03}+\omega(\hat{l}\partial_1+\hat{m}\partial_2+\hat{n}\partial_3)
\label{3D6op}
\end{equation}
and it yields the desired Klein--Gordon operator upon multiplication by the conjugate operator,
\begin{equation}
D_6\tilde{D}_6=\partial_{01}^2+\partial_{02}^2+\partial_{03}^2-\partial_1^2-\partial_2^2-\partial_3^2
\label{3D6sq}
\end{equation}
(while naively squaring $D_6$ itself would introduce extra cross-terms). This motivates the 6D Dirac equation
\begin{equation}
i\hbar D_6\psi = Qc\psi
\end{equation}
in which no separate $\gamma$-matrices are needed: the quaternionic Clifford structure is already doing that job. Here $\psi$ is a six-component Dirac spinor (six complex components). Importantly, the source on the right-hand side should not be interpreted as a 4D mass parameter; rather it is a more general charge $Q$ appropriate to the unified 6D gravi-weak phase, from which electric charge and mass emerge only after electroweak symmetry breaking.

For comparison with the usual 4D description, one can decompose $D_6$ into two natural 4D operators, $D_4$ and $D_4'$:
\begin{equation}
D_4=\hat{i}\partial_{01}+\omega(\hat{l}\partial_1+\hat{m}\partial_2+\hat{n}\partial_3)
\end{equation}
so that
\begin{equation}
D_4\tilde{D}_4=\partial_{01}^2-\partial_1^2-\partial_2^2-\partial_3^2
\end{equation}
and similarly
\begin{equation}
D_4'=\omega \hat{l}\partial_1+\hat{i}\partial_{01}+\hat{j}\partial_{02}+\hat{k}\partial_{03}
\label{eq:fourDirac}
\end{equation}
which again gives the appropriate Lorentzian signature. In this sense, both $D_4$ and $D_4'$ furnish Dirac operators for $M_4$ and $M_4'$. Moreover, Wilson \cite{Wilson6D} shows that using $SO(3,3)\sim SL(4,\mathbb R)$ and the structure of $SL(4,\mathbb R)$ generators (two copies of 4D spacetimes described via gamma matrices and quaternions), one can explicitly map quaternion vectors $\hat i,\hat j,\hat k,\hat l,\hat m,\hat n$ (and quaternion bivectors) to the usual gamma matrices. In particular, the identifications
$\hat{i} \mapsto \gamma^0,\omega\hat{l} \mapsto\gamma^1,\omega\hat{m}\mapsto\gamma^2,\omega\hat{n}\mapsto\gamma^3$
reduce the quaternionic form to the standard 4D Dirac operator.

A natural concern is the appearance of the split-imaginary $\omega$ in $D_6$. Dirac himself pursued a formulation using only real quaternions, rather than biquaternions. In our context, however, the presence of $\omega$ is not merely acceptable but conceptually aligned with the goal: we seek a quantum theory not fundamentally built on classical time, and the noncommutative/time-extended structure implicit in $D_6$ is a feature. Moreover, we have argued earlier that elementary particles (e.g.\ the electron) do not ``experience'' spacetime as classical/real in the same way as macroscopic objects do. One nevertheless recovers a classical spacetime interval and a classical Klein--Gordon operator after forming $x_6\tilde x_6$ and $D_6\tilde D_6$, as in (\ref{3ve6sq}) and (\ref{3D6sq}). This matches the broader theme that spacetime geometry and particle dynamics should be encoded by a common algebraic structure; complex split biquaternions provide a natural candidate. One may also argue that the anti-Hermitian sector of the complex split-biquaternionic structure corresponds to an antimatter-dominated mirror copy of our universe, separating around electroweak symmetry breaking.

There is also substantial prior work on $(3,3)$ geometry and on 6D Dirac theory. Particularly relevant is the paper by Patty and Smalley (1985) \cite{rf3patty}, where a $(3+3)$ spacetime is shown to admit a decomposition into six copies of $(3+1)$ subspaces. Six-dimensional extensions also arise in superluminal generalisations of special relativity: a 6D spacetime can be the minimal setting supporting both superluminal and subliminal $(3+1)$ branches \cite{rf3everett}. Lambek's work \cite{rf3lambek3} also studies quaternions with three temporal dimensions and emphasises the usefulness of two commuting quaternionic vector structures (the same two sets appearing in (\ref{3ve6})). Interpreting three noncommuting ``times'' remains an open physical question; one speculative idea is that they might relate to the three fermion generations (with possible implications for mixing and oscillations). Split-signature 6D spacetimes have been explored further by Cole \cite{rf3cole} and Teli \cite{rf3teli}. Early quaternionic approaches to quantum theory go back at least to Conway (1948) \cite{rf3conway1948}. Kritov (2021) \cite{rf3kritov} shows how $Cl(3,0)$ can be used to obtain two 4D spacetimes with relatively flipped signatures. Dartora and Cabrera (2009) \cite{rf3dartora} analyse the Dirac equation with $SO(3,3)$ symmetry and a non-chiral ``electroweak'' theory. Podolanski (1950) \cite{rf3podo} explicitly frames the geometry of the Dirac equation as six-dimensional. Venancio and Batista (2020) \cite{rf3batista} develop a two-component spinorial formalism using quaternions in 6D. Boyling and Cole (1993) \cite{rf3boyling} study the $(3+3)$ Dirac equation and discuss spatial and temporal spin-$1/2$; see also Brody and Graefe (2011) \cite{rf3brody}. In twistor-theoretic contexts, 6D spacetime has been suggested by Sparling \cite{rf3sparling} and analysed by Mason et al.\ \cite{rf3mason}. For a concrete proof that two copies of 4D Minkowski spacetimes $SL(2,C)$ sit inside $SL(4,R)\sim SO(3,3)$, see Tables 2 and 3 in Section 3 of Wilson \cite{Wilson6D}. The link between Dirac theory and extra time dimensions appears already in 1933 \cite{Schoutena, Schoutenb}. Finally, having two extra timelike dimensions offers an intriguing route toward understanding quantum nonlocality: events that are timelike separated in 6D can appear spacelike separated in the 4D projection, giving an effective ``nonlocal'' 4D appearance \cite{singh6d}. 

From the standpoint of (\ref{3ve6}), the two quaternionic imaginary triples are parity reverses of each other. After symmetry breaking, each triple is naturally associated with an $SU(2)$ automorphism group, with chirality: one identifies an $SU(2)_L$ for one triple and an $SU(2)_R$ for the other. In our picture, the former corresponds to the weak sector living on the second 4D spacetime, while the latter corresponds to the gravitational sector on our 4D spacetime. Equivalently, one may regard $Cl(3)$ as a direct sum of two copies of $Cl(2)$. In many algebraic unification schemes one uses one $Cl(2)$ to build the Lorentz algebra $SL(2,C)$ (three rotations and three boosts), while the second $Cl(2)$ is used only for internal weak isospin rotations---leaving its corresponding ``boosts'' unexplained. In the present interpretation, those missing boosts are precisely accounted for by recognising that the second $Cl(2)$ also describes a second 4D Minkowski spacetime (with flipped signature), whose curvature is sourced by the weak interaction.

Given these points, the case for treating the weak interaction as geometry of a flipped-signature 4D spacetime, and for embedding gravi-weak unification in a $(3+3)$ six-dimensional phase, becomes compelling. Once the Dirac operator is recognised as a (split-)quaternionic gradient operator on $(3,3)$ spacetime, it is also less mysterious why it is so central in geometry and why it is naturally connected to gravitational actions: curvature is quadratic in the connection, and the connection is built from (or tightly linked to) the spacetime gradient operator. In our broader unification proposal, the Lagrangian is bilinear in the Dirac operator on an octonionic space; correspondingly, the unified Lagrangian takes the form of a higher-dimensional generalisation of Einstein gravity. 

\subsection{Chirality, \texorpdfstring{$SL(2,\mathbbm{C})$}{SL(2,C)}, and complex conjugation in the two-leaf picture}
\label{sec:chirality-sl2c-twoleaves}

A potential source of confusion in the two-leaf interpretation is that in ordinary 4D Lorentzian
geometry the ``left'' and ``right'' $SL(2,\mathbbm{C})$ factors are \emph{not} attached to two different
base manifolds: they arise as the two chiral projections of the \emph{same} Lorentz algebra/connection
after complexification. Complex conjugation then relates the two chiral sectors because Lorentzian
self-duality is intrinsically complex.

To see this explicitly, recall (\hyperref[sec:motivationforsdasd]{Appendix~\ref*{sec:motivationforsdasd}}) that in 4D the Hodge operator maps 2-forms to 2-forms,
$\star:\Lambda^{2}\to \Lambda^{2}$, and in Lorentzian signature one has $\lambda=\pm i$ in
\eqref{starsquare}. Consequently $\star^{2}=-\mathbbm{1}$ on $\Lambda^{2}$ and the eigenspaces
$\Lambda^{\pm}$ satisfy
\begin{align}
  \star \omega^{(\pm)} = \pm i\,\omega^{(\pm)}\,,
  \qquad
  \Lambda^{2}=\Lambda^{+}\oplus\Lambda^{-}\,.
\end{align}
The projectors \eqref{selfantiselfproj} then reduce to
\begin{align}
  \mathcal{P}_{\pm}
  =\frac12\left(\mathbbm{1}\mp i\,\star\right),
  \qquad
  \omega^{(\pm)}=\mathcal{P}_{\pm}\omega
  =\frac12\left(\omega\mp i\,\star\omega\right).
  \label{eq:proj-lorentzian}
\end{align}
Because the decomposition uses $\pm i$, complex conjugation swaps the two eigenspaces. In particular,
if $\omega$ is real (in the sense that its components are real in a real coframe), then
\begin{align}
  \overline{\omega^{(+)}}=\omega^{(-)}\,,
  \label{eq:conj-swaps-sd-asd}
\end{align}
since $\overline{i}=-i$ and $\star$ is real-linear.

The same mechanism applies to Lorentz-algebra--valued objects. Writing the internal dual on Lorentz
indices as $(\star X)^{IJ}:=\frac12\,\epsilon^{IJ}{}_{KL}X^{KL}$, any real $\mathfrak{so}(3,1)$-valued
connection 1-form $A^{IJ}$ admits the chiral split
\begin{align}
  A^{(\pm)IJ}
  :=
  \mathcal{P}_{\pm}A^{IJ}
  =
  \frac12\left(A^{IJ}\mp i\,\star A^{IJ}\right),
  \qquad
  A^{IJ}=A^{(+)IJ}+A^{(-)IJ},
  \label{eq:chiral-split-connection}
\end{align}
and (under the usual Lorentzian reality conditions ensuring a real tetrad/metric) one has
\begin{align}
  \overline{A^{(+)IJ}}=A^{(-)IJ}.
  \label{eq:conj-swaps-connection}
\end{align}
Thus, in standard 4D chiral formulations the statement ``left and right $SL(2,\mathbbm{C})$ are
complex conjugates'' is fundamentally the statement that the \emph{two chiral projections of the same
real Lorentz connection} are exchanged by complex conjugation.\\

\noindent By contrast, in the present work we \emph{geometrise} these two chiral halves by distributing them across two
overlapping 4D Lorentzian leaves. Concretely, we place the self-dual sector (gravity) on
$M_{\mathcal I}$ and the anti-self-dual sector (weak) on $M_{\mathcal{II}}$ of flipped signature, and we
relate the two sectors on the overlap/interface by gluing constraints (e.g.\ \eqref{eq:Sglue}). This
raises a genuine new question: complex conjugation is an antilinear operation on fields/representations,
but it is \emph{not} by itself a canonical map between fields defined on two different base manifolds.
If the two leaves were completely independent spin manifolds, there would be no preferred notion of
``complex conjugation across leaves''.

Accordingly, the statement that the $SL(2,\mathbbm{C})$ sector on $M_{\mathcal I}$ and the $SL(2,\mathbbm{C})$
sector on $M_{\mathcal{II}}$ are related by complex conjugation should be understood as requiring
\emph{additional structure} that identifies the relevant real structures between the two leaves on their
overlap. In our setting such structure is available for two related reasons:

\begin{enumerate}
\item[(i)] The two leaves are not unrelated: they arise from symmetry breaking of a single real 6D
$(3,3)$ parent theory. At the 6D level the Lorentz group and its Clifford algebra admit a natural real
structure (cf.\ \hyperref[sec:6dsymmetricphase]{Appendix~\ref*{sec:6dsymmetricphase}}, where $Cl_{3,3}\cong Mat(8,\mathbbm{R})$), and the two 4D leaves inherit
their spin data from this common parent.
\item[(ii)] The overlap/interface $S$ provides a place where one can \emph{impose} an identification of the
chiral bundles. For example, a strengthened (or interpreted) gluing condition may relate the pullbacks
of the chiral 2-forms (and likewise the connections) by combining the $SU(2)$ frame rotation
$U_{\phi}$ with complex conjugation:
\begin{align}
  \left.\tilde\Sigma^{(-)i}\right|_{S}
  \;\sim\;
  U_{\phi}\cdot \left.\overline{\Sigma^{(+)i}}\right|_{S},
  \qquad
  \left.\tilde A^{i}\right|_{S}
  \;\sim\;
  U_{\phi}\cdot \left.\overline{A^{(+)i}}\right|_{S},
  \label{eq:conj-gluing-schematic}
\end{align}
where the symbol ``$\sim$'' indicates that the precise form depends on the chosen reality conditions
and on how the leaf data are pulled back to the overlap.
\end{enumerate}

Finally, it is important to separate two logically distinct operations. The relative signature flip between
the two leaves is a real-linear statement about the induced quadratic form; by itself it does not
generate complex conjugation. Rather, signature flip (in conjunction with the parity-reversed quaternionic
triples in \eqref{3ve6}) motivates assigning opposite chiral sectors to the two leaves. Complex
conjugation then enters because Lorentzian chirality is defined using $\pm i$ (cf.\ \eqref{eq:proj-lorentzian}),
so the two chiral halves are exchanged by the inherited real structure. In this sense, the
left/right (Weyl) conjugacy is best regarded as a relic of the underlying real 6D structure in conjunction with
the choice of how the two induced 4D spin structures are identified on the overlap.

\subsection{CPT, Connes time, and a possible CPT-symmetric completion at the electroweak transition}
\label{appA:CPT_Connes_time}

A natural follow-up question to the two-leaf chiral construction is: \emph{what becomes of the
``other'' chiral half on each 4D spacetime}?  In the main text and in \hyperref[sec:chirality-sl2c-twoleaves]{Appendix~\ref*{sec:chirality-sl2c-twoleaves}} we have adopted
a deliberately asymmetric \emph{packaging} in which the self-dual sector is geometrised on our
(gravitation-dominated) leaf, while the anti-self-dual sector is geometrised on the second
(weak-dominated) leaf of flipped signature, with the two related on the overlap/interface by gluing
constraints (e.g. (\ref{eq:Sglue})).

This subsection clarifies (i) what this does and does not imply about discrete symmetries at the
level of the \emph{dynamics} versus at the level of a \emph{realised cosmological history}, and
(ii) how one may consistently formulate a CPT-symmetric ``partner branch'' (a backward-moving
universe in the sense of \emph{Connes time}) at the electroweak transition, in the spirit of the
CPT-symmetric cosmological scenario of Boyle--Finn--Turok~\cite{BoyleFinnTurok2018CPT}.

\paragraph{Dynamics versus history: what it means to ``violate CPT at creation''.}
One should distinguish two logically different notions:
\begin{enumerate}
\item \textbf{CPT invariance of the fundamental laws (action/Hamiltonian).}
In ordinary local, Lorentz-invariant QFT, CPT is a theorem about the microscopic dynamics.
In the present work the post-transition leaf theories are written in standard local field-theoretic
form (Plebanski-type gravity sector on one leaf and Yang--Mills--Higgs sector on the other), so it
is natural to \emph{expect} CPT to be a symmetry of the effective dynamics on each leaf, once the
precise action of $C$, $P$, $T$ on the leaf data and on the interface/gluing constraints is fixed.

\item \textbf{CPT invariance of a chosen solution/boundary condition (cosmological history).}
Even if the underlying laws are CPT invariant, a \emph{particular} solution can pick an arrow of
time (or select one of multiple branches) and hence fail to be invariant under $T$ or CPT as a
\emph{history}.  In that sense one can speak of a ``CPT-asymmetric creation event'' when the
electroweak transition is treated not merely as a dynamical phase transition but effectively as a
\emph{branching/boundary} that selects only one time orientation.
\end{enumerate}
The second notion is the one relevant to the ``only a forward-moving universe is created'' picture.

\paragraph{Connes time and the meaning of ``backward-moving''.}
As emphasized in the main text (cf. \hyperref[sec:natureoftimeconnestime]{Section~\ref*{sec:natureoftimeconnestime}}), the presence of multiple timelike \emph{coordinates} in the $(3,3)$
phase is \emph{not} to be interpreted as ``time flowing in several directions''; rather the role
of a fundamental flow parameter is assigned to Connes time $\tau$, distinct from coordinate
time(s).  In 4D physics we typically (and tacitly) identify ``flowing time'' with a coordinate, but
in the present programme the intended identification is instead
\[
\text{flowing time} \equiv \tau \quad \text{(Connes time),}
\]
with coordinate time(s) playing a kinematical role.\footnote{This viewpoint is already implicit in
the 6D Dirac-operator motivation of \hyperref[sec:significanceof6Dspacetime]{Appendix~\ref*{sec:significanceof6Dspacetime}}, and it is explicitly stated in the discussion of
Connes time in the main text.}
Hence a ``backward-moving universe'' is to be understood as a second branch in which the fundamental
evolution parameter is reversed, $\tau \mapsto -\tau$, rather than as a naive reversal of a
particular coordinate.

\paragraph{Where are the ``missing'' chiral halves?}
In a standard 4D Lorentzian chiral formulation, both chiral projections exist as parts of the same
complexified Lorentz connection, and (under Lorentzian reality conditions) they are exchanged by
complex conjugation, cf.\ \eqref{eq:chiral-split-connection}--\eqref{eq:conj-swaps-connection} in \hyperref[sec:chirality-sl2c-twoleaves]{Appendix~\ref*{sec:chirality-sl2c-twoleaves}}.  In the present work we have instead chosen
to \emph{distribute} these chiral halves across two leaves.

There are then (at least) two consistent readings:

\begin{enumerate}
\item \textbf{Single-history reading (no extra universe).}
On this reading, the ``other'' chiral half on a given leaf is not an \emph{additional} independent
interaction; rather it is constrained by the same type of reality/compatibility conditions that, in
ordinary chiral GR, relate $A^{(+)}$ and $A^{(-)}$ (and similarly for the $\Sigma$'s).  One may
choose a chiral description for convenience (or because it matches the symmetry-breaking pattern),
without implying that the conjugate sector corresponds to an independently realised macroscopic
dynamics on that \emph{same} leaf.

\item \textbf{Two-branch (CPT-completed) reading.}
On this reading, the asymmetric chiral assignment in our observed post-transition branch is viewed
as a \emph{choice of history} that can be completed to a globally CPT-symmetric scenario by adding a
second branch in which the roles of the chiral sectors (and correspondingly the roles of the two
leaves) are interchanged, in conjunction with charge conjugation and reversal of Connes time.
\end{enumerate}

The second reading is particularly natural in view of the remark in \hyperref[sec:significanceof6Dspacetime]{Appendix~\ref*{sec:significanceof6Dspacetime}} that the anti-Hermitian
sector of the split-biquaternionic structure may be interpreted as an antimatter-dominated mirror
copy separating around electroweak symmetry breaking (cf.\ the discussion following (\ref{eq:fourDirac}).

\paragraph{A schematic CPT map in the two-leaf language.}
A minimal way to implement a CPT-completed picture is to postulate that the electroweak transition
creates \emph{two} branches related by an antilinear involution combining:
(i) reversal of Connes time $\tau \mapsto -\tau$,
(ii) exchange of the leaf roles (gravity-leaf $\leftrightarrow$ weak-leaf),
(iii) complex conjugation of the chiral variables (as in \hyperref[sec:chirality-sl2c-twoleaves]{Appendix~\ref*{sec:chirality-sl2c-twoleaves}}), and
(iv) charge conjugation on the matter sector.\\

\noindent Schematically, one may write a map of the form
\begin{align}
\label{eq:CPT_two_leaf_map}
\mathcal{CPT}:\quad
\Big(\tau;\; M_I,\; \Sigma^{(+)}, A^{(+)};\; M_{II},\; \widetilde{\Sigma}^{(-)}, \widetilde{A}^{(-)};\; \Psi\Big)
\;\longmapsto\; \hspace{12em} \nonumber \\
\Big(-\tau;\; M_I,\; \Sigma^{(-)}, A^{(-)};\; M_{II},\; \widetilde{\Sigma}^{(+)}, \widetilde{A}^{(+)};\; \Psi^{c}\Big),
\end{align}
where $\Psi$ denotes collectively the matter fields (once included), $\Psi^c$ their charge conjugates,
and the precise form of the leaf exchange and of the pullback to the overlap/interface must be chosen
so as to respect the gluing constraints (e.g. Eqn. (\ref{eq:Sglue})) up to the relevant $SU(2)$ frame rotation(s).
The intent of \eqref{eq:CPT_two_leaf_map} is not to fix conventions, but to emphasise the structural
point: in the two-leaf packaging, a CPT transformation naturally mixes (a) the chiral split
$self$/$anti$-$self$ dual, (b) the antilinear real structure (complex conjugation), and (c) the
identification of data across leaves.

With this understanding, the question ``what happens to the anti-self-dual part in our 4D spacetime,
and to the self-dual part in the other 4D spacetime?'' admits a clean answer in the CPT-completed
reading: they are realised as the corresponding dynamical sectors in the CPT-partner branch.

\paragraph{Relation to the Boyle--Finn--Turok CPT-symmetric universe.}
Boyle--Finn--Turok propose that the universe after the Big Bang is the CPT image of the universe
before it, yielding a universe/anti-universe pair and selecting a preferred QFT vacuum on such a
CPT-symmetric spacetime~\cite{BoyleFinnTurok2018CPT}.  The proposal being contemplated here is
analogous in \emph{structure} but differs in the choice of the ``branching surface'': instead of the
Big Bang hypersurface, one may take the electroweak transition (the $(3,3)\rightarrow$ two-leaf
bifurcation) as the event at which a CPT-related pair of branches is created.
On this view, a matter-dominated forward-$\tau$ branch and an antimatter-dominated backward-$\tau$
branch can coexist without requiring any microscopic CPT violation of the post-transition effective
laws; rather the full two-branch cosmological history is CPT symmetric, while each individual branch
exhibits an emergent arrow of time.

\paragraph{Comments on CP and T violation and the strong sector.}
Finally, we record a point of principle relevant to phenomenology:
if the post-transition effective dynamics is CPT invariant in the usual sense, then any nonzero
CP-odd observable in the weak/Yukawa sector corresponds to T violation as well (in the standard local
QFT setting).  A basis-independent quantitative measure of quark-sector CP violation is the
Jarlskog invariant, e.g.\ $J=\Im(V_{ud}V_{cs}V_{us}^\ast V_{cd}^\ast)$, or equivalently the
commutator-invariant built from Yukawa matrices.

In the present paper (which is bosonic and does not yet incorporate the fermion/Yukawa sector), no
numerical prediction for CP/T violation can be extracted \emph{within the BF/MM reduction alone};
such predictions necessarily belong to the extended framework in which the Yukawa textures and their
phases are derived.  By contrast, the QCD sector is vector-like and (in the absence of a $\theta$
term) preserves CP and T; empirically any strong CP violation is known to be extremely small.  Hence,
within a CPT-symmetric completion of the present two-leaf scenario, one expects the dominant observed
CP/T violation to remain tied to the electroweak/Yukawa sector, while the CPT-partner branch carries
the CPT-conjugate pattern.

\pagebreak
\section{Clifford Algebra, Chirality, and Electroweak Theory}\label{sec:6dsymmetricphase}
Clifford algebra in signature $(3,3)$ provides a natural and unified algebraic setting for the orthogonal group $SO(3,3)$, its double cover $\mathrm{Spin}(3,3)$, and the associated spinorial and chiral structures. Within the real Clifford algebra $Cl_{3,3}$, the grade‑1 elements provide an embedding of the vector space $\mathbb{R}^{3,3}$ with its bilinear form, the grade‑2 bivectors realize the Lorentz Lie algebra $\mathfrak{so}(3,3)$, and the even subalgebra $Cl^0_{3,3}$ contains the spin group $\mathrm{Spin}(3,3)$ as products of unit vectors.\\

This section exploits these structural features of $Cl_{3,3}$ to construct the six-dimensional framework used later in the electroweak theory. First, the graded structure of $Cl_{3,3}$ is described explicitly, in conjunction with the realization of the Lorentz algebra $\mathfrak{so}(3,3)$ in terms of bivectors and their spinorial action. Second, nilpotent and idempotent elements are constructed and used to obtain minimal left ideals, which serve as the spinor spaces carrying the spinorial representation of $SO(3,3)$. Finally, a chirality operator and associated projectors are defined, leading to a decomposition of six-dimensional spinors into left- and right-handed components adapted to the symmetry breaking pattern of the electroweak sector.

\subsection{Cl(3,3) and Lorentz Transformations in 6D}
Clifford Algebras facilitate the introduction of Spin groups in arbitrary dimensions and signatures. The familiar $Cl_{1,3}$ spacetime algebra is a linear span of 16 basis elements, including the identity. These are $\{\gamma_{A=1,2,3,... 16}\}=\{\mathbbm{1}_{4}, \gamma_{0},\gamma_{i}, \gamma_{0}\gamma_{i},\gamma_{i}\gamma_{j}, \gamma_{5}\gamma_{0}, \gamma_{5}\gamma_{i}, \gamma_{5}\}$. \\

The unification of gravity and weak isospin groups in a chiral manner that also represents all the left and right acting symmeteries requires a total of 6 copies of the underlying Clifford algebras. Therefore, the choice of a larger $Cl_{3,3}$ algebra is particularly convenient because it contains $Cl_{1,3}$ as subalgebras accommodating the familiar four-dimensional spacetime structure and also provides enough algebraic room to encode the enlarged symmetry content required by the six-dimensional electroweak-gravity framework. The construction of nilpotents, primitive idempotents, and minimal left ideals in $Cl_{3,3}$ then provides a canonical route to defining spinor spaces and their chiral decompositions, while the bivector generators realize the full 15‑parameter Lorentz group SO(3,3) in a purely algebraic way. \\

\noindent The $Cl_{3,3}$ algebra comprises 64 $\Gamma_{i}$'s in seven sets organized by grade:
\begin{align}
    [1]&: \text{Grade=0}&: \{ \mathbbm{1}_{8} \} \\
    [6]&: \text{Grade=1}&: \{  \Gamma^{\mu} \} \\
    [15]&: \text{Grade=2}&: \{  \Sigma^{\mu\nu} = \frac{i}{2}[\Gamma^{\mu},\Gamma^{\nu} ] \}  \\
    [20]&: \text{Grade=3}&: \left\{  \frac{i}{3!} \epsilon_{\alpha \beta \gamma \mu \nu \sigma} \Gamma^{\mu} \Gamma^{\nu} \Gamma^{\sigma} \right\}  \\
    [15]&: \text{Grade=4}&: \left\{  \frac{i}{4!} \epsilon_{\alpha \beta \gamma \mu \nu \sigma \rho} \Gamma^{\gamma} \Gamma^{\mu} \Gamma^{\nu} \Gamma^{\sigma} \right\}  \\
    [6]&: \text{Grade=5}&: \left\{  \frac{i}{5!} \epsilon_{\alpha \beta \gamma \mu \nu \sigma \rho} \Gamma^{\beta} \Gamma^{\gamma} \Gamma^{\mu} \Gamma^{\nu} \Gamma^{\sigma} \right\} \\
    [1]&: \text{Grade=6}&: \left\{  \frac{i}{6!} \epsilon_{\alpha \beta \gamma \mu \nu \sigma \rho} \Gamma^{\alpha} \Gamma^{\beta} \Gamma^{\gamma} \Gamma^{\mu} \Gamma^{\nu} \Gamma^{\sigma} \right\} 
\end{align}

The Clifford Algebra $Cl_{3,3}$ is isomorphic to $Mat(8,\mathbbm{R})$ \cite{lounesto2001clifford}, the real algebra of $8\times 8$ matrices. We can construct six such grade-1 $\Gamma_{i}$'s, which are the generators of Cl(3,3)
\begin{align}
    [\hat{\Gamma}_{3,3}] = \left\{ \widetilde{\Gamma}_{1}, \widetilde{\Gamma}_{2}, \widetilde{\Gamma}_{3}, \Gamma_{1}, \Gamma_{2}, \Gamma_{3}\right\}
\end{align}
such that,
\begin{align}
    \{\Gamma^{\mu}, \Gamma^{\nu} \} = 2 \eta^{\mu \nu} \mathbbm{1}_{8}
\end{align}
where, $\eta_{\mu \nu} = \text{diag}(+1,+1,+1,-1,-1,-1)$. We can see that the $\Gamma_{i}$'s satisfy,
\begin{align}
    \widetilde{\Gamma}_{i}^{2}&=\mathbbm{1}_{8}, \\
    \Gamma_{i}^{2}&=\mathbf{i}_{8}^{2}
\end{align}
Furthermore,
\begin{align}
    \widetilde{\Gamma}_{i} &= -\widetilde{\Gamma}_{i}^{\dagger} \hspace{2em} (i=1,2,3)\\
    \Gamma_{i} &= +\Gamma_{i}^{\dagger} \hspace{2em} (i=1,2,3)
\end{align}
We'll be employing only the first three sets of $\Gamma_{i}$'s. Notably, the 15 grade-2 $\Gamma_{i}$'s can be utilized as generators of the $SO(3,3)$ Lorentz Algebra. \\
Furthermore, in Cl(3,3), we introduce the nilpotents of index 2 as specific linear combinations of the grade-1 generators as
\begin{align}
    \alpha_{1}^{\pm} = \frac{1}{2} \left( \widetilde{\Gamma}_{1} \pm \Gamma_{1}\right) \\
    \alpha_{2}^{\pm} = \frac{1}{2} \left( \widetilde{\Gamma}_{2} \pm \Gamma_{2}\right) \\
    \alpha_{3}^{\pm} = \frac{1}{2} \left( \widetilde{\Gamma}_{3} \pm  \Gamma_{3}\right)
\end{align}
These nilpotents are the building blocks for constructing the idempotents, which are used to generate minimal left ideals. The idempotents are defined as
\begin{align}
    \epsilon_{1}^{+} &= \alpha_{1}^{-}\alpha_{1}^{+} = \frac{1}{2} \left( 1 + \widetilde{\Gamma}_{1} \Gamma_{1}\right) \\
    \epsilon_{2}^{+} &= \alpha_{2}^{-}\alpha_{2}^{+} = \frac{1}{2} \left( 1 +  \widetilde{\Gamma}_{2} \Gamma_{2}\right) \\
    \epsilon_{3}^{+} &= \alpha_{3}^{-}\alpha_{3}^{+} = \frac{1}{2} \left( 1 + \widetilde{\Gamma}_{3} \Gamma_{3}\right)
\end{align}
where, the set $\{\widetilde{\Gamma}_{1} \Gamma_{1}, \widetilde{\Gamma}_{2} \Gamma_{2}, \widetilde{\Gamma}_{3} \Gamma_{3} \}$ are mutually commuting elements, and each of them square to 1. Also, $\epsilon_{1}^{\pm}\epsilon_{1}^{\mp}=\epsilon_{2}^{\pm}\epsilon_{2}^{\mp}=\epsilon_{3}^{\pm}\epsilon_{3}^{\mp}=0$.

A primitive idempotent $e$ in $Cl(3,3)$ has $k=q-r_{q-p}=3$ factors, where $r_{i}$ is the Radon-Hurwitz number defined as $r_{0}=0$, $r_{1}=1$, $r_{2}=r_{3}=2$, $r_{4}=r_{5}=r_{6}=r_{7}=3$, and $r_{i+8}=r_{i}+4$ \cite{ablamowicz2013clifford}. Therefore, the Clifford algebra $Cl_{3,3}$ comprises a set of $N=2^{k}=8$ primitive mutually annihilating idempotents whose sum is 1 \cite{ablamowicz2016structure}\cite{ABLAMOWICZ1998510}:
\begin{align}
    \epsilon_{1}^{+}\epsilon_{2}^{+}\epsilon_{3}^{+} +
    \epsilon_{1}^{+}\epsilon_{2}^{+}\epsilon_{3}^{-} +
    \epsilon_{1}^{+}\epsilon_{2}^{-}\epsilon_{3}^{+} +
    \epsilon_{1}^{+}\epsilon_{2}^{-}\epsilon_{3}^{-} +
    \epsilon_{1}^{-}\epsilon_{2}^{+}\epsilon_{3}^{+} +
    \epsilon_{1}^{-}\epsilon_{2}^{+}\epsilon_{3}^{-} +
    \epsilon_{1}^{-}\epsilon_{2}^{-}\epsilon_{3}^{+} +
    \epsilon_{1}^{-}\epsilon_{2}^{-}\epsilon_{3}^{-} =1\\
    \left(Cl(3,3)\right) \nonumber
\end{align}
These primitive idempotents decompose the Clifford algebra into a direct sum of $N=2^{k}=8$ minimal left ideals,
\begin{align}
    Cl(3,3) = Cl(3,3) \hspace{0.5em} \left(\epsilon_{1}^{+}\epsilon_{2}^{+}\epsilon_{3}^{+} \oplus
    \epsilon_{1}^{+}\epsilon_{2}^{+}\epsilon_{3}^{-} \oplus
    \epsilon_{1}^{+}\epsilon_{2}^{-}\epsilon_{3}^{+} \oplus
    \epsilon_{1}^{+}\epsilon_{2}^{-}\epsilon_{3}^{-} \oplus \right. \nonumber \\
    \left( \epsilon_{1}^{-}\epsilon_{2}^{+}\epsilon_{3}^{+} \oplus
    \epsilon_{1}^{-}\epsilon_{2}^{+}\epsilon_{3}^{-} \oplus
    \epsilon_{1}^{-}\epsilon_{2}^{-}\epsilon_{3}^{+} \oplus
    \epsilon_{1}^{-}\epsilon_{2}^{-}\epsilon_{3}^{-} \right)\\
    = \bigoplus_{a=1}^{8} Cl(3,3) e_{a} \hspace{15em}
\end{align}

\noindent Each of these can be interpreted as an irreducible spinor space. In particular, the dimension and multiplicity of these minimal left ideals are compatible with the matrix isomorphism $Cl_{3,3} \simeq Mat(8,\mathbb{R})$ and the existence of an 8‑component spinor representation. From the physical point of view, this decomposition provides an algebraic realization of the different spinor sectors needed to describe left- and right-acting symmetries in the six-dimensional theory.\\

\noindent The Lorentz Transformations in SO(3,3) comprises 15 independent parameters: 3 spatial rotations, 3 temporal rotations, and 9 distinct boosts. To see this, we'll construct the Lorentz generators in SO(3,3) using the six grade-1 bases of Cl(3,3) associated with the metric $\eta_{\mu \nu} = \text{diag}(+1,+1,+1,-1,-1,-1)$,
\begin{align}
    [\hat{\Gamma}_{3,3}] = \left\{ \widetilde{\Gamma}_{1}, \widetilde{\Gamma}_{2}, \widetilde{\Gamma}_{3}, \Gamma_{1}, \Gamma_{2}, \Gamma_{3}\right\}
\end{align}
Once these grade-1 generators have been fixed, the Lorentz generators in $\mathfrak{so}(3,3)$ are naturally identified with the grade-2 elements,
\begin{align}
    \Sigma^{\mu\nu} = \frac{i}{4}[\hat{\Gamma}^{\mu},\hat{\Gamma}^{\nu} ] \label{genlorentz6d}
\end{align}
which form a 15-dimensional subspace of $Cl^0_{3,3}$. These grade-2 elements (bivectors) act on spinors via the commutator and generate infinitesimal Lorentz transformations, while finite transformations are implemented in spinorial representation via exponentiation,
\begin{align}
    \Lambda_{\frac{1}{2}} = \text{exp}\left[-\frac{i}{2}\omega_{\mu\nu} \Sigma^{\mu\nu} \right] \label{lorentztrnsf6d}
\end{align}

Although the indefinite orthogonal group SO(3,3) is non-compact, it consists of two compact SO(3) subgroups which correspond to temporal and spatial rotations.\\
In the spinorial representation, rotations within the time-like sector are generated by $\widetilde{\Sigma}^{ij} = \frac{i}{4}\big[\widetilde{\Gamma}^{i},\widetilde{\Gamma}^{j}\big]$,
yielding three generators $\widetilde{\Sigma}_i=\{\widetilde{\Sigma}_1,\widetilde{\Sigma}_2,\widetilde{\Sigma}_3\}$ satisfying $[\widetilde{\Sigma}_i,\widetilde{\Sigma}_j]= i\epsilon_{ijk}\widetilde{\Sigma}_k$.
These span an $\mathfrak{so}(3)$ subalgebra associated with the time-like directions.

Similarly, rotations within the space-like sector are generated by
$\Sigma^{ij} = \frac{i}{4}\big[\Gamma^{i},\Gamma^{j}\big]$,
producing $\Sigma_i=\{\Sigma_1,\Sigma_2,\Sigma_3\}$ satisfying
$[\Sigma_i,\Sigma_j]= i\epsilon_{ijk}\Sigma_k$,
which generate the second $\mathfrak{so}(3)$ subalgebra.

The nine boost generators mixing time-like and space-like directions are
$\Pi^{ai} = \frac{i}{4}\big[\widetilde{\Gamma}^{a},\Gamma^{i}\big]$, $(a,i=1,2,3)$, which includes both $\widetilde{\Gamma}_i$ and $\Gamma_i$. 

Collectively, $\{\widetilde{\Sigma}_i,\Sigma_i,\Pi^{ai}\}$ furnish the full $\mathfrak{so}(3,3)$ Lorentz algebra in the spinorial representation.

\subsection{Chirality in 6D}
For even-dimensional spacetimes,\footnote{Refer \cite{kugo1983supersymmetry}\cite{WETTERICH1983177}\cite{shirokov2013pauli} for a discussion on Dirac Spinors in arbitrary N-dimensions.} the Clifford algebra $Cl_{p,q}$ always contains a distinguished even multivector, the volume element $\omega$, obtained as the ordered product of an orthonormal basis of vectors. In the case of $Cl_{3,3}$, where $p+q=6$ and $p-q\equiv 0 \bmod 4$, the volume element $\omega$ (which we denote as $\Gamma_{\mathbf{Ch}}$) belongs to the center of the even subalgebra $Cl^0_{3,3}$, anticommutes with all grade‑1 elements, and satisfies $\omega^2 = +1$.\\

\noindent\textit{Remark.}  The full algebra $Cl_{3,3}$ has a trivial center $Z(Cl_{3,3}) = \mathbb{R} \mathbbm{1}$ (because higher-grade elements fail to commute with the odd generators). However, the pseudoscalar $\omega$ enlarges the center of the even subalgebra to $Z(Cl^0_{3,3}) = \mathbb{R} \mathbbm{1} \oplus \mathbb{R} \omega \cong \mathbb{C}$.\\

\noindent The volume element's action in an irreducible spinor representation defines a chirality operator $\Gamma_{\mathbf{Ch}}$ with eigenvalues $\pm 1$, and the associated projectors ($\mathcal{P}_{L}$ and $\mathcal{P}_{R}$) split the 8-dimensional Dirac spinor space into two 4-dimensional Weyl modules $S_{\pm}$, yielding the chiral spinor representation of $SO(3,3)$. Therefore, chirality and Weyl spinors emerge in a purely algebraic fashion from the internal structure of $Cl_{3,3}$.\footnote{Spinorial variables have also been used more dynamically in quantum-gravity constructions, where frame fields are realized as fermion bilinears and chiral behavior appears in lattice phases \cite{diakonov2011towards}\cite{vladimirov2012phase}.}\\

\noindent In the present formalism, the pseudoscalar or chirality operator is explicitly expressed as,
\begin{align}
    \Gamma_{\mathbf{Ch}} = \mathbf{i}^{\frac{d}{2}-1}\Gamma_{0}\Gamma_{1}\Gamma_{2}\dots \Gamma_{d-1}
\end{align}
For the SO(3,3) case, the pseudoscalar reduces to the expression
\begin{align}
    \Gamma_{\mathbf{Ch}}=-\widetilde{\Gamma}_{1} \widetilde{\Gamma}_{2} \widetilde{\Gamma}_{3} \Gamma_{1}, \Gamma_{2} \Gamma_{3} \label{chiralgamma6}
\end{align}
The properties satisfied by $\Gamma_{\mathbf{Ch}}$ in complexified $Cl(3,3)$ are as follows:
\begin{align}
    \Gamma_{\mathbf{Ch}}=\Gamma_{\mathbf{Ch}}^{\dagger} \\
    \Gamma_{\mathbf{Ch}}^{2}=1 \\
    \left\{ \Gamma_{\mathbf{Ch}},\hat{\Gamma}_{3,3} \right\}=0 \label{chirgammaanticomm}\\
    \left[ \Gamma_{\mathbf{Ch}}, \Sigma^{\mu\nu}\right]=0 \label{chirlorencomm}
\end{align}
where, $\Sigma^{\mu\nu}$ are the generators (\ref{genlorentz6d}) of the $\mathfrak{so}(3,3)$ Lorentz Algebra.\\

As a consequence of (\ref{chirlorencomm}), which follows from (\ref{chirgammaanticomm}), the two eigenspinors of $\Gamma_{\mathbf{Ch}}$ transform without mixing. Therefore, we will define the chiral projectors using (\ref{chiralgamma6}),
\begin{align}
    \mathcal{P}_{L} = \frac{1}{2} \left( \mathbbm{1} - \Gamma_{\mathbf{Ch}}\right) \\
    \mathcal{P}_{R} = \frac{1}{2} \left( \mathbbm{1} + \Gamma_{\mathbf{Ch}}\right)
\end{align}
such that, $\mathcal{P}_{L}^{2}=\mathcal{P}_{L}$, $\mathcal{P}_{R}^{2}=\mathcal{P}_{R}$, and $\mathcal{P}_{L}\mathcal{P}_{R}=\mathcal{P}_{R}\mathcal{P}_{L}=0$. The eigenspinors $\Phi_{L}$ and $\Phi_{R}$ are the chiral spinors given by,
\begin{align}
    \mathcal{P}_{L}\Phi &=  \Phi_{L} \hspace{2em} \Leftrightarrow \hspace{2em} \Gamma_{\mathbf{Ch}}\Phi_{L}=-\Phi_{L} \label{chprojL6d}\\
    \mathcal{P}_{R}\Phi &=  \Phi_{R} \hspace{2em} \Leftrightarrow \hspace{2em} \Gamma_{\mathbf{Ch}}\Phi_{R}=\Phi_{R} \label{chprojR6d}
\end{align}
Hence, we can decompose a Dirac spinor $\Phi$ into its left- and right-handed components using (\ref{chiralgamma6}),
\begin{align}
    \Phi &= \mathcal{P}_{L}\Phi + \mathcal{P}_{R}\Phi \\
    &= \Phi_{L} + \Phi_{R} \\
    &= \begin{pmatrix}
    0  \\
    \Phi_{L}
    \end{pmatrix} + \begin{pmatrix}
    \Phi_{R}  \\
    0
    \end{pmatrix}
\end{align}
Furthermore, the left- and right-handed fields $\Phi_{L}$ and $\Phi_{R}$ transform in the following way under rotations $\vartheta$'s and boosts $\beta$'s:
\begin{align}
    \Phi_{L} \xrightarrow{} \text{exp}\left[ - \hspace{0.2em}\frac{1}{2} \left(\hat{\widetilde{\beta}} \cdot \hat{\Pi} \right) - \frac{\mathbf{i}}{2}\hspace{0.2em}\left( \Vec{\widetilde{\vartheta}} \cdot \Vec{\widetilde{\Sigma}}\right)  - \frac{\mathbf{i}}{2}\hspace{0.2em}\left( \Vec{\vartheta} \cdot \Vec{\Sigma}\right) \right]  \Phi_{L} \\
    \Phi_{R} \xrightarrow{} \text{exp}\left[ + \hspace{0.2em}\frac{1}{2} \left(\hat{\widetilde{\beta}} \cdot \hat{\Pi} \right) - \frac{\mathbf{i}}{2}\hspace{0.2em}\left( \Vec{\widetilde{\vartheta}} \cdot \Vec{\widetilde{\Sigma}}\right)  - \frac{\mathbf{i}}{2}\hspace{0.2em}\left( \Vec{\vartheta} \cdot \Vec{\Sigma}\right) \right]  \Phi_{R}
\end{align}
where,
\begin{align}
    \hat{\widetilde{\beta}} \cdot \hat{\Pi} &= \widetilde{\omega}_{1i} \hspace{0.2em} \widetilde{\Gamma}^{1} \hspace{0.2em} \widetilde{\Gamma}^{i} + \hspace{0.2em} \widetilde{\omega}_{2i} \hspace{0.2em} \widetilde{\Gamma}^{2} \hspace{0.2em} \widetilde{\Gamma}^{i} + \hspace{0.2em} \widetilde{\omega}_{3i} \hspace{0.2em} \widetilde{\Gamma}^{3} \hspace{0.2em} \widetilde{\Gamma}^{i} \\
    \Vec{\widetilde{\vartheta}} \cdot \Vec{\widetilde{\Sigma}} &= \frac{1}{2} \hspace{0.2em} \widetilde{\omega}_{ij} \hspace{0.2em} \widetilde{\Gamma}^{i} \hspace{0.2em} \widetilde{\Gamma}^{j} \\
    \Vec{\vartheta} \cdot \Vec{\Sigma} &= \frac{1}{2} \hspace{0.2em} \omega_{ij} \hspace{0.2em} \Gamma^{i} \hspace{0.2em} \Gamma^{j}
\end{align}

From a geometric viewpoint, the same volume element $\omega$ that defines chirality on spinors also induces a Hodge duality operator on bivectors, leading to a decomposition of the space of 2‑forms into self-dual and anti-self-dual parts under the associated Hodge star. In signature $(3,3)$, this splitting is closely associated with the decomposition of the complexified Lie algebra $\mathfrak{so}(3,3)_\mathbb{C} \simeq \mathfrak{sl}(4,\mathbb{R})_\mathbb{C}$ into two simple ideals. Therefore, both the chiral splitting of spinors and the self/anti-self-dual decomposition of 2-forms result from a common Clifford-algebraic origin. Therefore, this intrinsic chiral structure of $Cl_{3,3}$ provides the correct algebraic environment in which to realize left- and right-handed multiplets and to encode their distinct transformation properties under the enlarged Lorentz and internal symmetry groups.

\subsection{Electroweak Theory and the Lorentz algebra in the second 4D Spacetime}\label{sec:electroweak2ndspacetime}
To construct the Lorentz generators in the second spacetime with flipped Lorentzian signature, we will use the following Gamma matrices:
\begin{align}
    ^{(0,1)}[\hat{\Gamma}_{3,3}] = \left\{ \widetilde{\Gamma}_{1}, \widetilde{\Gamma}_{2}, \widetilde{\Gamma}_{3}, \Gamma_{1} \right\}
\end{align}
Then, the time-like dimensions are invariant under SO(3), giving us the same rotation generators:
\begin{align}
    \widetilde{\Sigma}^{ij} = \frac{i}{4}\big[\widetilde{\Gamma}^{i},\widetilde{\Gamma}^{j}\big]
\end{align}
giving us,
\begin{align}
   \widetilde{\Sigma}_{i}=\left\{\widetilde{\Sigma}_{1},\widetilde{\Sigma}_{2},\widetilde{\Sigma}_{3}   \right\}
\end{align}
Following Electroweak Unification, we can use the gauge symmetry $SU(2)\otimes U(1)$ by adding an overall factor,
\begin{align}
    \Psi \xrightarrow[]{} \text{exp}\left[ i\Phi + \frac{i}{2} \vartheta_{i}\cdot \widetilde{\Sigma}^{i} \right] \Psi 
\end{align}
The Dirac Operator is now written as \cite{rf3dartora},
\begin{align}
    \mathscr{D} = i \hspace{0.6em} ^{(0,1)}\Gamma^{\mu} \left[ ^{(0,1)}\nabla_{\mu} + \text{g}W_{\mu}^{i}\frac{\widetilde{\Sigma}^{i}}{2} +  \text{g}'B_{\mu}\frac{Y}{2} \right] 
\end{align}
where, we define the following orthogonal fields:
\begin{align}
    Z_{\mu} &= \cos{\theta} \cdot W_{\mu}^{3}-\sin{\theta}\cdot B_{\mu} \\
    A_{\mu} &= \sin{\theta}\cdot W_{\mu}^{3}+\cos{\theta}\cdot B_{\mu} \\
    W_{\mu}^{\pm} &= W_{\mu}^{1} \pm iW_{\mu}^{2}
\end{align}
where, $\tan{\theta} = g'/g$ .

\pagebreak
\section{Fundamental Automorphism Invariance and the Emergence of Diffeomorphism Symmetry}\label{sec:algebraautomorphisms}
This appendix formulates, in a more precise way, the idea that the deepest symmetry of the octonionic pre-geometric theory is neither ordinary spacetime diffeomorphism invariance nor ordinary Yang--Mills gauge invariance, but a more primitive covariance under admissible automorphisms of the underlying noncommutative, nonassociative coordinate algebra. The purpose is not to claim that this framework has already been fully derived, but to state a mathematically clean organizing principle which may underlie the graviweak construction discussed in the main text.

\subsection{Pre-geometric data and admissible automorphisms}

Let the pre-geometric configuration be described by a tuple
\[
(\mathcal A,\mathcal E,D,J,\Gamma,\nabla,S),
\]
where $\mathcal A$ is the fundamental coordinate algebra, $\mathcal E$ is the matter module, $D$ is a Dirac-type operator, $J$ and $\Gamma$ denote the real structure and grading when these are defined, $\nabla$ is the generalized connection, and $S$ is the underlying action functional. The point of departure is that $\mathcal A$ is not assumed to be commutative, and in the octonionic setting it is not even assumed to be associative. Classical spacetime coordinates are therefore not fundamental variables.

An \emph{admissible automorphism} is a bijection $\varphi:\mathcal A\to\mathcal A$ such that
\[
\varphi(ab)=\varphi(a)\varphi(b), \qquad \varphi(a^*)=\varphi(a)^*,
\]
and such that $\varphi$ preserves the distinguished algebraic structures used in the construction: the trace pairing, the relevant octonionic multiplication rules, the chosen real structure, and the symmetry-breaking data that identify the physically relevant branch. If $U_\varphi$ denotes the induced action of $\varphi$ on $\mathcal E$, then
\[
\Psi \mapsto \Psi^\varphi := U_\varphi\Psi,
\qquad
D \mapsto D^\varphi := U_\varphi D U_\varphi^{-1},
\qquad
\nabla \mapsto \nabla^\varphi := U_\varphi \nabla U_\varphi^{-1}.
\]

\begin{principle}[fundamental automorphism invariance]
The pre-geometric laws are invariant under all admissible automorphisms:
\[
S[\Psi,\nabla,D;\mathcal A] = S[\Psi^\varphi,\nabla^\varphi,D^\varphi;\mathcal A],
\qquad \forall\,\varphi\in\Autad(\mathcal A).
\]
\end{principle}

This should be viewed as the noncommutative, nonassociative lift of the usual notion of general covariance. In an ordinary manifold description the relevant symmetry is invariance under changes of coordinates. Here, by contrast, the primitive object is the coordinate algebra itself, so the natural replacement for a change of coordinates is an automorphism of that algebra.

\subsection{Infinitesimal form and derivations}

The infinitesimal version of this principle is conveniently expressed in terms of derivations. An admissible derivation is a linear map $X:\mathcal A\to\mathcal A$ obeying
\[
X(ab)=X(a)b+aX(b),
\qquad
X(a^*)=X(a)^*,
\]
in conjunction with compatibility conditions with the octonionic structure and the trace pairing. The admissible derivations form a Lie algebra $\Derad(\mathcal A)$.

For $X\in\Derad(\mathcal A)$ one defines
\[
\delta_X a := X(a),
\qquad
\delta_X \Psi := \rho(X)\Psi,
\qquad
\delta_X \nabla := [\rho(X),\nabla],
\qquad
\delta_X D := [\rho(X),D],
\]
where $\rho(X)$ denotes the induced action of $X$ on the matter module. The infinitesimal covariance condition is then
\[
\delta_X S = 0,
\qquad \forall\,X\in\Derad(\mathcal A).
\]

In the associative limit, this naturally separates into a geometric sector and an internal sector: outer derivations are associated with generalized coordinate transformations, whereas inner derivations are associated with generalized gauge transformations. In the genuinely nonassociative regime, this distinction is expected to be less rigid. The point is precisely that, at the fundamental level, spacetime covariance and internal gauge symmetry need not be independent inputs; they can arise as different classical faces of one deeper automorphism principle.

\subsection{Emergent classical limit and the graviweak sector}

Suppose now that coarse-graining, symmetry breaking, or dynamical condensation selects an approximately associative and approximately commutative branch of the theory. Then one expects an effective factorization of the schematic form
\[
\mathcal A_{\rm pre}
\longrightarrow
C^{\infty}(M_4)\otimes \mathcal A_{\rm int},
\]
where $M_4$ is an emergent four-manifold and $\mathcal A_{\rm int}$ carries the residual internal symmetry data. At this level the automorphism principle should bifurcate as
\[
\Autad(\mathcal A_{\rm pre})
\rightsquigarrow
\Diff(M_4)\ltimes \mathcal G_{\rm int}.
\]
This formula is schematic, but it captures the central claim: ordinary diffeomorphism invariance and ordinary Yang--Mills gauge symmetry are not assumed independently at the fundamental level; rather, they appear as two low-energy descendants of a single pre-geometric covariance principle.

For the graviweak construction in the main text, the relevant broken phase begins with the chiral decomposition of the higher-dimensional gauge algebra,
\[
\mathfrak{so}(3,3)
\longrightarrow
\mathfrak{su}(2)_L \oplus \mathfrak{su}(2)_R \oplus \mathfrak p,
\]
where $\mathfrak p$ denotes the coset directions. The connection correspondingly decomposes into left and right chiral pieces together with coset components. Restricting to the emergent four-dimensional branch, the right-handed part defines a connection
\[
A_R^i \in \Omega^1(M_4,\mathfrak{su}(2)_R),
\qquad
F_R^i = dA_R^i + \frac12\epsilon^i{}_{jk} A_R^j\wedge A_R^k.
\]
The relevant two-forms are self-dual variables,
\[
\Sigma_R^i = P^i{}_{IJ}\,e^I\wedge e^J,
\]
where $P^i{}_{IJ}$ projects two-forms onto the self-dual subspace and $e^I$ is the emergent tetrad.

An effective gravitational branch is then expected to take the constrained BF/Plebanski form
\[
S_R[\Sigma_R,A_R,\phi]
= \int_{M_4}
\Sigma_{R i}\wedge F_R^i
- \frac12\phi_{ij}\,\Sigma_R^i\wedge\Sigma_R^j
+ \Lambda\,\Sigma_{R i}\wedge\Sigma_R^i + \cdots,
\]
where $\phi_{ij}$ imposes the simplicity constraints which select the nondegenerate tetrad branch. Once those constraints hold, the metric is reconstructed as
\[
g_{\mu\nu}=e^I{}_{\mu}e^J{}_{\nu}\eta_{IJ},
\]
and one recovers the self-dual formulation of general relativity.

The important conceptual point is that the broken $SU(2)_R$ sector is \emph{not} to be identified with $\Diff(M_4)$ itself. Rather, it supplies the self-dual gravitational gauge variables whose constrained geometric branch admits a spacetime interpretation. Diffeomorphism invariance then appears only after the tetrad or metric has emerged as the appropriate collective variable.

\subsection{\texorpdfstring{How diffeomorphism invariance emerges from the $SU(2)_R$ branch}
{How diffeomorphism invariance emerges from the SU(2)R branch}}

The sense in which diffeomorphism invariance emerges from the $SU(2)_R$ branch can be made more explicit at the level of the constrained BF variables. Let $\xi$ be a vector field on the emergent manifold $M_4$. The Lie derivative of the right-handed connection can be written as
\[
\Lie_\xi A_R^i = \iota_\xi F_R^i + D_R(\iota_\xi A_R^i),
\]
where $D_R$ is the $SU(2)_R$ covariant derivative and $\iota_\xi$ denotes contraction with $\xi$. Likewise, for the self-dual two-forms one has
\[
\Lie_\xi \Sigma_R^i
= \iota_\xi (D_R\Sigma_R^i)
+ D_R(\iota_\xi \Sigma_R^i)
- \epsilon^i{}_{jk}(\iota_\xi A_R^j)\Sigma_R^k.
\]
These identities show that the infinitesimal action of a spacetime diffeomorphism can be rewritten as a combination of an internal gauge transformation and terms proportional to the field equations. Hence, on the gravitational branch, diffeomorphism symmetry is encoded in the constrained $SU(2)_R$ gauge system in a field-dependent way.

This is the precise sense in which one may say that classical spacetime covariance emerges from the right-handed gauge sector. One should not say that
\[
SU(2)_R \cong \Diff(M_4),
\]
which is false. One should instead say that the right-handed gauge variables admit a constrained BF/Plebanski branch for which the on-shell gauge-covariant transformations reproduce the ordinary diffeomorphism redundancy of emergent spacetime geometry. The chain of ideas is therefore
\[
\Autad(\mathcal A_{\rm pre})
\longrightarrow
\text{chiral gauge sectors}
\longrightarrow
SU(2)_R\ \text{gravitational branch}
\longrightarrow
(e^I{}_{\mu},g_{\mu\nu})
\longrightarrow
\Diff(M_4).
\]

\subsection{Conjectural statement and open tasks}

The above discussion may be summarized as follows.

\begin{conjecture}
There exists a pre-geometric octonionic phase, governed by a noncommutative and nonassociative coordinate algebra $\mathcal A_{\rm pre}$, such that the exact symmetry principle is invariance under admissible automorphisms of $\mathcal A_{\rm pre}$. After symmetry breaking and emergence of an approximately commutative branch, this principle bifurcates into ordinary spacetime diffeomorphism invariance together with the residual internal gauge symmetries. In particular, the broken $SU(2)_R$ sector descends to the self-dual gravitational variables whose constrained BF/Plebanski branch reproduces general relativity.
\end{conjecture}

Several steps remain to be derived in full detail:
\begin{itemize}
\item a precise classification of the admissible automorphisms and derivations of the relevant split-bioctonionic coordinate algebra;
\item an explicit emergence map from the pre-geometric algebra to the effective commutative branch $C^{\infty}(M_4)\otimes \mathcal A_{\rm int}$;
\item a derivation of the simplicity constraints and the nondegenerate tetrad branch from the underlying automorphism-invariant dynamics;
\item a full treatment of matter couplings and of how the left and right chiral sectors glue across the common four-dimensional interface.
\end{itemize}

These are significant open tasks. Nevertheless, the automorphism principle provides a precise way to formulate the intuition that gauge symmetry and general covariance should not be regarded as separate primitive ingredients, but as emergent descendants of one deeper nonassociative notion of covariance.

\bigskip

\bigskip
\pagebreak

\bibliography{ref}

\end{document}